\documentclass[12pt,a4paper]{article}
\usepackage{amsmath,xcolor}
\usepackage[font={small,it}]{caption}
\usepackage[top=1.2in, bottom=1.2in, left=1in, right=1in]{geometry}
\DeclareCaptionStyle{italic}[justification=centering]{labelfont={bf},textfont={it},labelsep=colon}
\usepackage{graphicx,psfrag,epsf}
\usepackage{enumerate}
\usepackage{natbib}
\usepackage{url}
\usepackage{booktabs, subfig, bm, paralist,mathpazo,tikz,todonotes,longtable,microtype,dsfont,rotating} 
\usepackage[pdftex,colorlinks=true]{hyperref}
\definecolor{darkblue}{rgb}{0,0,.6}
\hypersetup{citecolor=darkblue,linkcolor=darkblue,urlcolor=darkblue}
\newcommand{\X}{\mathcal{X}}
\newcommand{\argmax}{\operatornamewithlimits{argmax}}
\newcommand{\blind}{0}
\definecolor{coralred}{rgb}{1.0, 0.25, 0.25}

\graphicspath{{plots/}}
\DeclareMathOperator*{\argmin}{\arg\!\min}
\newsavebox\CBox
\def\textBF#1{\sbox\CBox{#1}\resizebox{\wd\CBox}{\ht\CBox}{\textbf{#1}}}

\definecolor{a0}{rgb}{0.0, 0.5, 0.0}
\definecolor{bistre}{rgb}{0.24, 0.17, 0.12}
\definecolor{amethyst}{rgb}{0.6, 0.4, 0.8}
\definecolor{blue-violet}{rgb}{0.54, 0.17, 0.89}
\definecolor{Rcolor}{RGB}{150,160,190}
\definecolor{blush}{rgb}{0.87, 0.36, 0.51}
\definecolor{brightturquoise}{rgb}{0.03, 0.91, 0.87}
\definecolor{burntorange}{rgb}{0.8, 0.33, 0.0}

\date{}
\AtBeginDocument{}
\begin{document}

\def\spacingset#1{\renewcommand{\baselinestretch}
{#1}\small\normalsize} \spacingset{1}

\if0\blind
{
  \title{\bf Dynamic functional time-series forecasts of foreign exchange implied volatility surfaces}
  \author{Han Lin Shang\thanks{Address: Department of Actuarial Studies and Business Analytics, Level 7, 4 Eastern Road, Macquarie University, Sydney, NSW 2109, Australia; Telephone: +61(2) 9850 4689; Email: hanlin.shang@mq.edu.au; ORCID: \url{https://orcid.org/0000-0003-1769-6430}.}
  \hspace{.2cm}\\
    Department of Actuarial Studies and Business Analytics \\
    Macquarie University \\
     \\
    Fearghal Kearney\thanks{Address: Queen's Management School, Queen's University Belfast, Belfast, United Kingdom; Telephone: +44(0) 28 9097 4795; Email: f.kearney@qub.ac.uk; ORCID: \url{https://orcid.org/0000-0002-3251-8707}.} \\
    Queen's Management School, \\
    Queen's University Belfast}
  \maketitle
} \fi

\if1\blind
{
  \title{\bf Dynamic functional time-series forecasts of foreign exchange implied volatility surfaces}
  \maketitle
} \fi

\bigskip

\begin{abstract}
This paper presents static and dynamic versions of univariate, multivariate, and multilevel functional time-series methods to forecast implied volatility surfaces in foreign exchange markets. We find that dynamic functional principal component analysis generally improves out-of-sample forecast accuracy. More specifically, the dynamic univariate functional time-series method shows the greatest improvement. Our models lead to multiple instances of statistically significant improvements in forecast accuracy for daily EUR-USD, EUR-GBP, and EUR-JPY implied volatility surfaces across various maturities, when benchmarked against established methods. A stylised trading strategy is also employed to demonstrate the potential economic benefits of our proposed approach.
\end{abstract}


\newpage
\spacingset{1.5}

\section{Introduction}
A Foreign Exchange (FX) option is a contract that gives the buyer the right, but not the obligation, to buy or sell a particular currency at a pre-specified rate (exercise price) at a pre-specified future date. Implied volatility (IV) is where the market's expectation of future realised volatility is backed out of traded option prices. Models including \cite{BS73}, assume that IV is constant across all option contracts for the same underlying currency. However, constant IV is not what is observed empirically, and IV varies according to both option delta and time-to-expiry. This concept results in non-flat IV surfaces. Further, the shape of these IV surfaces change from day to day, with \cite{DGS08, KST08} highlighting how important it is to model daily changes in IV surfaces. In order to model and predict these daily IV surface changes we propose a dynamic functional time-series approach, considering daily IV surfaces as the time-series of observations.

Theoretical general equilibrium models have been used by \cite{DV00}, \cite{GT03}, \cite{GLR03} and \cite{BG15}, among others, to explain why we observe non-flat IV surfaces. This strand of literature links agent learning and uncertainty in options markets to macroeconomic and dividend fundamentals to explain IV surface shape. Latent factors are then said to drive changes in the shape of the IV surface over time. \cite{BG14, BG15} also conclude that IV predictability is driven by the learning behaviour of agents in options markets. They highlight that as options constitute forward-looking market information, agents' expectations of future economic scenarios can be determined by interpreting time-to-expiry as investor forecast horizons. \cite{GG06} find predictable patterns in how the S\&P 500 IV surface changes over time, however they conclude that their results do not constitute first-order evidence against the efficient market hypothesis. \cite{BG14} conclude that if the efficient market hypothesis is imposed, the discovery of such predictable patterns could be associated with microstructural imperfections and time-varying risk premia. Following \cite{BG14}, we hypothesise that isolated pockets of option market inefficiency exist, and that these inefficiencies could lead to the presence of predictable patterns for segments of the IV surface. We seek to identify these patterns using functional time-series techniques.

In our analysis, we consider IV smiles of a single maturity as a univariate functional time-series. Nevertheless, IV also varies considerably across time to maturity, as highlighted by \cite{BP09}, among others. Therefore, when forecasting the IV surface (essentially multiple IV smiles of differing maturities for the same underlying), it could be advantageous to incorporate information from other maturity series simultaneously. Our multivariate and multilevel functional time-series methods follow this principle, allowing us to extend the previously proposed univariate functional time-series method. Both multivariate and multilevel functional time-series methods capture correlation among IV smiles of differing maturities. We illustrate their use in producing forecasts of daily FX IV surfaces comprised of a number of maturities. To the best of our knowledge, no study to date evaluates and compares forecast errors associated with univariate, multivariate, and multilevel functional time-series methods. We attempt to fill this gap using both static and dynamic functional principal component approaches.

Functional time-series analysis has become increasingly popular in both implied and realised volatility modelling and forecasting. For instance, \cite{BHK09} test stochastic similarities and differences between IV smiles of one- and three-month maturity equity options by conducting a functional principal component analysis. Further, \cite{MSS11} utilise a functional model to characterise high frequency S\&P 500 volatility trajectories. They find that when combined with forecasting techniques, their model accurately predicts future volatility levels. \cite{KMC15} use the smoothness in the IV smile to obtain a measure of IV steepness in oil options using functional data analysis. In a closely related study to ours, \cite{KCM18} model and forecast foreign exchange IV smile separately using static univariate functional time-series techniques, outperforming benchmark methods from \cite{GG06} and \cite{CT10, CT11}.

Functional principal component analysis is the functional counterpart of traditional discrete principal component analysis. Functional principal component analysis reduces the dimensionality of the data to a small number of latent components, in order to effectively summarise the main features of the data. While discrete principal component analysis searches for an optimal set of orthonogal eigenseries that describe the highest level of variance in a discrete data set, functional principal component analysis seeks to identify eigenfunctions that best describe a set of continuous/functional observations. Advances of functional principal component analysis date back to the early 1940s when \cite{Karhunen46} and \cite{Loeve46} independently developed a theory on the optimal series expansion of a continuous stochastic process. Later, \cite{Rao58} and \cite{Tucker58} provided applications of the Karhunen-Lo\`{e}ve expansion to functional data, by applying multivariate principal component analysis to observed function values. For additional methodological details of functional principal component analysis, we direct the interested reader to survey articles by \cite{Hall11, Shang14} and \cite{WCM16}.

Our paper extends \cite{KCM18} through firstly utilising dynamic functional time-series specifications, to see if their ability to dynamically update their predictions can help improve forecast accuracy. This dynamic approach captures the inherent time-series relationship present between daily observations of IV surfaces. When there is moderate to high temporal dependence, such as between daily observations of IV surfaces, the extracted principal components obtained from the (static) functional principal component analysis are not consistent. This inconsistency leads to inefficient estimators. To overcome this issue, we consider a dynamic approach that extracts principal components based on the long-run covariance between daily observations of IV surfaces, instead of using variance alone. Long-run covariance and spectral density estimation enjoy a vast literature in the case of finite-dimensional time-series, beginning with the seminal work of \cite{Bartlett50} and \cite{Parzen57}, and still the most commonly used techniques resort to smoothing the periodogram at frequency zero by employing a smoothing weight function and bandwidth parameter. Data-driven bandwidth selection methods have received a great deal of attention (see, e.g., \cite{Andrews91} for the univariate time-series context and \cite{RS17} for the functional time-series context). Similar to a univariate time-series framework, the long-run covariance includes the variance function as a component, but it also measures temporal auto-covariance at different positive and negative lags. The estimation of the long-run covariance is the sum of the empirical auto-covariance functions and is often truncated at some finite lag in practice. As shown in \cite{Shang19} and \cite{MGG20}, a dynamic functional principal component analysis is a more suitable decomposition than a static functional principal component. \cite{Shang19} demonstrate this for mortality forecasting, while \cite{MGG20} show this for yield curve forecasting. However, we highlight its use for forecasting FX IV surfaces.

Our second contribution is the use of the multivariate functional principal component regression \citep[see, e.g.,][]{SY17} and multilevel functional principal component regression models \citep[see, e.g.,][]{SSB+16} to model and forecast FX IV surfaces. These multivariate and multilevel approaches overcome the issue that two-dimensional univariate approaches have when modelling IV surfaces, where they proceed by essentially slicing the surface and modelling the smiles of different maturities separately. There are, however, intuitive and theoretical links between smiles of different maturities for the same IV surface, so we adopt the multivariate and multilevel extensions to explicitly incorporate information derived from all smiles simultaneously. Both multilevel and multivariate regression models can jointly model and forecast multiple functional time-series that are potentially correlated with each other. By taking into account this correlation between IV smiles of differing maturities, the point and interval forecast accuracies of the IV surfaces are improved.

The remainder of the paper is organised as follows. In Section~\ref{sec:2}, we introduce the FX options data set. In Section~\ref{sec:3}, we provide a background to the univariate, multivariate, and multilevel functional time-series methods, and present extensions based on the dynamic functional principal component analysis. In Section~\ref{sec:4}, we outline our forecast evaluation procedure and present several forecast error measures. In Section~\ref{sec:5}, we present comparisons of in-sample model fitting and out-of-sample forecast accuracy among the functional time-series approaches and traditional IV surface forecasting benchmarks. In our comparisons, we consider the benchmark methods of \cite{CT11} and \cite{GG06}, as well as the random walk (RW) model and an autoregressive model of order 1 (AR(1)). Using the model confidence set procedure of \cite{HLN11}, we determine a set of superior forecasting models based on the out-of-sample accuracy of predicting daily IV surfaces. Further, we present a stylised trading strategy based on the proposed dynamic functional time-series approach. As a robustness check, we also consider an SPX S\&P 500 equity options dataset to further showcase the use of the functional time-series methods. We outline our conclusions in Section~\ref{sec:6}, along with some thoughts on how the methods presented here can be extended.

\section{Data description}\label{sec:2}

We mirror \cite{KCM18} in utilising aggregated market participant contributed IV quotes, thus circumventing the need to adopt a specific options pricing model or impose no-arbitrage assumptions or calibration methods to back out IV from option prices. The use of contributed IV quotes is a common convention in FX options markets, as unlike equity or commodity options, the contracts are most actively traded over-the-counter between market participants, as opposed to being listed on an exchange. We treat these retrieved IV quotes as economic variables of interest in their own right and model how they change over time. Our data set comprises at-the-money, risk reversal, and butterfly composition IV quotes for Euro/United States Dollar (EUR-USD), Euro/British Pound (EUR-GBP), and Euro/Japanese Yen (EUR-JPY) currency options obtained from Bloomberg. These four currencies account for almost 77\% of total global foreign exchange market turnover according to the \citeauthor{BIS16}'s \citeyearpar{BIS16} report. The same four currencies are also considered by \cite{CT11}.

IV quotes associated with delta values of 5, 10, 15, 25, 35, 50, 65, 75, 85, 90 and 95 are available from Bloomberg. However, in line with \cite{CT11} and \cite{KCM18}, we restrict our forecast prediction to the segments of the surface with the greatest trading intensity; namely, delta values of 10, 25, 50, 75 and 90. Therefore, our models and forecasts focus on these specific delta values. In addition, for each FX rate, we observe three maturities; one month (1M), six months (6M) and two years (2Y). Our data set runs from January 1, 2008 to December 30, 2016.

\begin{figure}[!htbp]
\centering
\includegraphics[width=18.5cm]{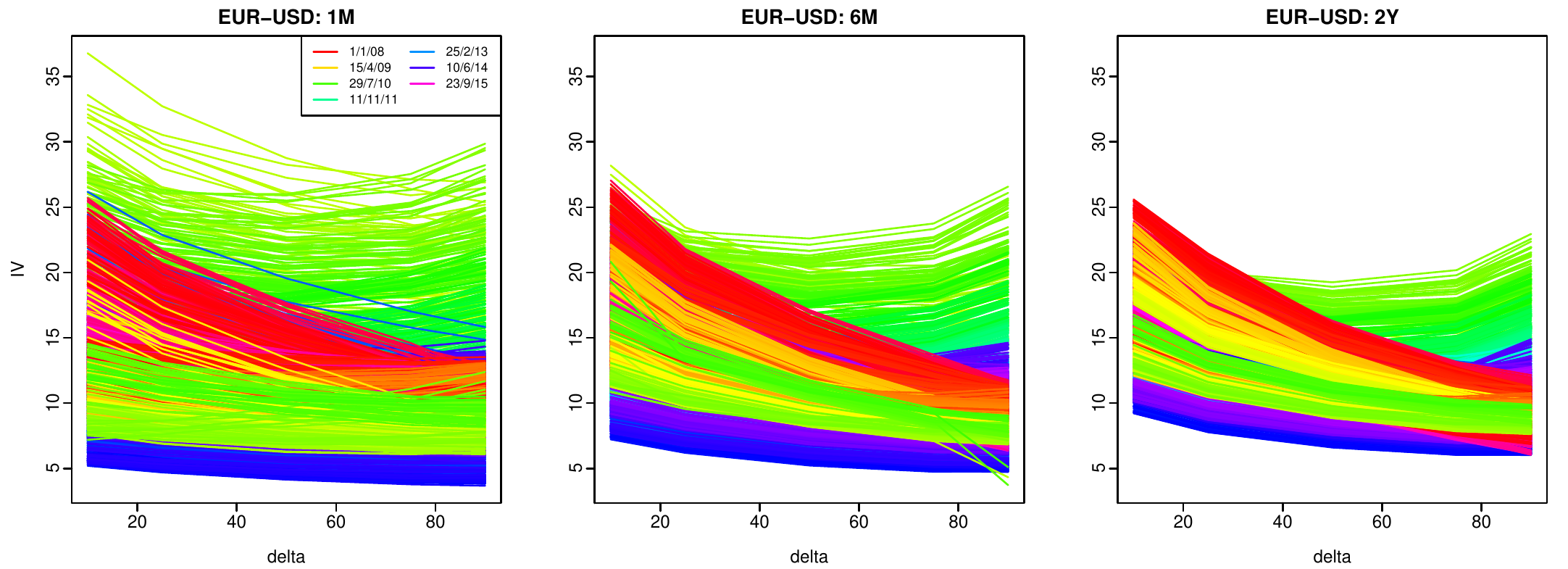}
\caption{\small{This figure presents rainbow plots of the EUR-USD IV smile for the 1-month, 6-month and 2-year maturities from January 1, 2008 to December 30, 2016. The daily smile observations are ordered chronologically according to the colours of the rainbow. The oldest curves are shown in red, with the most recent curves presented in violet.}}\label{fig:1}
\end{figure}

Figure~\ref{fig:1} presents rainbow plots of the EUR-USD IV smiles for the three maturities. The rainbow plot is a plot of all the IV data in our data set, with a rainbow colour palette used to indicate when the observations occur \citep[see also][]{HS10}. Days in the distant past are shown in red, with the most recent days presented in violet. We observe that as the maturity period increases from 1M to 2Y, the spread of IV decreases.

\begin{figure}[!htbp]
\centering
\includegraphics[width = \textwidth]{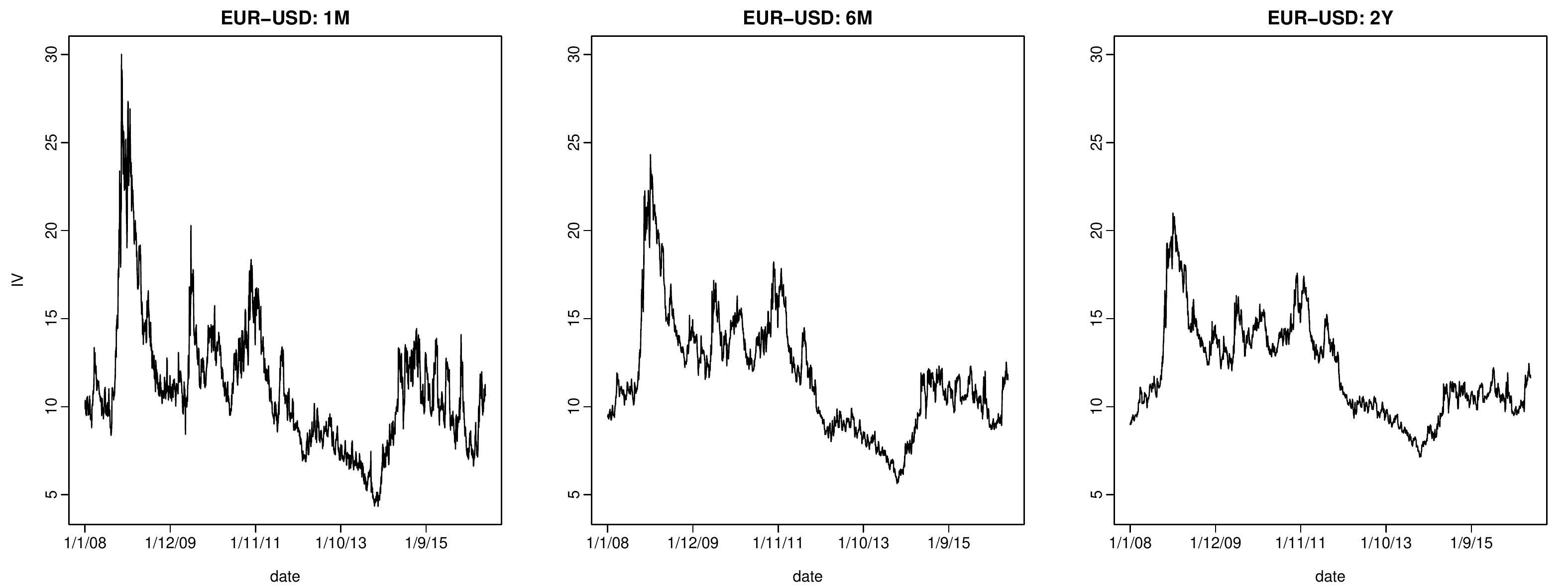}
\caption{\small{This figure presents univariate time-series plots of the EUR-USD IV for the 1-month, 6-month and 2-year maturities averaged over the five delta values of 10, 25, 50, 75 and 90, from January 1, 2008 to December 30, 2016.}}\label{fig:2}
\end{figure}

Figure~\ref{fig:2} presents plots of implied volatilities (IVs) averaged across delta values. They show that the 1M-averaged IVs reached a peak on October 27, 2008 and a trough on August 1, 2014. Both the 6M- and 2Y-averaged IVs reached their peaks on December 17, 2008 and troughs on July 3, 2014.

\section{Methodology}\label{sec:3}

\subsection{Univariate functional time-series method}\label{sec:3.1}

\subsubsection{Functional principal component analysis}\label{sec:fpca}

Let $\{\X_i(t)\}_{i\in \mathbb{Z}}$ be an arbitrary functional time-series defined on a common probability space $(\Omega, A, P)$. It is assumed that the observations $\X_i$ are elements of the Hilbert space $\mathcal{H} = \mathcal{L}^2(\mathcal{I})$ equipped with the inner product $\langle w, v\rangle = \int_{\mathcal{I}}w(t)v(t)dt$, where $t$ represents a continuum and $\mathcal{I}\subset R$ denotes a function support range (i.e., range of delta values), and $R$ is the real line. Each function is a square-integrable function satisfying $\left\|\X_i\right\|^2 = \int_{\mathcal{I}}\X_i^2(t)dt<\infty$ and associated norms. The notation $\X\in L_{\mathcal{H}}^p(\Omega, A, P)$ is used to indicate $\text{E}(\|\X\|^p)<\infty$ for some $p>0$. When $p=1$, $\X(t)$ has the mean curve $\mu(t) = \text{E}[\X(t)]$; when $p=2$, a non-negative definite covariance function is given by 
\begin{align}
c(t,s) &:= \text{Cov}[\X(t), \X(s)] \notag\\
&= \text{E}\left\{[\X(t) - \mu(t)][\X(s) - \mu(s)]\right\}, \label{eq:1}
\end{align}
for all $t, s\in \mathcal{I}$. The covariance function $c(t,s)$ in~\eqref{eq:1} allows the covariance operator of $\X$, denoted by $\mathcal{K}$, to be defined as
\begin{equation*}
\mathcal{K}(\phi)(s) = \int_{\mathcal{I}}c(t,s)\phi(t)dt.
\end{equation*}
Via Mercer's lemma, there exists an orthonormal sequence $(\phi_k)$ of continuous functions in $\mathcal{L}^2(\mathcal{I})$ and a non-increasing sequence $\lambda_k$ of positive numbers, such that
\begin{equation*}
c(t,s) = \sum_{k=1}^{\infty}\lambda_k \phi_k(t) \phi_k(s), \qquad t, s \in \mathcal{I}.
\end{equation*}
By the separability of Hilbert spaces, the Karhunen-Lo\`{e}ve expansion of a stochastic process $\X(t)$ can be expressed as
\begin{align}
\X(t) &= \mu(t) + \sum_{k=1}^{\infty}\beta_k\phi_k(t), \notag \\
&= \mu(t) + \sum^K_{k=1}\beta_{k}\phi_k(t) + e(t), \label{eq:2}
\end{align}
where principal component scores $\beta_k$ are given by the projection of $[\X(t) - \mu(t)]$ in the direction of the $k^{\text{th}}$ eigenfunction $\phi_k$, i.e., $\beta_k = \langle [\X(t) - \mu(t)], \phi_k(t)\rangle$; $e(t)$ denotes the model truncation error function with a mean of zero and a finite variance; and $K<n$ is the number of retained components. Equation~\eqref{eq:2} facilitates dimension reduction as the first $K$ terms often provide a good approximation to the infinite sums, and thus the information inherited in $\bm{\X}(t)=\{\X_1(t),\dots,\X_n(t)\}$ can be adequately summarised by the $K$-dimensional vector $\left(\bm{\beta}_1,\dots,\bm{\beta}_K\right)$. These scores constitute an uncorrelated sequence of random variables with zero mean and variance $\lambda_k$, and they can be interpreted as the weights of the contribution of the functional principal components $\phi_k$ to $[\X(t) - \mu(t)]$.

The estimation accuracy of the functional time-series method crucially relies on the optimal selection of $K$. There are at least five conventional approaches; namely, eigenvalue ratio tests \citep{AH13}, cross-validation \citep{RS05}, Akaike's information criterion \citep{Akaike74}, the bootstrap method \citep{HV06} and explained variance \citep{Chiou12}. Here, the value of $K$ is determined as the minimum that reaches a certain level of the cumulative percentage of variance (CPV) explained by the $K$ leading components such that
\begin{equation*}
K = \argmin_{K:K\geq 1}\left\{\sum^K_{k=1}\widehat{\lambda}_k\Bigg/\sum^{\infty}_{k=1}\widehat{\lambda}_k\mathds{1}_{\left\{\widehat{\lambda}_k>0\right\}}\geq P_1\right\}, 
\end{equation*}
where $P_1 = 0.99$, $\mathds{1}_{\left\{\widehat{\lambda}_k>0\right\}}$ is to exclude possible zero eigenvalues, and $\mathds{1}\{\cdot\}$ represents the binary indicator function. As a sensitivity test, we also consider the number of retained components $K=4$, since there are 5 discrete delta values.

\subsubsection{Long-run covariance and its estimation}\label{sec:long-run}

While Section~\ref{sec:fpca} presents a method based on the functional principal component analysis extracted from the variance function, these \textit{static} functional principal components are not consistent in the presence of moderate to strong temporal dependence. This issue motivates the development of dynamic functional principal component analysis, such as in \cite{HKH15} and \cite{RS17}, where functional principal components are extracted from the long-run covariance function.

We implement a stationarity test of \cite{HKR14}, and from the $p$-values reported in Table~\ref{tab:3} in Section~\ref{sec:5.2}, we conclude that all series are stationary. For a stationary functional time-series $\{\X_i(t)\}_{i\in \mathbb{Z}}$, the long-run covariance function $C(t,s)$ is defined as
\begin{align*}
C(t,s) &= \sum_{\ell=-\infty}^{\infty}\gamma_{\ell}(t,s) \\
\gamma_{\ell}(s,t) &= \text{Cov}[\X_0(s), \X_{\ell}(t)]
\end{align*}
and is a well-defined element of $\mathcal{L}^2(\mathcal{I})^2$ for a compact support interval $\mathcal{I}$, under mild dependence and moment conditions. By assuming $\X$ is a continuous and square-integrable covariance function, the function $\mathcal{K}$ induces the kernel operator $\mathcal{L}^2(\mathcal{I})\mapsto \mathcal{L}^2(\mathcal{I})$. Through right integration, $C(t,s)$ defines a Hilbert-Schmidt integral operator on $\mathcal{L}^2(\mathcal{I})$ given by
\begin{equation*}
\mathcal{K}(\phi)(s) = \int_{\mathcal{I}}C(t,s)\phi(t)dt,
\end{equation*}
whose eigenvalues and eigenfunctions are related to the dynamic functional principal components defined in \cite{HKH15}. \cite{HKH15} provide asymptotically optimal finite-dimensional representations of the sample mean of a functional time-series.

In practice, we need to estimate $C(t, s)$ from a finite sample $\bm{\X}(t) = \{\X_1(t), \dots, \X_n(t)\}$. Given its definition as a bi-infinite sum, a natural estimator of $C(t, s)$ is
\begin{equation}
\widehat{C}_{h,q}(t,s) = \sum^{\infty}_{\ell=-\infty}W_q\Big(\frac{\ell}{h}\Big)\widehat{\gamma}_{\ell}(t,s), \label{eq:3}
\end{equation}
where $h$ is the bandwidth parameter, and
\[ \widehat{\gamma}_{\ell}(t,s) = \left\{ \begin{array}{ll}
         \frac{1}{n}\sum_{j=1}^{n-\ell}\left[\X_j(t)-\overline{\X}(t)\right]\left[\X_{j+\ell}(s)-\overline{\X}(s)\right] & \mbox{if $\ell \geq 0$};\\
        \frac{1}{n}\sum_{j=1-\ell}^n\left[\X_j(t)-\overline{\X}(t)\right]\left[\X_{j+\ell}(s)-\overline{\X}(s)\right] & \mbox{if $\ell < 0$},\end{array} \right. \] 
is an estimator of $\gamma_{\ell}(t,s)$. $W_q$ is a symmetric weight function with bounded support of order $q$, and the following properties:
\begin{enumerate}
\item[1)] $W_q(0) = 1$
\item[2)] $W_q(u) = W_q(-u)$
\item[3)] $W_q(u) = 0$ if $|u|>m$ for some $m>0$
\item[4)] $W_q(u)$ is continuous on $[-m, m]$
\item[5)] $c$ exists and satisfies
\begin{equation*}
0<c=\lim_{t\mapsto 0}t^{-q}[W_q(t)-1]<\infty.
\end{equation*} 
\end{enumerate}
The kernel sandwich estimator in~\eqref{eq:3} is discussed in \cite{HKR13}, \cite{PT13}, \cite{RS17}, and \cite{MGG20}, among others. As with the kernel sandwich estimator, a crucial consideration here is the estimation of the bandwidth parameter $h$. One approach is to use a data-driven approach, such as the plug-in algorithm proposed in \cite{RS17}. \cite{RS17} prove that the estimator in~\eqref{eq:3} is a consistent estimator of the true and unknown long-run covariance. They also find that the estimated dynamic functional principal components and associated scores extracted from the estimated long-run covariance are consistent too.

\subsubsection{Dynamic functional principal component analysis}\label{sec:dfpca}

Using the estimated long-run covariance function $\widehat{C}(t,s)$, we apply functional principal component decomposition to extract the dynamic functional principal components and their associated scores. The sample mean and sample covariance are given by
\begin{align*}
\widehat{\mu}(t) &= \frac{1}{n}\sum^n_{i=1}\X_i(t), \\
\widehat{C}(t,s) &= \sum^{\infty}_{k=1}\widehat{\lambda}_k\widehat{\phi}_k(t)\widehat{\phi}_k(s),
\end{align*}
where $\widehat{\lambda}_1>\widehat{\lambda}_2>\cdots \geq 0$ are the sample eigenvalues of $\widehat{C}(t,s)$, and $[\widehat{\phi}_1(t), \widehat{\phi}_2(t), \dots]$ are the corresponding orthogonal sample eigenfunctions. Via the Karhunen-Lo\`{e}ve expansion, the realisations of the stochastic process $\X$ can be written as
\begin{equation*}
\X_i(t) = \widehat{\mu}(t)+\sum^{\infty}_{k=1}\widehat{\beta}_{i,k}\widehat{\phi}_k(t),\qquad i=1,2\dots,n,
\end{equation*}
where $\widehat{\beta}_{i,k}$ is the $k$\textsuperscript{th} estimated principal component score for the $i$\textsuperscript{th} observation. 

Dynamic functional principal component analysis has three properties:
\begin{enumerate}
\item[(a)] It minimizes the mean integrated squared error of the reconstruction error function over the whole functional data set.
\item[(b)] It provides an effective way of extracting a large amount of long-run covariance.
\item[(c)] Its dynamic principal component scores are uncorrelated.
\end{enumerate}

\subsubsection{Univariate functional time-series forecasting method}\label{sec:fts}

We obtain an IV smile of a specific maturity by interpolating the moneyness level (in terms of delta) of the option contract. We extract estimated functional principal component functions $\bm{B} = \big\{\widehat{\phi}_1(t), \dots, \widehat{\phi}_K(t)\big\}$, using the empirical covariance function. Conditioning on the past functions $\bm{\X}(t)$ and the estimated functional principal components $\bm{B}$, the $h$-step-ahead point forecast of $\X_{n+h}(t)$ can be expressed as
\begin{equation*}
\widehat{\X}_{n+h|n}(t) = \text{E}\left[\X_{n+h}(t)|\bm{\X}(t),\bm{B}\right] =\widehat{\mu}(t) + \sum^K_{k=1}\widehat{\beta}_{n+h|n, k}\widehat{\phi}_k(t),
\end{equation*}
where $\widehat{\beta}_{n+h|n, k}$ denotes time-series forecasts of the $k^{\text{th}}$ principal component scores. The forecasts of these scores can be obtained via a univariate time-series forecasting method, such as the autoregressive integrated moving average (ARIMA)$(p,d,q)$ model \citep[see][for a review]{OF13}. We use the automatic algorithm of \cite{HK08} to choose the optimal orders of $p$, $q$, and $d$. In the supplement, we also present the results using exponential smoothing, which is a generalisation of the ARIMA model \citep[see, e.g.,][]{KS19}. The forecast accuracy results are qualitatively similar between the two univariate time-series methods.

\subsection{Multivariate functional time-series method}\label{sec:mfts}

Section~\ref{sec:3.1} introduces a method for modelling and forecasting one functional time-series or multiple functional time-series individually. However, explicitly modelling the correlation between multiple time-series may improve forecast accuracy. Therefore, we introduce a multivariate functional time-series method to jointly model and forecast multiple functional time-series that could be correlated with each other.

Let $\X_i^{1}(t)$ be the IV smile for 1M-maturity options, $\X_i^{2}(t)$ the IV smile for 6M-maturity options, and $\X_i^{3}(t)$ the IV smile for 2Y-maturity options. As our multiple functional time-series have the same function support, we consider data where each observation consists of $\omega\geq 2$ functions, i.e., $\left[\X^{1}(t),\dots, \X^{\omega}(t)\right]\in R^{\omega}$, where $t\in \mathcal{I}$ and $\omega = 3$ in our data set.

The multivariate functional time-series are stacked in a vector with $[\X^{1}(t),\dots,\X^{3}(t)]\in R^{3}$. We assume that $\bm{\nu}(t):=\{\text{E}[\X^{1}(t)], \dots, \text{E}[\X^{3}(t)]\}\in R^{3}$. For $s, t\in \mathcal{I}$, the theoretical cross-covariance function can be defined with elements
\begin{align*}
c_{fg}(t, s) &:= \text{Cov}\left[\X^{f}(t), \X^{g}(s)\right] \\
&=\text{E}\left\{\left[\X^{f}(t) - \nu^{f}(t)\right]\left[\X^{g}(s) - \nu^{g}(s)\right]\right\},
\end{align*}
where $\nu^{f}(t)$ and $\nu^{g}(s)$ denote the mean functions for the $f^{\text{th}}$ and $g^{\text{th}}$ series, respectively. By assuming that $\X$ is a continuous and square-integrable covariance function, the function $\mathcal{K}$ induces the kernel operator given by
\begin{equation*}
\mathcal{K}(\phi)^{f}(t) = \sum^{\omega}_{g=1}\int_{\mathcal{I}}c_{fg}(t,s)\phi^{g}(s)ds, \qquad \X^{f}\in \mathcal{L}^2(\mathcal{I}).
\end{equation*}
Via Mercer's lemma, there exists an orthonormal sequence $(\phi_k)$ of continuous functions in $\mathcal{L}^2(\mathcal{I})$ and a non-increasing sequence $\lambda_k$ of positive numbers, such that
\begin{equation*}
c_{fg}(t,s) = \sum^{\infty}_{k=1}\lambda_k \phi_k^{f}(t)\phi_k^{g}(s).
\end{equation*}
By the separability of Hilbert space, the Karhunen-Lo\`{e}ve expansion of a stochastic process $\X^{f}(t)$ can be expressed as
\begin{equation*}
\X_i^{f}(t) \approx \nu^{f}(t) + \sum^{K}_{k=1}\beta_{i,k}^{f}\phi_k^{f}(t),\qquad i=1,\dots,n,
\end{equation*}
where $K$ denotes the number of retained functional principal components. Its matrix formulation is
\begin{equation*}
\bm{\X}^*_i(t) = \bm{\nu}(t) + \bm{\Phi}(t)\bm{\beta}_i^{\top},
\end{equation*}
where $\bm{\X}_1(t) = [\X_1^1(t),\X_1^2(t),\X_1^{3}(t)]$ and $\bm{\X}(t)=[\bm{\X}_1(t),\dots,\bm{\X}_n(t)]$ denote stacked historical functions. Moreover, $\bm{\beta}_i = \left(\beta_{i,1}^{1},\dots,\beta_{i,K}^{1}, \beta_{i,1}^{2},\dots,\beta_{i,K}^{2}, \beta_{i,1}^{3},\dots,\beta_{i,K}^{3}\right)$ is the vector of the basis expansion coefficients, $\bm{\nu}(t) = [\nu^{1}(t), \nu^{2}(t), \nu^{3}(t)]$, and
\begin{equation*}
\bm{\Phi}(t) = \left( \begin{array}{ccccccccc}
\phi_1^{1}(t) & \cdots & \phi_{K}^{1}(t) & 0 & \cdots & 0 & 0 & \cdots & 0 \\
0 & \cdots & 0 & \phi_1^{2}(t) & \cdots & \phi_{K}^{2}(t) & 0 & \cdots & 0 \\
 0 & \cdots & 0 & 0 & \cdots & 0 & \phi_1^{3}(t) & \cdots & \phi_{K}^{3}(t)  \end{array} \right)_{3\times (K\times 3)}.
\end{equation*}

Conditioning on the past functions $\bm{\X}^*(t)$ and the estimated functional principal components $\bm{\Phi}(t)$, the $h$-step-ahead point forecast of $\bm{\X}_{n+h}^*(t)$ can be expressed as
\begin{align*}
\widehat{\bm{\X}}^*_{n+h|n}(t) &= \text{E}\left[\bm{\X}^*_{n+h}(t)|\bm{\X}^*(t), \bm{\Phi}(t)\right] \\
&= \widehat{\bm{\nu}}(t) + \widehat{\bm{\Phi}}(t)\bm{\widehat{\beta}}_{n+h|n}^{\top},
\end{align*}
where $\widehat{\bm{\beta}}_{n+h|n}$ denotes the time-series forecasts of the principal component scores corresponding to the three maturities. In our implementation, each series is standardised to have equal scale before being combined into a long vector \citep[see, e.g.,][]{CCY14}.

\subsection{Multilevel functional time-series method}\label{sec:multilevel}

The multilevel functional data model bears a strong resemblance to a two-way functional analysis of variance model studied by \cite{MVB+03} and \cite{CF10}. It is a special case of the general ``functional mixed model" proposed in \cite{MC06} and applied in \cite{Shang16} and \cite{SSB+16}. As the IV surface comprises distinct IV smiles for multiple maturities of the same currency pair, the basic idea is to decompose the related smiles into an average IV smile across the entire surface $\mu^{(c)}(t)$, a maturity-specific deviation from the averaged IV smile $\eta^j(t)$, a common trend across maturities $R^{(c)}_i(t)$, and a maturity-specific residual trend $U_i^j(t)$. The common trend and maturity-specific residual trend are modelled by projecting them onto the eigenvectors of covariance functions of the aggregate and maturity-specific centred stochastic processes, respectively. The curve at observation $i=1,2,\dots,n$ can be written as
\begin{align}
\X_i^j(t) &= \mu^{(c)}(t) + \eta^j(t) + R_i^{(c)}(t) + U_i^j(t),\qquad t\in \mathcal{I}, \label{eq:4}\\
&= \mu^j(t) + R_i^{(c)}(t) + U_i^j(t), \label{eq:5}
\end{align}
where $j$ corresponds to the 1-month, 6-month and 2-year maturities, respectively. To ensure model and parameter identifiability, the observations of the two stochastic processes $\bm{R}^{(c)}(t) = \big\{R_1^{(c)}(t),\dots,R_n^{(c)}(t)\big\}$ and $\bm{U}^j(t) =\big\{U_1^j(t), \dots, U_n^j(t)\big\}$ are uncorrelated since we implement two-stage functional principal component decomposition. 

Since the centred stochastic processes $R^{(c)}(t)$ and $U^j(t)$ are unknown in practice, the population eigenvalues and eigenfunctions can only be approximated at best through a set of realisations $\bm{R}^{(c)}(t)$ and $\bm{U}^j(t)$. From the covariance of $\bm{R}^{(c)}(t)$, we can extract a set of functional principal components and the associated scores, along with a set of residual functions. From the covariance function of the residual functions, we can then extract a second set of functional principal components and their associated scores. While the first functional principal component decomposition captures the common trend shared across various maturities of the same currency, the second functional principal component decomposition captures the maturity-specific residual trend.

The sample versions of the aggregate mean function, maturity-specific mean function deviation, a common trend, and maturity-specific residual trend for a set of functional data can be estimated by
\begin{align}
\X_i^{(c)}(t) &= \sum^J_{j=1}\X^j(t)/J, \qquad \widehat{\mu}^{(c)}(t) = \sum^n_{i=1}\X_i^{(c)}(t)/n \label{eq:6}\\
\widehat{\mu}^j(t) &= \sum^n_{i=1}\X_i^j(t)/n, \qquad \widehat{\eta}^j(t) = \widehat{\mu}^j(t) - \widehat{\mu}^{(c)}(t) 
\end{align}
\begin{align}
\widehat{R}_i^{(c)}(t) &= \sum^{\infty}_{k=1}\widehat{\beta}^{(c)}_{i,k}\widehat{\phi}^{(c)}_k(t) \approx \sum^K_{k=1}\widehat{\beta}^{(c)}_{i,k}\widehat{\phi}_k^{(c)}(t) \\
\widehat{U}_i^j(t) &= \sum^{\infty}_{l=1}\widehat{\gamma}_{i,l}^j\widehat{\psi}_l^j(t) \approx \sum^L_{l=1}\widehat{\gamma}_{i,l}^j\widehat{\psi}_l^j(t),\label{eq:9}
\end{align}
where $\left\{\X_1^{(c)}(t),\dots,\X_n^{(c)}(t)\right\}$ represents a time-series of functions for the currency exchange averaged across different maturities; $\widehat{\mu}(t)$ represents the simple average of the averaged currency exchange, whereas $\widehat{\mu}^j(t)$ represents the simple average of the currency exchange at the $j^{\text{th}}$ maturity; $\big\{\widehat{\bm{\beta}}_k = (\widehat{\beta}_{1,k},\dots,\widehat{\beta}_{n,k}); k=1,\dots,K\big\}$ represents the $k^{\text{th}}$ sample principal component scores of $\widehat{\bm{R}}(t)=\{\widehat{R}_1^{(c)}(t),\dots, \widehat{R}_n^{(c)}(t)\}$; and $\bm{\Phi} = \big\{\widehat{\phi}_1(t),\dots,\widehat{\phi}_K(t)\big\}$ denotes the corresponding orthogonal sample eigenfunctions in a square-integrable function space.

Substituting equations~\eqref{eq:6} to \eqref{eq:9} into equations~\eqref{eq:4}--\eqref{eq:5}, we obtain
\begin{equation*}
\X_i^j(t) = \widehat{\mu}^{(c)}(t) + \widehat{\eta}^j(t) + \sum^K_{k=1}\widehat{\beta}_{i,k}^{(c)}\widehat{\phi}_k^{(c)}(t) + \sum^L_{l=1}\widehat{\gamma}_{i,l}^j\widehat{\psi}^j_l(t) + e_i^j(t),
\end{equation*} 
where $e_i^j(t)$ denotes measurement error with a finite variance $(\sigma^2) ^j$.

To select the optimal number of components, we use a CPV criterion to determine $K$ and $L$. The optimal numbers of $K$ and $L$ are determined by
\begin{align*}
K &= \argmin_{K:K\geq 1}\left\{\sum^K_{k=1}\widehat{\lambda}_k\Big/\sum^{\infty}_{k=1}\widehat{\lambda}_k\mathds{1}_{\left\{\widehat{\lambda}_k>0\right\}}\geq P_1\right\}, \\
L &= \argmin_{L:L\geq 1} \left\{\sum^L_{l=1}\widehat{\lambda}_l^j\Big/\sum^{\infty}_{l=1}\widehat{\lambda}_l\mathds{1}_{\left\{\widehat{\lambda}_l>0\right\}}\geq P_2\right\},
\end{align*}
where $\mathds{1}_{\{\cdot\}}$ denotes a binary indicator function. Following \cite{Chiou12}, we chose $P_1 = P_2 = 0.9$.

From the estimated eigenvalues, we can estimate the proportion of variability explained by aggregated data \citep[also known as the variance explained by the within-cluster variability, see for example,][]{DCC+09}. A possible measure of within-cluster variability is given by
\begin{align}
\frac{\int_{\mathcal{I}}\text{Var}[\bm{R}^{(c)}(t)]dt}{\int_{\mathcal{I}}\text{Var}[\bm{R}^{(c)}(t)]dt + \int_{\mathcal{I}}\text{Var}[\bm{U}^j(t)]dt}\approx \frac{\sum^{K}_{k=1}\widehat{\lambda}_k}{\sum^{K}_{k=1}\widehat{\lambda}_k+\sum^L_{l=1}\widehat{\lambda}_l}, \label{eq:10}
\end{align}
where $\text{Var}(\cdot)$ denotes the population variance, which can be approximated by the estimated and retained eigenvalues. When the common trend can explain the primary mode of total variability, the value of the within-cluster variability is close to 1.

Conditioning on observed data $\bm{\X}^j(t) = \{\X_1^j(t),\dots,\X_n^j(t)\}$ and basis functions $\bm{\Phi}$ and $\bm{\Psi}=\big\{\widehat{\psi}_1^j(t),\dots,\widehat{\psi}_L^j(t)\big\}$, the $h$-step-ahead forecasts can be obtained by
\begin{align*}
\widehat{\mathcal{X}}_{n+h|n}^{j}(t) &= \text{E}\left[\X_{n+h}^j(t)\Big|\mu^{(c)}(t),\eta^j(t),\bm{\Phi},\bm{\Psi},\bm{\X}^j(t)\right]\\
 &= \widehat{\mu}^{(c)}(t) + \widehat{\mu}^{j}(t) + \sum^K_{k=1}\widehat{\beta}_{n+h|n,k}^{(c)}\widehat{\phi}_k^{(c)}(t)  + \sum^L_{l=1}\widehat{\gamma}_{n+h|n,l}^{j}\widehat{\psi}_l^{j}(t), 
\end{align*}
where $\widehat{\beta}^{(c)}_{n+h|n,k}$ and $\widehat{\gamma}_{n+h|n,l}^j$ are the forecasted principal component scores, obtained from a univariate time-series forecasting method, such as an ARIMA model.

While Sections~\ref{sec:mfts} and~\ref{sec:multilevel} present static multivariate and multilevel functional time-series methods, dynamic counterparts can be produced using dynamic functional principal components and their associated scores. These quantities are constructed based on an estimated long-run covariance function in~\eqref{eq:3}.

\section{Forecast evaluation}\label{sec:4}

\subsection{Expanding-window approach}

An expanding-window analysis of time-series models is commonly used to assess model and parameter stability over time. It determines the constancy of a model's parameters by computing parameter estimates, and their forecasts over an expanding window of a fixed size through the sample \citep[see][Chapter 9 for details]{ZW06}. Using the first 1,827 observations from January 1, 2008 to December 31, 2014 for each currency pair, we produce one-step-ahead point forecasts. Through an expanding window approach, we re-estimate the parameters in all models using the first 1,828 observations from January 1, 2008 to January 1, 2015. One-step-ahead forecasts are then produced for each model. We iterate this process by increasing the sample size by one trading day until we reach the end of the data period. This process produces 522 one-step-ahead forecasts from January 1, 2015 to December 30, 2016. We compare these forecasts with the holdout samples to determine out-of-sample point forecast accuracy. This evaluation is in line with \cite{CT10}, who adopt a recursive one-day-ahead strategy scheme. Also, we consider 518 five-step-ahead forecasts and 513 ten-step-ahead forecasts to assess forecast accuracy at comparably longer horizons. We choose to expand the training set and re-estimate all the models daily to incorporate all available information into our prediction. This approach more accurately simulates the action likely to be taken by a market practitioner who aims to predict the following day's movement. 

To control for sensitivity to specific out-of-sample periods, various window lengths are tested: In the case of one-step-ahead prediction, we consider 522 days (out-of-sample: January 1, 2015 to December 30, 2016), 261 days (out-of-sample: January 1, 2016 to December 30, 2016), 131 days (out-of-sample: July 1, 2016 to December 30, 2016). The out-of-sample forecasts between the end of the in-sample period and December 30, 2016 are obtained using an expanding-window scheme.

\subsection{Forecast error measures}

A number of criteria exist for evaluating point forecast accuracy \citep[see][]{AC92, HK06}. They can be grouped into four categories:
\begin{inparaenum}
\item[(1)] scale-dependent metrics,
\item[(2)] percentage-error metrics,
\item[(3)] relative-error metrics, which average the ratios of the errors from a designated method to the errors of a na\"{i}ve method, and
\item[(4)] scale-free error metrics, which express each error as a ratio to the average error from a baseline method.
\end{inparaenum}
Here, we use the following three accuracy measures:
\begin{enumerate}
\item[(1)] Mean absolute forecast error (MAFE); a measure of the absolute difference between the forecast, $\widehat{\X}_{\kappa}(t_{\tau})$, and the corresponding observation, $\X_{\kappa}(t_{\tau})$. It measures the average error magnitude in the forecasts, regardless of error direction and serves to aggregate the errors into a single measure of predictive power. The MAFE can be expressed as
\begin{equation*}
\text{MAFE}_{\kappa} = \frac{1}{R}\sum^R_{\tau=1}\left|\X_{\kappa}(t_{\tau}) - \widehat{\X}_{\kappa}(t_{\tau})\right|, \qquad
\overline{\text{MAFE}} = \frac{1}{N}\sum^N_{\kappa=1}\text{MAFE}_{\kappa},
\end{equation*}
where $R$ denotes the number of discretised points in a curve and $N$ denotes the number of observations in the forecasting period; $\X_{\kappa}(t_{\tau})$ denotes the holdout sample for the $\tau^{\text{th}}$ delta and $\kappa^{\text{th}}$ curve in the forecasting period; and $\widehat{\X}_{\kappa}(t_{\tau})$ denotes the point forecast of the holdout sample. 
\item[(2)] Mean squared forecast error (MSFE); a measure of the squared difference between the predicted and realised values, which again serves to aggregate the errors into a single measure of predictive power. The MSFE can be expressed as
\begin{equation*}
\text{MSFE}_{\kappa} = \frac{1}{R}\sum^R_{\tau=1}\left[\X_{\kappa}(t_{\tau}) - \widehat{\X}_{\kappa}(t_{\tau})\right]^2, \qquad
\overline{\text{MSFE}} = \frac{1}{N}\sum^N_{\kappa=1}\text{MSFE}_{\kappa}.
\end{equation*}
\item[(3)] Mean mixed error (MME); an asymmetric loss function. MME(U) penalises under-predictions more heavily, while MME(O) penalises over-predictions more heavily. This feature is crucial for investors in options markets, as an under (over)-prediction of IV is more likely to be of greater concern to a seller (buyer) than a buyer (seller). This measure has previously been used in studies that evaluate volatility forecasting techniques, such as \cite{BF96}, \cite{FIK09}, and \cite{KCM18}. With a slight abuse of notation $\kappa$, the MME(U) and MME(O) are given as
\begin{align*}
\text{MME(U)} &= \frac{1}{N\times R}\left[\sum_{\kappa = \eta_1^O}^{\eta_N^O}\sum^R_{\tau=1}\left|\X_{\kappa}(t_{\tau}) - \widehat{\X}_{\kappa}(t_{\tau})\right|+\sum_{\kappa = \eta_1^U}^{\eta_N^U}\sum^R_{\tau=1}\sqrt{\left|\X_{\kappa}(t_{\tau}) - \widehat{\X}_{\kappa}(t_{\tau})\right|}\right], \\
\text{MME(O)} &= \frac{1}{N\times R}\left[\sum_{\kappa = \eta_1^U}^{\eta_N^U}\sum^R_{\tau=1}\left|\X_{\kappa}(t_{\tau}) - \widehat{\X}_{\kappa}(t_{\tau})\right|+\sum_{\kappa = \eta_1^O}^{\eta_N^O}\sum^R_{\tau=1}\sqrt{\left|\X_{\kappa}(t_{\tau}) - \widehat{\X}_{\kappa}(t_{\tau})\right|}\right], 
\end{align*}
where $\eta_N^U$ denotes the number of under-predictions and $\eta_N^O$ denotes the number of over-predictions. $\big\{\eta_1^O,\dots,\eta_N^O\big\}$ represents the indices of the over-predictions, and $\big\{\eta_1^U,\dots,\eta_N^U\big\}$ represents the indices of the under-predictions.
\end{enumerate}

\section{Data analysis results}\label{sec:5}

\begin{figure}[!htbp]
\centering
\includegraphics[width=\textwidth]{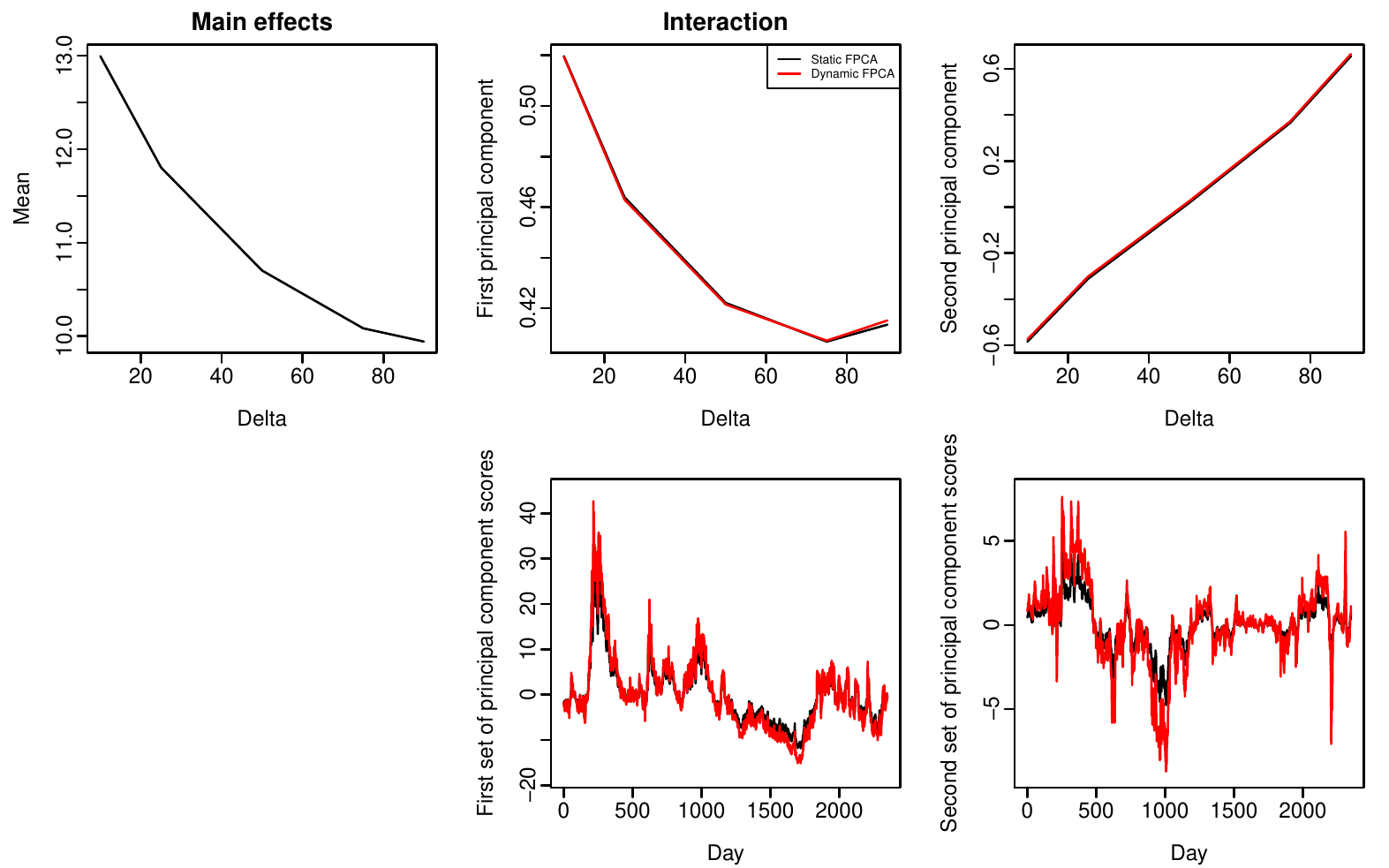}
\caption{\small{The figure presents the estimated mean function and the first two estimated static and dynamic functional principal components, along with their associated scores. For both approaches, the number of components, $K=2$, is retained as two, based on explaining at least 99\% of the total variance or long-run covariance.}}\label{fig:PC}
\end{figure}

This section presents the results of modelling IV using the three functional time-series models for the entire data sample from January 1, 2008 to December 30, 2016. Using the entire EUR-USD dataset, we present the first two estimated static and dynamic principal components and their associated scores in Figure~\ref{fig:PC}. While the static principal components are extracted based on sample variance, the dynamic principal components are extracted based on sample long-run covariance. When a time-series has moderate to strong temporal dependence, the extracted static and dynamic principal components are different, as shown in Figure~\ref{fig:PC}. Due to the variation in the extracted principal components, the associated scores for both approaches also exhibit visible differences.

In Section~\ref{sec:5.1}, the models are fitted in-sample to ascertain goodness-of-fit when seeking to capture the empirical dynamics of the IV surfaces for each currency pair during the entire sample period. In Section~\ref{sec:5.2}, we evaluate and compare out-of-sample forecast accuracy among the functional time-series methods. In Section~\ref{sec:5.3}, we examine statistical significance based on point forecast accuracy.

\subsection{In-sample model fitting}\label{sec:5.1}

Three functional time-series models are fitted to the underlying IV data, for each day over the full sample of January 1, 2008 to December 30, 2016. There are various ways of assessing the fit of a linear model, and we initially consider a criterion introduced by \citet[][Section 16.3]{RS05}. The approach borrowed from conventional linear models is to consider the squared correlation function:
\begin{equation*}
R^2(t) = 1 - \frac{\sum^n_{i=1}[\X_i(t)-\widehat{\X}_i(t)]^2}{\sum^n_{i=1}[\X_i(t) - \overline{\X}(t)]^2}, \qquad t\in \mathcal{I}.
\end{equation*}

\begin{figure}[!htbp]
\centering
\includegraphics[width = 5.75cm]{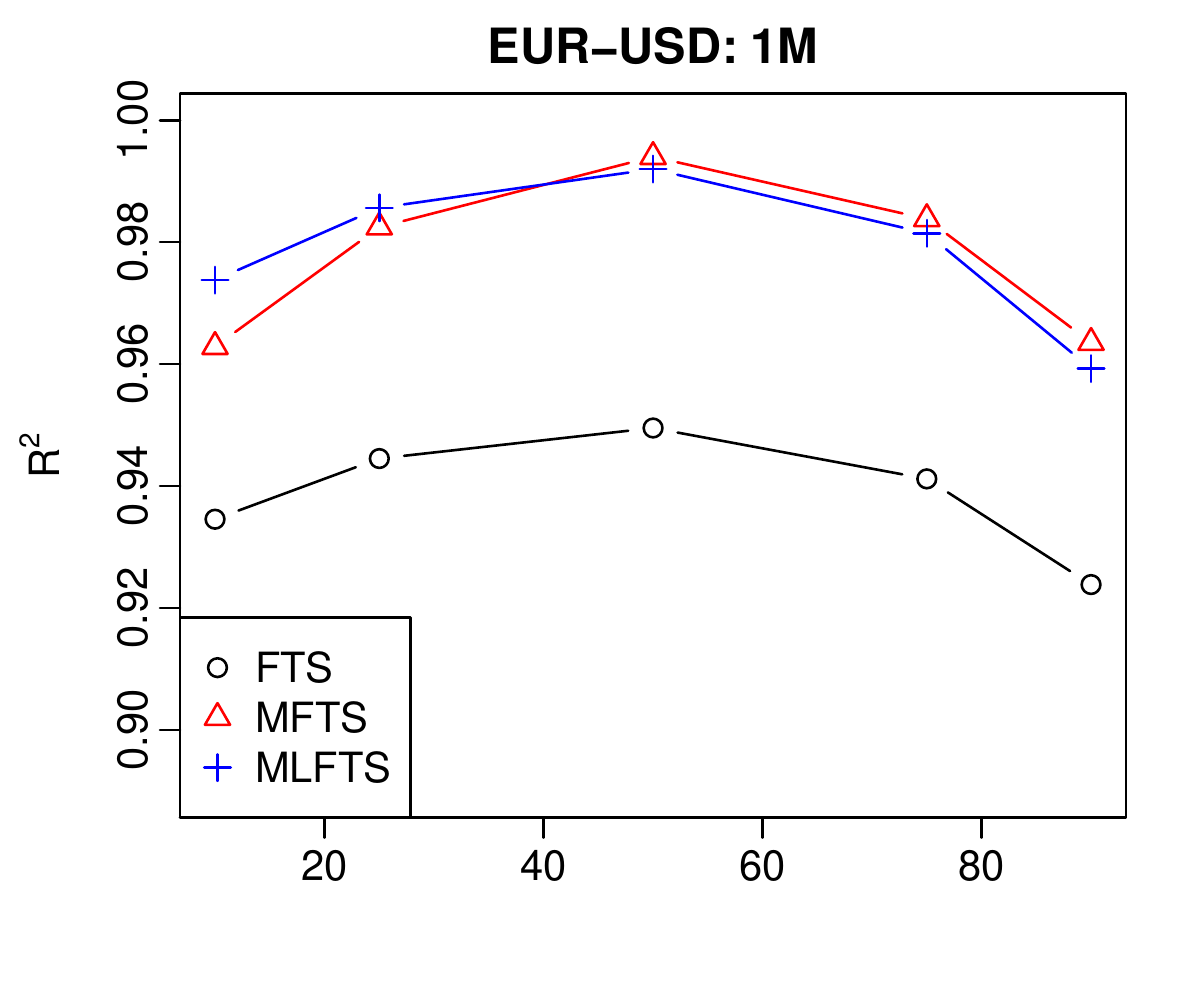}
\includegraphics[width = 5.75cm]{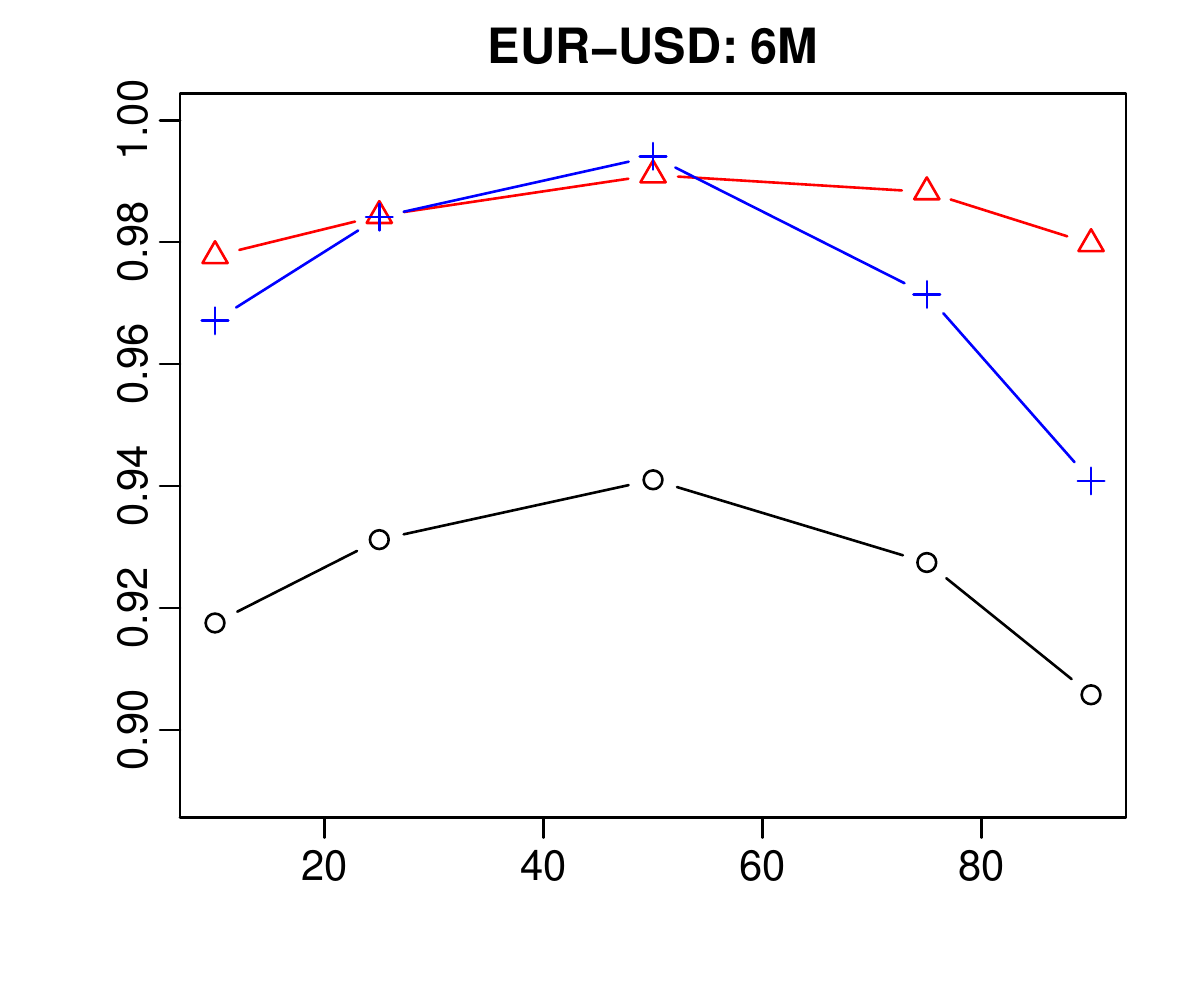}
\includegraphics[width = 5.75cm]{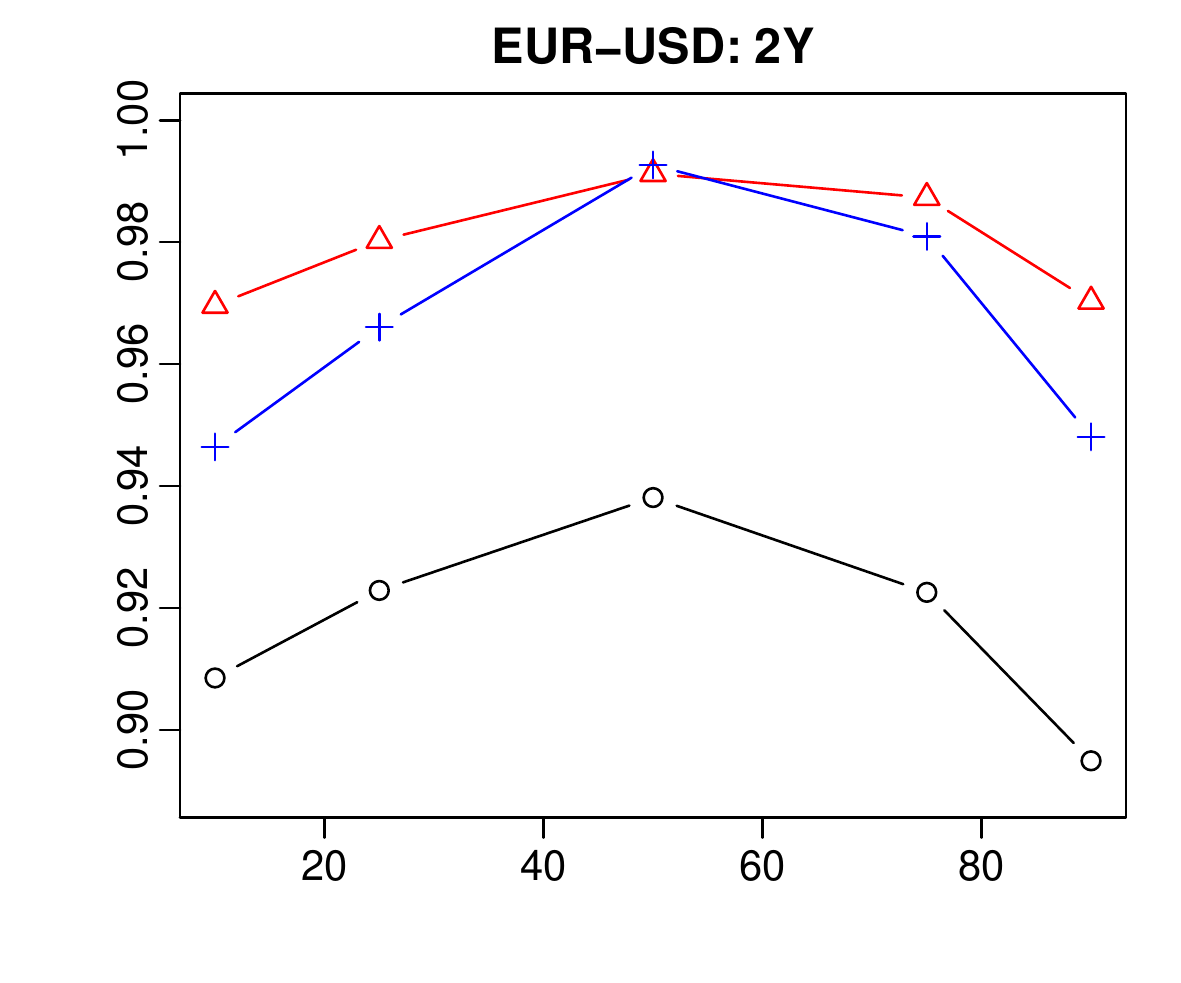}
\\
\includegraphics[width = 5.75cm]{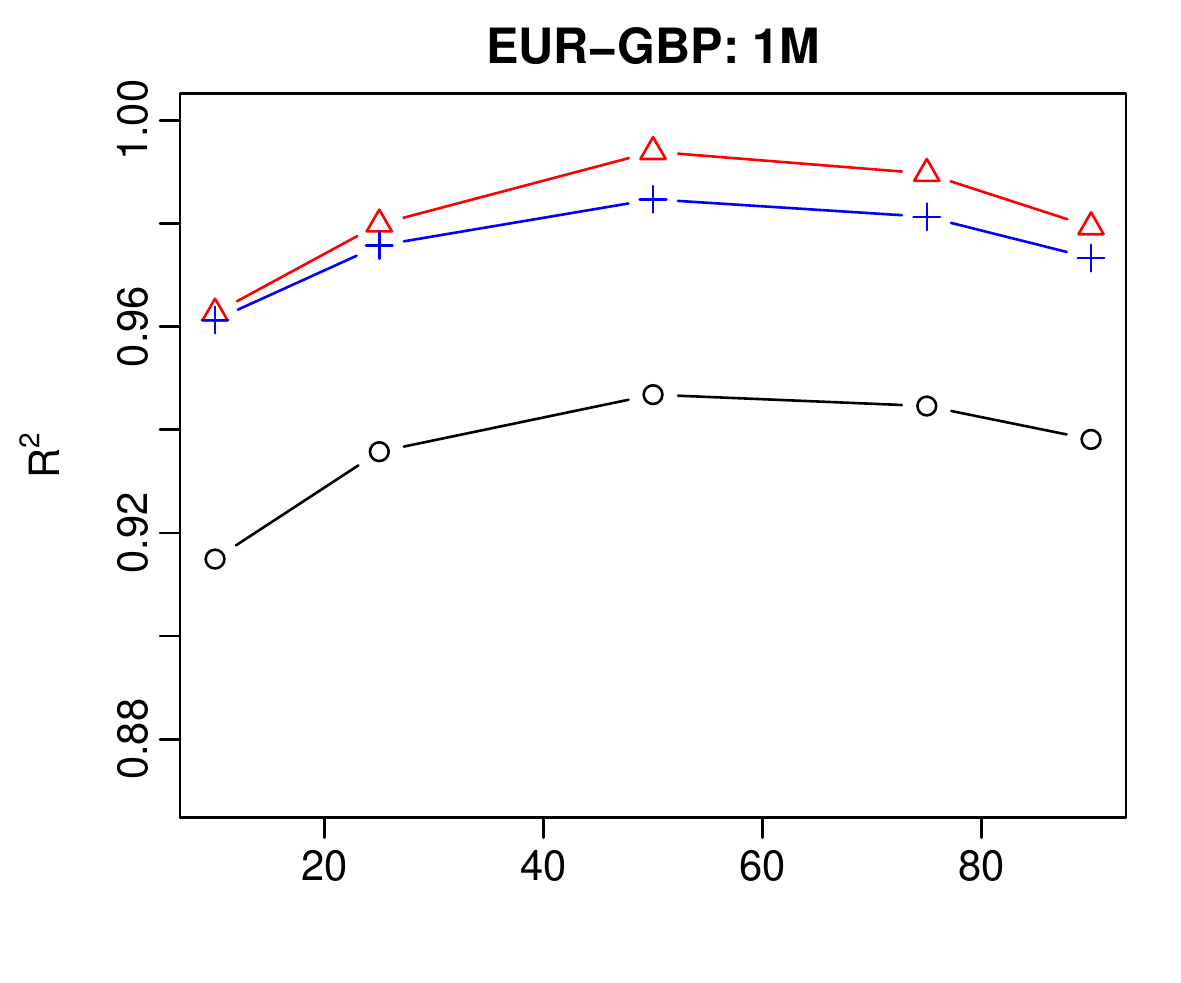}
\includegraphics[width = 5.75cm]{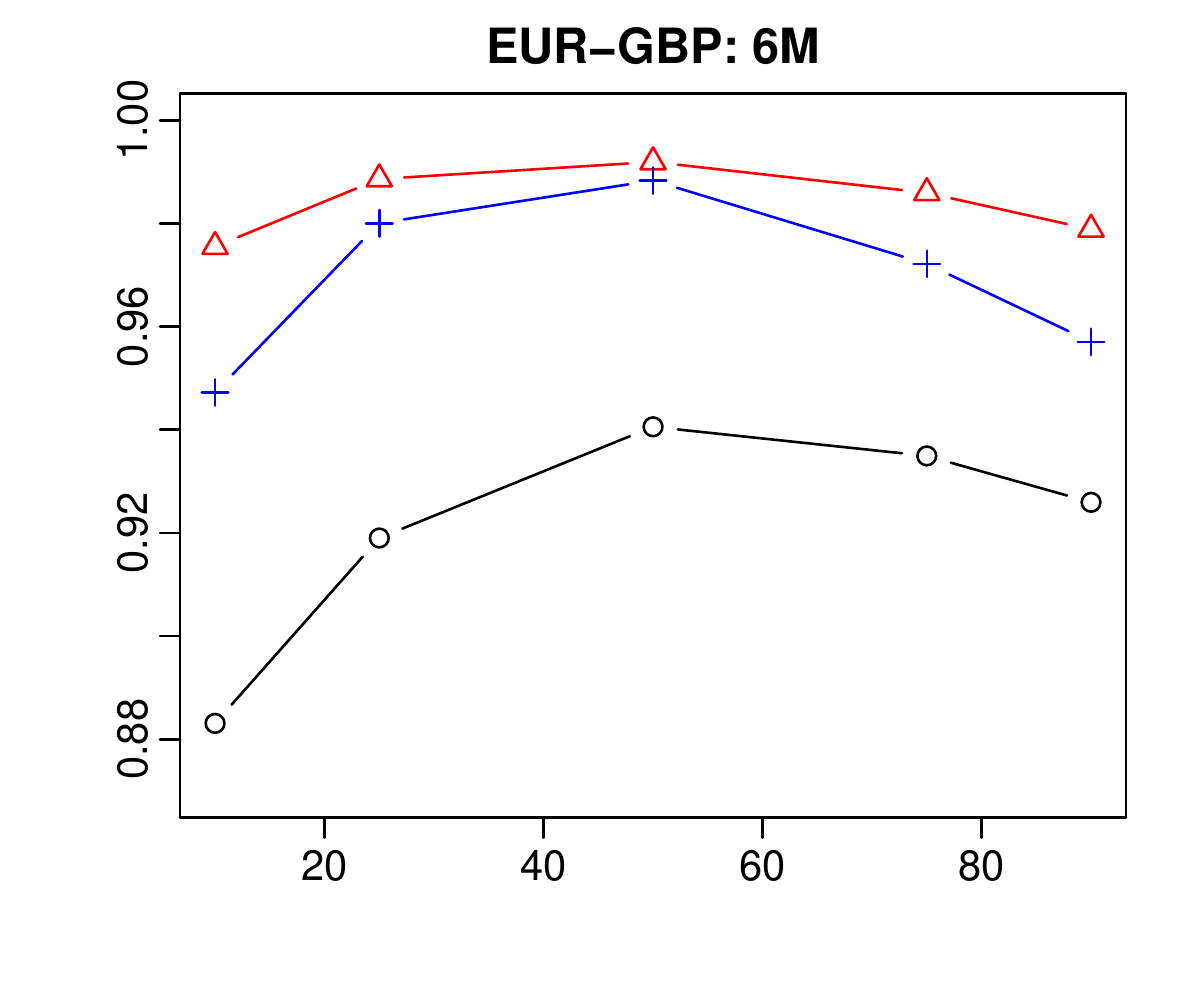}
\includegraphics[width = 5.75cm]{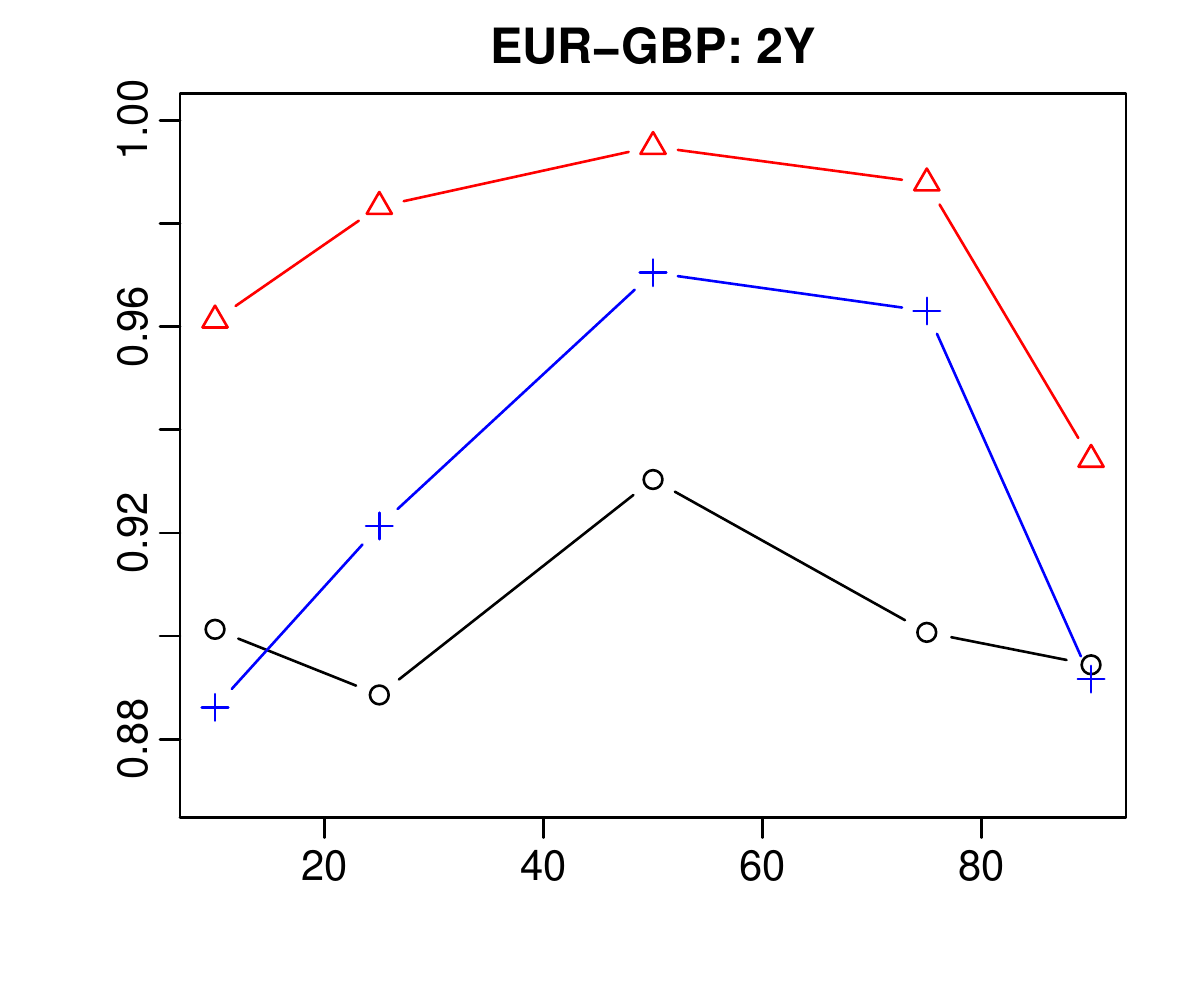}
\\
\includegraphics[width = 5.75cm]{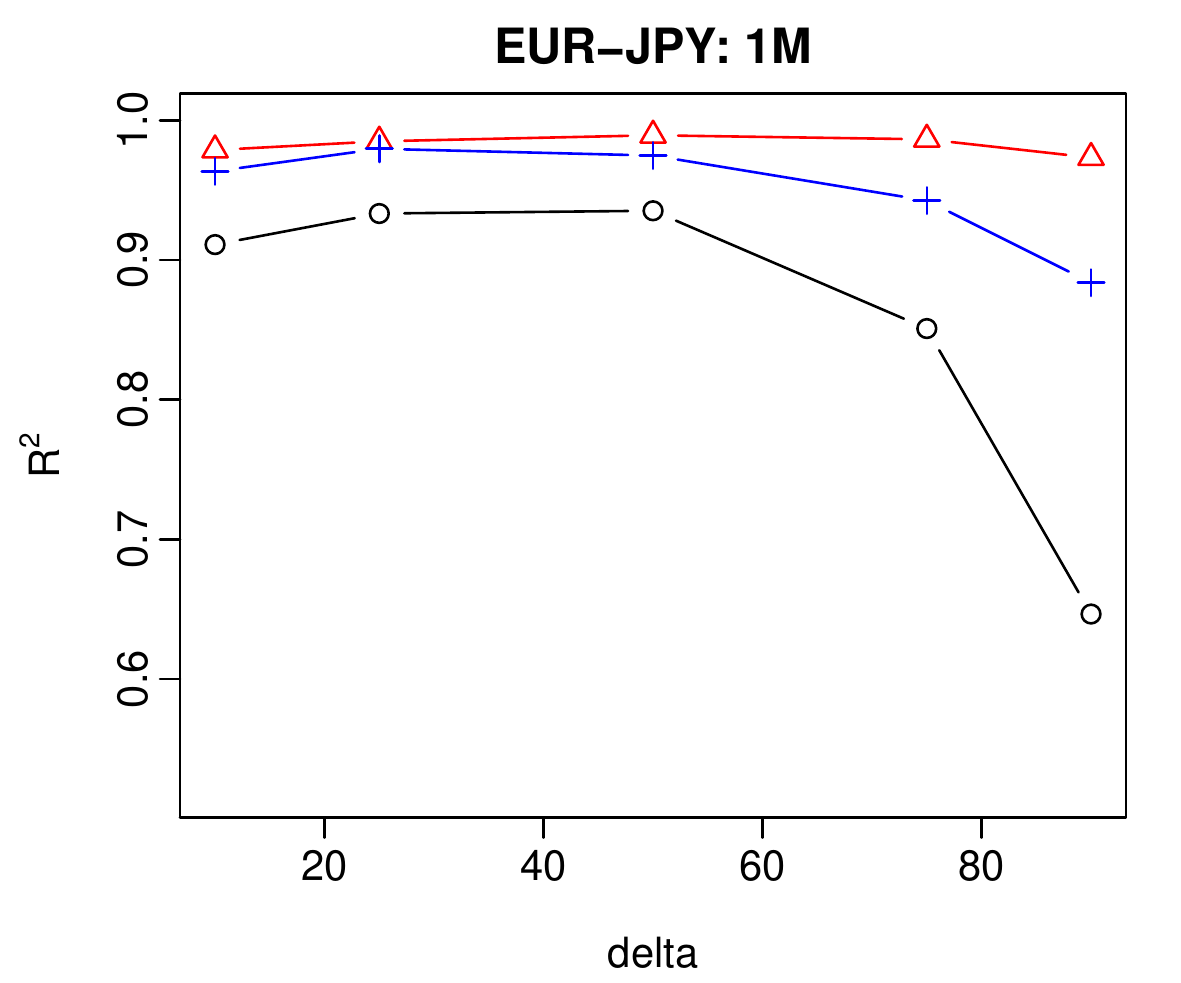}
\includegraphics[width = 5.75cm]{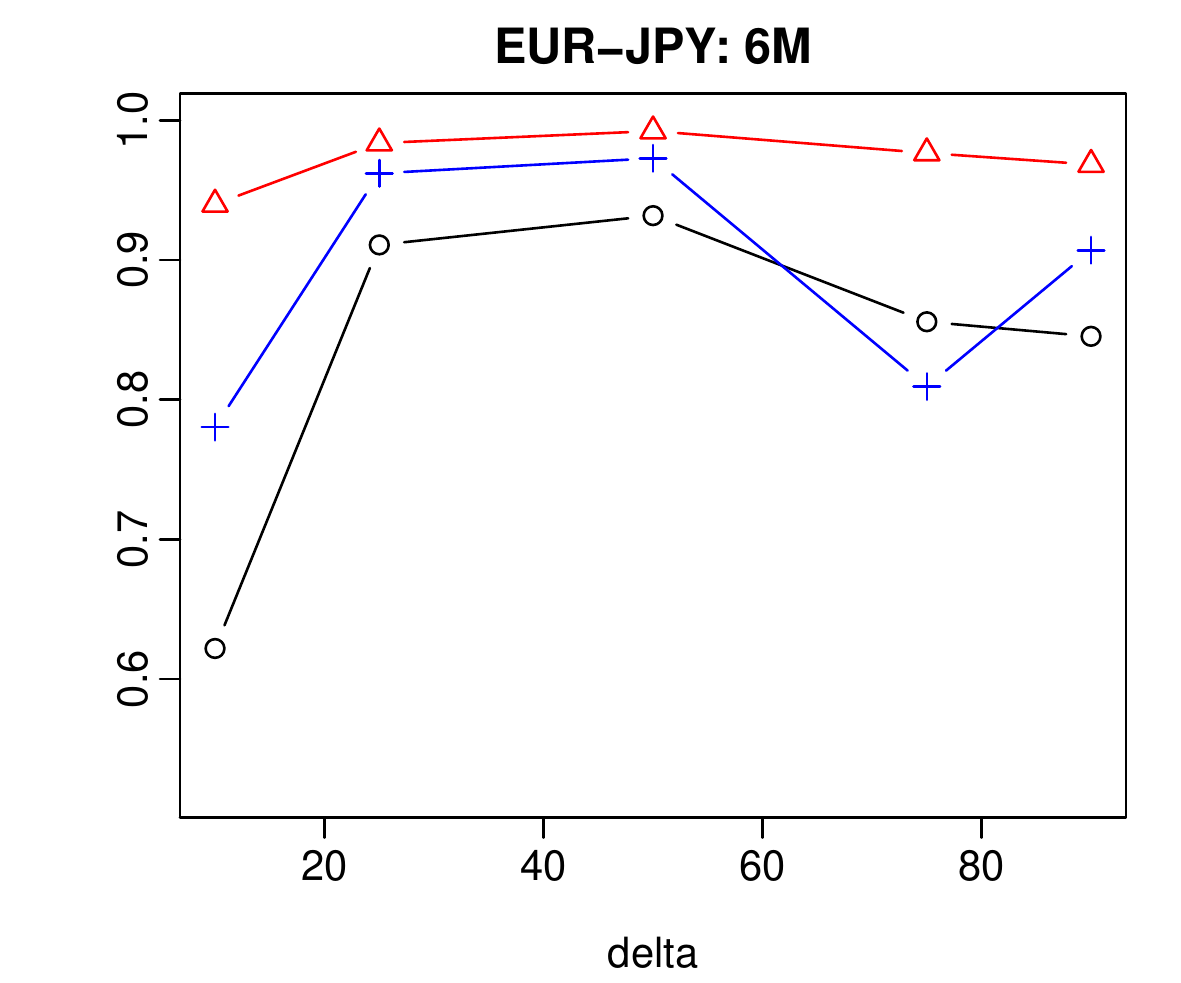}
\includegraphics[width = 5.75cm]{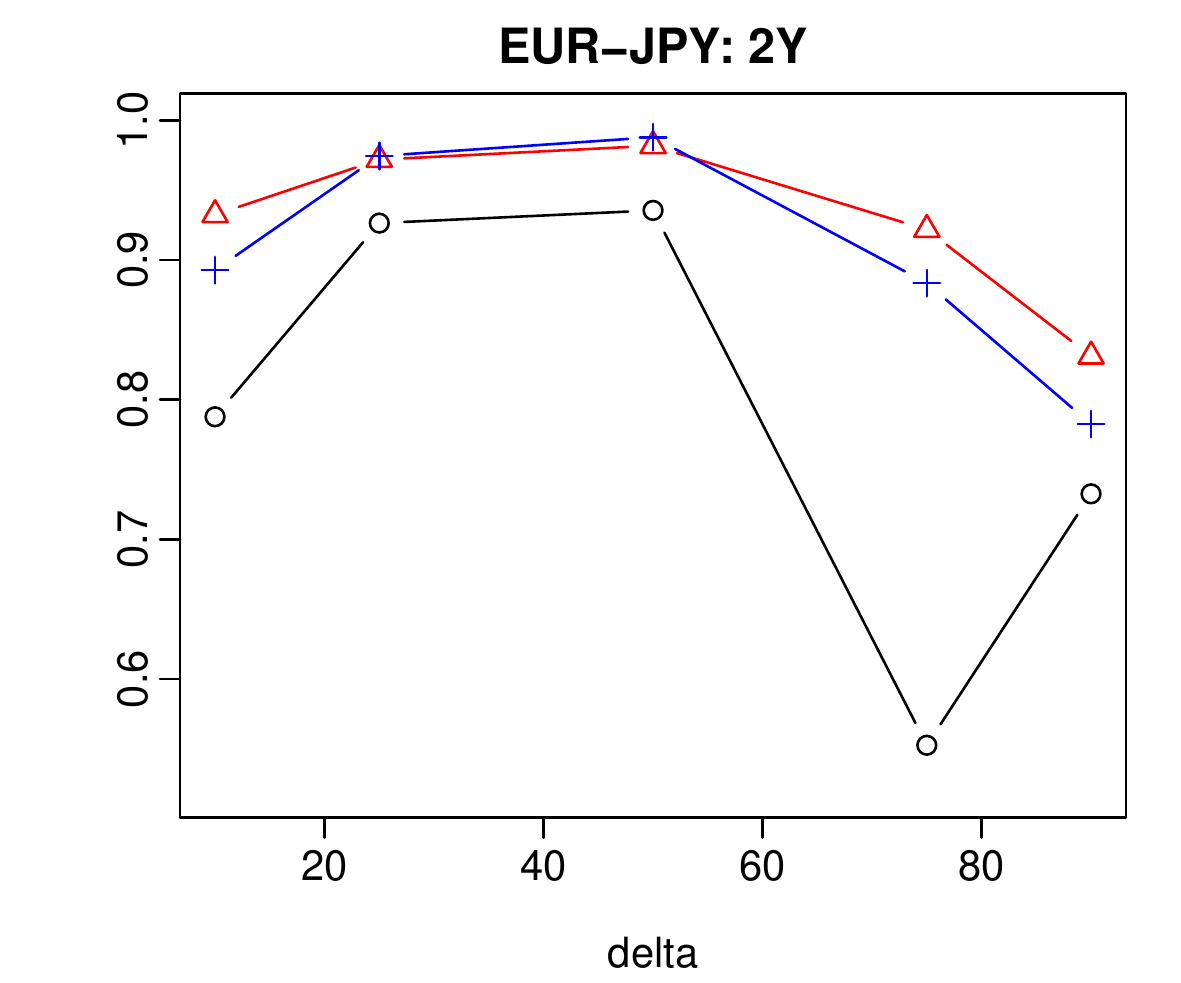}
\caption{\small{These figures present a comparison of goodness-of-fit among the three functional time-series methods over five delta values of 10, 25, 50, 75 and 90, as measured by $R^2(t)$. FTS, MFTS and MLFTS denote univariate, multivariate and multilevel functional time-series methods, respectively.}}\label{fig:model_fitting}
\end{figure}

In Figure~\ref{fig:model_fitting}, we present $R^2(t)$ over the five delta values, for the three maturities and three currencies.

Total $R^2$ is useful if one requires a single numerical measure of fit. It can be expressed as
\begin{equation*}
R^2 = \int_{t\in \mathcal{I}}R^2(t)dt. 
\end{equation*}

The total $R^2$ for the three maturities for each currency pair are listed in Table~\ref{tab:1}. \cite{KCM18} find that the univariate functional time-series model provides the best goodness-of-fit in comparison to the benchmark methods of \cite{GG06} and \cite{CT10, CT11}. Here, we improve on \cite{KCM18} by showing that the goodness-of-fit of the univariate functional time-series model they propose is outperformed by our multivariate and multilevel functional time-series models, as measured by total $R^2$. We also observe that the optimal $R^2$ values are often associated with the dynamic methods.

\begin{table}[!htbp]
\centering
\tabcolsep 0.55in
\caption{\small{This table presents a comparison of goodness-of-fit among the univariate, multivariate and multilevel functional time-series methods using static and dynamic functional principal component analyses, as measured by total $R^2$. The results are aggregated over the five delta values of 10, 25, 50, 75 and 90, over the January 2008 to December 2016 sample period. FTS, MFTS and MLFTS represent the univariate, multivariate and multilevel functional time-series models, respectively, with a leading `D' indicating that dynamic principal component evaluation is used. The largest $R^2$ values are highlighted in bold for each maturity and currency pair.}}\label{tab:1}
\renewcommand{\arraystretch}{0.8}
\begin{tabular}{@{}llccc@{}}
\toprule
& & \multicolumn{3}{c}{Maturity} \\
Currency & Method & 1M & 6M & 2Y \\\midrule
EUR-USD & FTS & 0.9387 & 0.9246 & 0.9174 \\ 
& MFTS & 0.9773 & 0.9842 & 0.9797 \\
& MLFTS & 0.9784 & 0.9715 & 0.9668 \\
& DFTS & 0.9976 & \textBF{0.9939} & 0.9877 \\
& DMFTS & 0.9849 & 0.9934 & 0.9914 \\
& DMLFTS & \textBF{0.9988} & 0.9885 & \textBF{0.9977} \\
\\
EUR-GBP & FTS & 0.9360 & 0.9207 & 0.9031 \\
& MFTS & 0.9811 & 0.9842 & 0.9723 \\
& MLFTS & 0.9752 & 0.9689 & 0.9265 \\
& DFTS & 0.9975 & 0.9929 & 0.9764 \\
& DMFTS & 0.9896 & 0.9942 & \textBF{0.9957} \\
& DMLFTS & \textBF{0.9979} & \textBF{0.9972} & 0.9796 \\
\\
EUR-JPY & FTS & 0.8553 & 0.8331 & 0.7869 \\
& MFTS & 0.9825 & 0.9721 & 0.9277 \\
& MLFTS & 0.9488 & 0.8863 & 0.9041 \\
& DFTS & 0.9032 & 0.8850 & 0.8337 \\
& DMFTS &0.9872 & \textBF{0.9929} & \textBF{0.9848} \\ 
& DMLFTS & \textBF{0.9892} & 0.9594 & 0.9811 \\
\bottomrule
\end{tabular}
\end{table}

A key measure of goodness-of-fit for the multilevel functional time-series method is the within-cluster variability defined in~\eqref{eq:10}. When the common factor explains the primary mode of total variability, the value of within-cluster variability is close to 1. We calculate this within-cluster variability for each currency pair and maturity period in Table~\ref{tab:2}.

\begin{table}[!htbp]
\tabcolsep 0.55in
\centering
\caption{\small{This table presents the computed within-cluster variability of the multilevel functional time-series method for each currency pair and each maturity period aggregated over the five delta values of 10, 25, 50, 75 and 90. The sample period is from January 2008 to December 2016. MLFTS and DMLFTS represent the static and dynamic multilevel functional time-series models, respectively. The largest values are highlighted in bold for each maturity and currency pair.}}\label{tab:2}
\begin{tabular}{@{}llccc@{}}
\toprule
Currency & Method & 1M & 6M & 2Y \\
\midrule
EUR-USD & MLFTS & 0.7507 & 0.8021 & \textBF{0.8360} \\
		& DMLFTS & \textBF{0.8384} & \textBF{0.8604} & 0.8251 \\
EUR-GBP & MLFTS & 0.7423 & 0.8087 & 0.7134 \\
		& DMLFTS & \textBF{0.7672} & \textBF{0.8197} & \textBF{0.7340} \\
EUR-JPY  & MLFTS & \textBF{0.6048} & 0.6918 & 0.7117 \\
		& DMLFTS & 0.6047 & \textBF{0.7258} & \textBF{0.7854} \\
\bottomrule
\end{tabular}
\end{table}

\subsection{Out-of-sample forecast evaluation}\label{sec:5.2}

It can be seen from the previous section that, the multivariate functional time-series and multilevel functional time-series models provide better in-sample fit when seeking to capture shape of the IV surface. We now turn our attention to out-of-sample forecasting. 

\subsubsection{Stationarity test of functional time-series}

Since the computation of long-run covariance is only meaningful if the functional time-series is stationary, we implement the stationarity test of \cite{HKR14} and its corresponding R function T\_stationary in the \textit{ftsa} package \citep{HS19}. Based on the $p$-values reported in Table~\ref{tab:3}, we conclude that all the functional time-series we consider are stationary.

\begin{table}[!htbp]
\tabcolsep 0.2in
\centering
\caption{\small{With the null hypothesis being that of stationarity, this table presents the probabilities of rejecting the null hypothesis ($p$-values) for each currency and maturity using the stationarity test of \cite{HKR14}.}}\label{tab:3}
\begin{tabular}{@{}llll@{}}
\toprule
Currency & 1M & 6M & 2Y \\
\midrule
EUR-USD & 0.533 & 0.424 & 0.415 \\
EUR-GBP & 0.771 & 0.747 & 0.487 \\
EUR-JPY & 0.455 & 0.646 & 0.363 \\
\bottomrule
\end{tabular}
\end{table}

\subsubsection{EUR-USD}

A summary of the out-of-sample forecast measures calculated under a recursive parameter estimation scheme and 522-day out-of-sample window length for EUR-USD are presented in Table~\ref{tab:4}. We consider the \citeauthor{CT11}'s \citeyearpar{CT11} method, the \citeauthor{GG06}'s \citeyearpar{GG06} method, RW and AR(1) as benchmakr methods. We use a generalised least squares (GLS) procedure to produce these \cite{GG06} and \cite{CT11} benchmarks, in line with \cite{BG14}. GLS is used to control for the option measurement errors identified by \cite{Hentschel03}. We direct the interested reader to Appendix B of \cite{GG06} for the technical detail of applying the GLS method suggested by \cite{Hentschel03} in the context of deterministic IV surface models.

\begin{center}
\tabcolsep 0.067in
\renewcommand{\arraystretch}{0.8}
\begin{small}
\begin{longtable}{@{}lllcccccccc@{}}
\caption{\small{To forecast EUR-USD IV from January 1, 2015 to December 30, 2016, we compute one-step-ahead, five-step-ahead and ten-step-ahead out-of-sample forecast performance measures for each maturity. The measures are aggregated over the five delta values of 10, 25, 50, 75 and 90. FTS, MFTS and MLFTS represent the univariate, multivariate and multilevel functional time-series models, respectively, with a leading `D' indicating that dynamic principal component evaluation is used. We use ARIMA$(p, d, q)$ with orders selected by the automatic algorithm of \cite{HK08} to forecast the principal component scores. CPV indicates the number of principal components $K$ is determined as the minimum that reaches a cumulative percentage of the variance of 99\%, whereas $K=4$ fixes this number at four. The smallest averaged error measures are highlighted in bold for each maturity and each horizon.}}\label{tab:4}
\\
\toprule
& &    & \multicolumn{4}{c}{CPV} & \multicolumn{4}{c}{$K = 4$} \\
Horizon & Maturity & Method& $\overline{\text{MAFE}}$ & $\overline{\text{MSFE}}$ & MME(U) & MME(O) & $\overline{\text{MAFE}}$ & $\overline{\text{MSFE}}$ & MME(U) & MME(O)  \\\hline
\endfirsthead
Horizon & Maturity & Method  & $\overline{\text{MAFE}}$ & $\overline{\text{MSFE}}$ & MME(U) & MME(O) & $\overline{\text{MAFE}}$ & $\overline{\text{MSFE}}$ & MME(U) & MME(O)   \\\hline
\endhead
\hline  \multicolumn{11}{r}{{Continued on next page}} \\ 
\endfoot
\endlastfoot
1 day &  1M & FTS & 0.5936 & 0.5709 & 0.6653 & 0.6356 & 0.5610 & 0.5155 & 0.6414 & 0.6062 \\ 
  & & MFTS & 0.6557 & 0.7748 & 0.7016 & 0.6878 & 0.5237 & 0.4776 & 0.6090 & 0.5749 \\ 
  & & MLFTS & 0.5074 & 0.4678 & 0.5910 & 0.5624 & 0.4429 & 0.3701 & 0.5314 & 0.5128 \\
  & & DFTS  & 0.4346 & 0.3586 & 0.5187 & 0.5094 & 0.3975 & 0.3177 & 0.4938 & 0.4669 \\ 
  & & DMFTS & 0.6426 & 0.7541 & 0.6830 & 0.6831 & 0.4587 & 0.3916 & 0.5503 & 0.5204 \\ 
  & & DMLFTS & 0.5493 & 0.5024 & 0.6100 & 0.6194 & 0.4023 & 0.3252 & 0.4978 & 0.4717 \\
  & & CT11 & 0.6040 & 0.6486 & 0.6479 & 0.6582 \\
  & & GG06 & 0.6675 & 0.7272 & 0.7097 & 0.7063 \\
  & & RW & 0.4071 & 0.3311 & 0.4983 & 0.4793 \\
  & & AR(1) & 0.4129 & 0.3328 & 0.5120 & 0.4802 \\
  \cmidrule{2-11}
   &  6M & FTS & 0.5026 & 0.3950 & 0.6258 & 0.5355 & 0.4580 & 0.3323 & 0.5941 & 0.4919 \\ 
  & & MFTS & 0.3708 & 0.2821 & 0.4685 & 0.4496 & 0.3791 & 0.2959 & 0.4699 & 0.4616 \\ 
  & & MLFTS & 0.3674 & 0.2772 & 0.4834 & 0.4253 & 0.2573 & 0.1585 & 0.3805 & 0.3277 \\
   & & DFTS & 0.3649 & 0.2462 & 0.4472 & 0.4719 & 0.2308 & 0.1557 & 0.3332 & 0.3142 \\ 
  & & DMFTS & 0.3748 & 0.2858 & 0.4439 & 0.4808 & 0.4013 & 0.3205 & 0.4610 & 0.5095 \\ 
  & & DMLFTS & 0.3961 & 0.2836 & 0.4638 & 0.5079 & 0.2288 & 0.1490 & 0.3320 & 0.3130 \\
  & & CT11 & 0.7322 & 0.8314 & 0.7871 & 0.7370 \\
  & & GG06 & 0.7557 & 0.9552 & 0.7966 & 0.7549 \\
  & & RW   & 0.2358 & 0.1775 & 0.3353 & 0.3175 \\
  & & AR(1) & 0.2395 & 0.1704 & 0.3361 & 0.3187 \\
  \cmidrule{2-11}
   & 2Y & FTS & 0.4629 & 0.3101 & 0.6122 & 0.4885 & 0.4162 & 0.2401 & 0.5815 & 0.4430 \\ 
  & & MFTS & 0.3360 & 0.1977 & 0.4880 & 0.3738 & 0.3374 & 0.1879 & 0.4970 & 0.3728 \\ 
  & & MLFTS & 0.3070 & 0.1772 & 0.4399 & 0.3631 & 0.2280 & 0.0953 & 0.3713 & 0.2892 \\
   & & DFTS & 0.3259 & 0.1527 & 0.4213 & 0.4423 & 0.1617 & 0.0582 & 0.2643 & 0.2556 \\ 
  & & DMFTS & 0.2713 & 0.1384 & 0.4015 & 0.3383 & 0.2600 & 0.1206 & 0.4082 & 0.3149 \\ 
  & & DMLFTS & 0.1960 & 0.0746 & 0.2949 & 0.3001 & 0.1576 & 0.0561 & 0.2604 & 0.2504 \\
  & & CT11 & 0.2548 & 0.1203 & 0.3585 & 0.3529 \\
  & & GG06 & 0.2435 & 0.1067 & 0.3521 & 0.3403 \\
  & & RW & 0.1593 & 0.0629 & 0.2599 & 0.2479 \\
  & & AR(1) & 0.1626 & 0.0643 & 0.2596 & 0.2466 \\
   \cmidrule{2-11}
   & Average & FTS & 0.5197 & 0.4253 & 0.6344 & 0.5532 & 0.4784 & 0.3626 & 0.6057 & 0.5137 \\ 
  & & MFTS & 0.4542 & 0.4182 & 0.5527 & 0.5037 & 0.4134 & 0.3205 & 0.5253 & 0.4698 \\
  & & MLFTS & 0.3939 & 0.3074 & 0.5048 & 0.4503 & 0.3094 & 0.2080 & 0.4277 & 0.3766\\
  & & DFTS & 0.3751 & 0.2525 & 0.4624 & 0.4745 & 0.2633 & 0.1772 & 0.3638 & 0.3456 \\ 
  & & DMFTS & 0.4295 & 0.3928 & 0.5095 & 0.5007 & 0.3733 & 0.2776 & 0.4732 & 0.4483 \\
  & & DMLFTS & 0.3805 & 0.2869 & 0.4562 & 0.4758 & \textBF{0.2629} & \textBF{0.1768} & \textBF{0.3634} & \textBF{0.3450} \\
  & & CT11 & 0.5303 & 0.5334 & 0.5978 & 0.5827  \\
  & & GG06 & 0.5556 & 0.5964 & 0.6195 & 0.6005 \\
  & & RW & 0.2674 & 0.1905 & 0.3645 & 0.3482 \\
  & & AR(1) & 0.2717 & 0.1892 & 0.3692 & 0.3452 \\
\midrule
5 day & 1M & FTS & 0.9598 & 1.5348 & 0.9398 & 0.9200 & 0.9481 & 1.5261 & 0.9327 & 0.9075 \\ 
  & & MFTS & 1.0091 & 1.8200 & 0.9640 & 0.9579 & 0.9365 & 1.5892 & 0.9172 & 0.8980 \\ 
  & & MLFTS & 0.9298 & 1.5650 & 0.9284 & 0.8769 & 0.9116 & 1.5657 & 0.9140 & 0.8607 \\
  & & DFTS & 0.9172 & 1.5958 & 0.9095 & 0.8703 & 0.9002 & 1.5563 & 0.8981 & 0.8554 \\ 
 & & DMFTS & 1.0162 & 1.8861 & 0.9672 & 0.9597 & 0.9270 & 1.6338 & 0.9075 & 0.8862 \\ 
  & & DMLFTS & 0.9652 & 1.6986 & 0.9411 & 0.9136 & 0.9092 & 1.5883 & 0.9047 & 0.8629 \\
  & & CT11 & 0.9953 & 1.7950 & 0.9612 & 0.9355 \\
  & & GG06 & 1.0148 & 1.8012 & 0.9831 & 0.9459 \\
  & & RW & 0.9100 & 1.5753 & 0.9080 & 0.8625 \\  
  & & AR(1) & 0.9203 & 1.5898 & 0.9206 & 0.8879 \\
   \cmidrule{2-11}
   & 6M & FTS & 0.6420 & 0.6596 & 0.7153 & 0.6685 & 0.6124 & 0.6189 & 0.6935 & 0.6397 \\ 
   & & MFTS & 0.5758 & 0.6414 & 0.6477 & 0.6105 & 0.5753 & 0.6440 & 0.6455 & 0.6109 \\ 
   & & MLFTS & 0.5640 & 0.6032 & 0.6400 & 0.6006 & 0.5123 & 0.5325 & 0.5942 & 0.5606 \\
   & & DFTS & 0.5732 & 0.6186 & 0.6290 & 0.6291 & 0.5042 & 0.5289 & 0.5791 & 0.5584 \\ 
   & & DMFTS & 0.5853 & 0.6757 & 0.6476 & 0.6248 & 0.5983 & 0.7091 & 0.6570 & 0.6339 \\ 
   & & DMLFTS & 0.5933 & 0.6528 & 0.6362 & 0.6540 & 0.5000 & 0.5248 & 0.5761 & 0.5533 \\
   & & CT11 & 0.8295 & 1.2002 & 0.8687 & 0.7917 \\
   & & GG06 & 0.8726 & 1.3253 & 0.8884 & 0.8330 \\
   & & RW & 0.5127 & 0.5441 & 0.5874 & 0.5657 \\
   & & AR(1) & 0.5140 & 0.5362 & 0.5943 & 0.5638 \\
   \cmidrule{2-11}
   & 2Y  & FTS & 0.5389 & 0.4449 & 0.6454 & 0.5729 & 0.5051 & 0.3882 & 0.6244 & 0.5399 \\ 
  & & MFTS & 0.4615 & 0.3782 & 0.5626 & 0.5124 & 0.4583 & 0.3668 & 0.5610 & 0.5095 \\ 
& & MLFTS & 0.4409 & 0.3503 & 0.5348 & 0.5045 & 0.3995 & 0.2974 & 0.4982 & 0.4710 \\
   & & DFTS & 0.4656 & 0.3595 & 0.5396 & 0.5515 & 0.3733 & 0.2662 & 0.4662 & 0.4575 \\ 
 & & DMFTS & 0.4259 & 0.3418 & 0.5196 & 0.4929 & 0.4194 & 0.3233 & 0.5154 & 0.4873 \\ 
& & DMLFTS & 0.3904 & 0.2840 & 0.4775 & 0.4760 & 0.3742 & 0.2669 & 0.4658 & 0.4595\\
& & CT11 & 0.4203 & 0.3196 & 0.5062 & 0.5005 \\
& & GG06 & 0.4220 & 0.3206 & 0.5088 & 0.5020 \\
& & RW & 0.3765 & 0.2713 & 0.4692 & 0.4593 \\
& & AR(1) & 0.3754 & 0.2647 & 0.4656 & 0.4543 \\
   \cmidrule{2-11}
   & Average & FTS & 0.7148 & 0.8862 & 0.7668 & 0.7215 & 0.6885 & 0.8444 & 0.7502 & 0.6957 \\ 
  & & MFTS & 0.6821 & 0.9465 & 0.7248 & 0.6936 & 0.6567 & 0.8667 & 0.7079 & 0.6728 \\
  & & MLFTS & 0.6449 & 0.8395 & 0.7011 & 0.6607 & 0.6078 & 0.7985 & 0.6688 & 0.6308\\
  & & DFTS & 0.6520 & 0.8580 & 0.6926 & 0.6837 & \textBF{0.5926} &  \textBF{0.7838} & \textBF{0.6478} & \textBF{0.6238} \\ 
  & & DMFTS & 0.6758 & 0.9679 & 0.7115 & 0.6925 & 0.6482 & 0.8887 & 0.6933 & 0.6691 \\
  & & DMLFTS & 0.6496 & 0.8785 & 0.6849 & 0.6812 & 0.5945 & 0.7933 & 0.6489 & 0.6252\\
  & & CT11 & 0.7484 & 1.1049 & 0.7787 & 0.7426  \\
  & & GG06 & 0.7698 & 1.1490 & 0.7934 & 0.7603 \\
  & & RW & 0.5997 & 0.7969 & 0.6549 & 0.6292 \\
  & & AR(1) & 0.6032 & 0.7969 & 0.6602 & 0.6353 \\
\midrule   
10 day & 1M & FTS & 1.2442 & 2.4842 & 1.1378 & 1.1328 & 1.2452 & 2.5329 & 1.1433 & 1.1267 \\ 
& & MFTS & 1.2935 & 2.8260 & 1.1776 & 1.1548 & 1.2534 & 2.6819 & 1.1543 & 1.1210 \\ 
& & MLFTS & 1.2405 & 2.6064 & 1.1631 & 1.0944 & 1.2527 & 2.7404 & 1.1746 & 1.0978 \\
 & & DFTS & 1.2862 & 2.8628 & 1.1875 & 1.1350 & 1.2747 & 2.8248 & 1.1792 & 1.1262 \\ 
& & DMFTS & 1.3352 & 3.0192 & 1.2095 & 1.1798 & 1.2834 & 2.8693 & 1.1732 & 1.1418 \\ 
& & DMLFTS & 1.2865 & 2.8253 & 1.1816 & 1.1383 & 1.2774 & 2.8527 & 1.1801 & 1.1283 \\
& & CT11 & 1.3039 & 2.9100 & 1.1966 & 1.1457 \\
& & GG06 & 1.2983 & 2.8572 & 1.1936 & 1.1419 \\
& & RW & 1.2786 & 2.8572 & 1.1815 & 1.1264\\
& & AR(1) & 1.2628 & 2.7316 & 1.1809 & 1.1291 \\
\cmidrule{2-11}     
     & 6M & FTS & 0.7615 & 0.9441 & 0.7948 & 0.7717 & 0.7454 & 0.9348 & 0.7850 & 0.7538 \\ 
& & MFTS & 0.7324 & 0.9967 & 0.7768 & 0.7300 & 0.7347 & 1.0048 & 0.7812 & 0.7280 \\ 
& & MLFTS & 0.7138 & 0.9389 & 0.7529 & 0.7239 & 0.6908 & 0.9378 & 0.7376 & 0.6979 \\
& & DFTS & 0.7386 & 1.0367 & 0.7624 & 0.7498 & 0.6849 & 0.9482 & 0.7300 & 0.6916 \\ 
& & DMFTS & 0.7525 & 1.0859 & 0.7918 & 0.7407 & 0.7635 & 1.1112 & 0.8037 & 0.7469 \\ 
& & DMLFTS & 0.7625 & 1.0729 & 0.7856 & 0.7664 & 0.6866 & 0.9424 & 0.7287 & 0.6965 \\
& & CT11 & 0.9473 & 1.6022 & 0.9614 & 0.8680 \\
& & GG06 & 1.0002 & 1.7614 & 0.9880 & 0.9216 \\
& & RW & 0.6843 & 0.9544 & 0.7323 & 0.6942 \\
& & AR(1) & 0.6863 & 0.9337 & 0.7347 & 0.6947  \\
\cmidrule{2-11}
     & 2Y  & FTS & 0.6001 & 0.5765 & 0.6752 & 0.6368 & 0.5793 & 0.5363 & 0.6631 & 0.6173 \\ 
& & MFTS & 0.5508 & 0.5400 & 0.6177 & 0.6021 & 0.5522 & 0.5417 & 0.6157 & 0.6062 \\ 
& & MLFTS & 0.5340 & 0.5083 & 0.5973 & 0.5971 & 0.5187 & 0.4907 & 0.5906 & 0.5795 \\
& & DFTS & 0.5820 & 0.5690 & 0.6359 & 0.6443 & 0.5085 & 0.4772 & 0.5824 & 0.5700 \\ 
& & DMFTS & 0.5399 & 0.5360 & 0.6076 & 0.5943 & 0.5374 & 0.5222 & 0.6016 & 0.5976 \\ 
& & DMLFTS & 0.5166 & 0.4836 & 0.5842 & 0.5821 & 0.5081 & 0.4709 & 0.5792 & 0.5743 \\
& & CT11 & 0.5416 & 0.5213 & 0.6071 & 0.6029 \\
& & GG06 & 0.5443 & 0.5316 & 0.6038 & 0.6091 \\
& & RW & 0.5058 & 0.4717 & 0.5829 & 0.5691 \\
& & AR(1) & 0.5120 & 0.5328 & 0.5812 & 0.5704 \\
\cmidrule{2-11}
     & Average & FTS & 0.8702 & 1.3441 & 0.8707 & 0.8480 & 0.8566 & \textBF{1.3347} & 0.8638 & 0.8326 \\ 
  & & MFTS & 0.8589 & 1.4542 & 0.8574 & 0.8290 & 0.8468 & 1.4095 & 0.8504 & 0.8184 \\
  & & MLFTS & 0.8294 & 1.3512 & 0.8377 & 0.8051 & 0.8207 & 1.3896 & 0.8343 & \textBF{0.7917} \\
  & & DFTS & 0.8689 & 1.4893 & 0.8619 & 0.8430 & 0.8227 & 1.4167 & 0.8305 & 0.7959 \\ 
  & & DMFTS & 0.8759 & 1.5470 & 0.8696 & 0.8383 & 0.8614 & 1.5009 & 0.8595 & 0.8288 \\
  & & DMLFTS & 0.8552 & 1.4606 & 0.8505 & 0.8289 & 0.8240 & 1.4220 & \textBF{0.8293} & 0.7997 \\
  & & CT11 & 0.9309 & 1.6778 & 0.9217 & 0.8722 \\
  & & GG06 & 0.9476 & 1.7167 & 0.9285 & 0.8909  \\
  & & RW & 0.8229 & 1.4278 & 0.8322 & 0.7966 \\
  & & AR(1) & \textBF{0.8204} & 1.3394 & 0.8323 & 0.7981 \\
\bottomrule
\end{longtable}
\end{small}
\end{center}

In Table~\ref{tab:4_moneyness}, we present a comparison of out-of-sample accuracy for forecasting EUR-USD IV for the five delta values considered. The errors are aggregated over three maturities of one month, six months and two years. In the supplement, we also present the results using exponential smoothing, which is a generalisation of the ARIMA model \citep[see, e.g.,][]{KS19}.

\begin{center}
\tabcolsep 0.08in
\renewcommand{\arraystretch}{0.95}
\begin{small}
\begin{longtable}{@{}llrrrrrrrrrr@{}}
\caption{\small{This table presents one-step-ahead, five-step-ahead and ten-step-ahead out-of-sample performance measures for forecasting EUR-USD IV from January 1, 2015 to December 30, 2016, for the five delta values of 10, 25, 50, 75 and 90. The measures are aggregated over three maturities of one month, six months and two years.}}\label{tab:4_moneyness} \\
\toprule
& & \multicolumn{5}{c}{$\overline{\text{MAFE}}$} & \multicolumn{5}{c}{$\overline{\text{MSFE}}$} \\
\underline{Horizon} & & \multicolumn{5}{c}{Delta} & \multicolumn{5}{c}{Delta} \\
$K$ & Method & 10 & 25 & 50 & 75 & 90 & 10 & 25 & 50 & 75 & 90 \\ 
\endfirsthead
\midrule
\underline{1 day} \\
  CPV & FTS & 0.7114 & 0.4847 & 0.3888 & 0.4507 & 0.5628 & 0.7330 & 0.3712 & 0.2442 & 0.2928 & 0.4853 \\ 
  & DFTS & 0.4811 & 0.3305 & 0.4047 & 0.2882 & 0.3711 & 0.3981 & 0.2273 & 0.2491 & 0.1484 & 0.2396 \\ 
  & MFTS & 0.6692 & 0.4438 & 0.2969 & 0.3430 & 0.5178 & 0.8482 & 0.3783 & 0.1634 & 0.2171 & 0.4841 \\ 
  & DMFTS & 0.6273 & 0.4518 & 0.2869 & 0.3087 & 0.4729 & 0.8019 & 0.3716 & 0.1610 & 0.1997 & 0.4296 \\ 
  & MLFTS & 0.5441 & 0.3520 & 0.2735 & 0.3400 & 0.4599 & 0.5283 & 0.2509 & 0.1544 & 0.2095 & 0.3939 \\ 
  & DMLFTS & 0.4819 & 0.3568 & 0.4003 & 0.2896 & 0.3738 & 0.4393 & 0.2621 & 0.2857 & 0.1691 & 0.2781 \\ 
  & CT11 & 0.8555 & 0.5321 & 0.2413 & 0.4563 & 0.5664 & 1.1277 & 0.5310 & 0.1411 & 0.3431 & 0.5244 \\ 
  & GG06 & 0.8299 & 0.5783 & 0.3702 & 0.4428 & 0.5565 & 1.2154 & 0.5403 & 0.2565 & 0.3739 & 0.5959 \\ 
  & RW & 0.4388 & 0.3692 & 0.3255 & 0.3133 & 0.3404 & 0.4113 & 0.2817 & 0.2264 & 0.2070 & 0.2762 \\ 
  & AR(1) & 0.4440 & 0.3738 & 0.3296 & 0.3178 & 0.3432 & 0.4117 & 0.2827 & 0.2277 & 0.2085 & 0.2651 \\ 
\\
  $K=4$ & FTS & 0.4916 & 0.5449 & 0.4583 & 0.4903 & 0.4069 & 0.4099 & 0.4552 & 0.3176 & 0.3459 & 0.2846 \\ 
  & DFTS & 0.3392 & 0.2759 & 0.2364 & 0.2252 & 0.2400 & 0.2977 & 0.1835 & 0.1303 & 0.1127 & 0.1618 \\ 
  & MFTS & 0.6084 & 0.4289 & 0.3088 & 0.3101 & 0.4107 & 0.6274 & 0.3191 & 0.1709 & 0.1721 & 0.3129 \\ 
  & DMFTS & 0.5603 & 0.4209 & 0.3045 & 0.2640 & 0.3171 & 0.5566 & 0.3144 & 0.1741 & 0.1357 & 0.2069 \\ 
  & MLFTS & 0.3777 & 0.3094 & 0.2601 & 0.2687 & 0.3310 & 0.3273 & 0.2048 & 0.1406 & 0.1404 & 0.2267 \\ 
  & DMLFTS & 0.3393 & 0.2734 & 0.2328 & 0.2233 & 0.2457 & 0.2972 & 0.1834 & 0.1286 & 0.1126 & 0.1622 \\ 
\underline{5 day} \\
CPV & FTS & 0.9794 & 0.7249 & 0.5826 & 0.5902 & 0.6907 & 1.5516 & 0.9166 & 0.5917 & 0.5661 & 0.7729 \\ 
  & DFTS & 0.8839 & 0.6681 & 0.6219 & 0.5129 & 0.5732 & 1.5173 & 0.9570 & 0.6965 & 0.4996 & 0.6195 \\ 
  & MFTS & 0.9677 & 0.7142 & 0.5403 & 0.5305 & 0.6581 & 1.7978 & 1.0189 & 0.5774 & 0.5340 & 0.8045 \\ 
  & DMFTS & 0.9600 & 0.7292 & 0.5470 & 0.5143 & 0.6284 & 1.8601 & 1.0722 & 0.6011 & 0.5332 & 0.7727 \\ 
  & MLFTS & 0.8848 & 0.6623 & 0.5344 & 0.5302 & 0.6126 & 1.4774 & 0.8905 & 0.5689 & 0.5334 & 0.7272 \\ 
  & DMLFTS & 0.8776 & 0.6773 & 0.6076 & 0.5076 & 0.5781 & 1.5378 & 0.9682 & 0.7178 & 0.5132 & 0.6555 \\ 
  & CT11 & 1.1232 & 0.7907 & 0.5209 & 0.6331 & 0.6741 & 2.2234 & 1.1967 & 0.5653 & 0.6793 & 0.8600 \\ 
  & GG06 & 1.1279 & 0.8204 & 0.5879 & 0.6073 & 0.7057 & 2.2554 & 1.2012 & 0.6532 & 0.6880 & 0.9473 \\ 
  & RW & 0.9040 & 0.7474 & 0.6212 & 0.5733 & 0.6028 & 1.5274 & 1.0120 & 0.6797 & 0.5659 & 0.6494 \\ 
  & AR(1) & 0.9129 & 0.7544 & 0.6269 & 0.5800 & 0.6089 & 1.5404 & 1.0255 & 0.6841 & 0.5663 & 0.6348 \\ 
\\
$K=4$ & FTS & 0.8566 & 0.7602 & 0.6258 & 0.6211 & 0.5789 & 1.3881 & 0.9659 & 0.6439 & 0.6149 & 0.6091 \\ 
  & DFTS & 0.8032 & 0.6491 & 0.5287 & 0.4782 & 0.5037 & 1.4280 & 0.9107 & 0.5774 & 0.4669 & 0.5360 \\ 
  & MFTS & 0.9356 & 0.7044 & 0.5430 & 0.5106 & 0.5900 & 1.6303 & 0.9723 & 0.5777 & 0.4918 & 0.6612 \\ 
  & DMFTS & 0.9296 & 0.7216 & 0.5543 & 0.4921 & 0.5434 & 1.7016 & 1.0426 & 0.6156 & 0.4850 & 0.5988 \\ 
  & MLFTS & 0.8251 & 0.6495 & 0.5280 & 0.4924 & 0.5441 & 1.4799 & 0.8797 & 0.5604 & 0.4739 & 0.5985 \\ 
  & DMLFTS & 0.8075 & 0.6514 & 0.5273 & 0.4796 & 0.5066 & 1.4564 & 0.9289 & 0.5816 & 0.4678 & 0.5320 \\ 
\underline{10 day} \\
 CPV & FTS & 1.2108 & 0.9144 & 0.7186 & 0.7019 & 0.7972 & 2.4328 & 1.4580 & 0.9144 & 0.8170 & 1.0525 \\ 
  & DFTS & 1.2000 & 0.9142 & 0.7899 & 0.6727 & 0.7678 & 2.7602 & 1.7044 & 1.1267 & 0.8382 & 1.0182 \\ 
  & MFTS & 1.2233 & 0.9244 & 0.7035 & 0.6604 & 0.7828 & 2.7716 & 1.6262 & 0.9481 & 0.8182 & 1.1070 \\ 
  & DMFTS & 1.2493 & 0.9541 & 0.7251 & 0.6650 & 0.7858 & 2.9874 & 1.7623 & 1.0130 & 0.8506 & 1.1217 \\ 
  & MLFTS & 1.1544 & 0.8845 & 0.6988 & 0.6678 & 0.7416 & 2.4566 & 1.5021 & 0.9409 & 0.8198 & 1.0366 \\ 
  & DMLFTS & 1.1880 & 0.9117 & 0.7609 & 0.6562 & 0.7591 & 2.6871 & 1.6501 & 1.1105 & 0.8283 & 1.0269 \\ 
  & CT11 & 1.3822 & 1.0004 & 0.6993 & 0.7721 & 0.8009 & 3.3731 & 1.8564 & 0.9667 & 0.9949 & 1.1981 \\ 
  & GG06 & 1.3937 & 1.0222 & 0.7478 & 0.7339 & 0.8403 & 3.4007 & 1.8671 & 1.0235 & 0.9891 & 1.3032 \\ 
  & RW & 1.2423 & 1.0022 & 0.8120 & 0.7488 & 0.8058 & 2.7958 & 1.7611 & 1.1154 & 0.9151 & 1.0515 \\ 
  & AR(1) & 1.2410 & 1.0041 & 0.8026 & 0.7364 & 0.7902 & 2.6958 & 1.7183 & 1.0729 & 0.8776 & 0.9875 \\ 
  \\
 $K=4$ & FTS & 1.1489 & 0.9336 & 0.7488 & 0.7232 & 0.7287 & 2.4680 & 1.4674 & 0.9427 & 0.8594 & 0.9357 \\ 
  & DFTS & 1.1438 & 0.9028 & 0.7116 & 0.6509 & 0.7044 & 2.6916 & 1.6516 & 1.0033 & 0.8073 & 0.9299 \\ 
  & MFTS & 1.2095 & 0.9249 & 0.7096 & 0.6482 & 0.7416 & 2.6913 & 1.6134 & 0.9562 & 0.7878 & 0.9986 \\ 
  & DMFTS & 1.2333 & 0.9538 & 0.7310 & 0.6518 & 0.7371 & 2.9213 & 1.7583 & 1.0243 & 0.8106 & 0.9899 \\ 
  & MLFTS & 1.1548 & 0.8841 & 0.6952 & 0.6439 & 0.7256 & 2.7230 & 1.5367 & 0.9401 & 0.7787 & 0.9697 \\ 
  & DMLFTS & 1.1496 & 0.9059 & 0.7115 & 0.6479 & 0.7052 & 2.7216 & 1.6731 & 1.0037 & 0.8017 & 0.9099 \\ 
  \bottomrule
\end{longtable}
\end{small}
\end{center}

From Tables~\ref{tab:4} and~\ref{tab:4_moneyness}, we observe that the proposed dynamic functional principal component analysis extension leads to the univariate functional time-series forecasting method being the most accurate method in general. This result is due to its ability to capture the time-series momentum present in the data accurately \citep[see also][]{MGG20}. The benchmark methods RW and AR(1) are quite hard to beat, for short-term forecasts. However, the advantages of implementing our functional time-series method becomes evident at longer horizons. Between the two methods of selecting the number of functional principal components, we find that $K=4$ often produces improved forecasts with smaller errors. The multivariate and multilevel functional time-series methods are designed to improve forecasts by modelling correlation between the three maturities. However, we find that the independent functional time-series method improves the most in terms of forecast accuracy when we consider the dynamic functional principal component analysis. In the trading strategy in Section~\ref{sec:trading}, we consider this dynamic independent functional time-series method.

\subsubsection{EUR-GBP}

Similarly, a summary of the out-of-sample forecast measures calculated under a recursive estimation scheme and 522-day out-of-sample window length for EUR-GBP is presented in Table~\ref{tab:5}.
%As benchmark methods, we also consider the \citeauthor{CT11}'s \citeyearpar{CT11} method, the \citeauthor{GG06}'s \citeyearpar{GG06} method, RW and AR(1). 

\begin{center}
\tabcolsep 0.067in
\renewcommand{\arraystretch}{0.95}
\begin{small}
\begin{longtable}{@{}lllcccccccc@{}}
\caption{\small{This table presents one-step-ahead, five-step-ahead and ten-step-ahead out-of-sample performance measures for forecasting EUR-GBP IV from January 1, 2015 to December 30, 2016, for maturities of one month, six months, and two years. The measures are aggregated over the five delta values of 10, 25, 50, 75 and 90. FTS, MFTS and MLFTS represent the univariate, multivariate and multilevel functional time-series models, respectively, with a leading `D' indicating that dynamic principal component evaluation is used. When forecasting principal component scores, we use ARIMA$(p, d, q)$ with orders selected by the automatic algorithm of \cite{HK08}. CPV indicates that the number of principal components $K$ is determined as the minimum that reaches a cumulative percentage of variance level of 99\%, whereas $K=4$ fixes this number at four. The smallest averaged error measures are highlighted in bold for each maturity and each horizon.}}\label{tab:5}\\
\toprule
& &    & \multicolumn{4}{c}{CPV} & \multicolumn{4}{c}{$K = 4$} \\
Horizon & Maturity & Method& $\overline{\text{MAFE}}$ & $\overline{\text{MSFE}}$ & MME(U) & MME(O) & $\overline{\text{MAFE}}$ & $\overline{\text{MSFE}}$ & MME(U) & MME(O)  \\\hline
\endfirsthead
Horizon & Maturity & Method  & $\overline{\text{MAFE}}$ & $\overline{\text{MSFE}}$ & MME(U) & MME(O) & $\overline{\text{MAFE}}$ & $\overline{\text{MSFE}}$ & MME(U) & MME(O)   \\\hline
\endhead
\hline  \multicolumn{11}{r}{{Continued on next page}} \\ 
\endfoot
\endlastfoot
1 day & 1M & FTS  & 0.6010 & 1.3218 & 0.6702 & 0.5914 & 0.5678 & 1.1207 & 0.6382 & 0.5724 \\ 
& & MFTS & 0.5742 & 1.0926 & 0.6275 & 0.5979 & 0.5784 & 1.1059 & 0.6310 & 0.6008 \\ 
& & MLFTS & 0.5583 & 1.0616 & 0.6182 & 0.5796 & 0.4850 & 0.7443 & 0.5461 & 0.5387 \\
& & DFTS & 0.4664 & 0.6863 & 0.5330 & 0.5269 & 0.4445 & 0.6653 & 0.5145 & 0.5034 \\ 
& & DMFTS & 0.6140 & 1.0346 & 0.6648 & 0.6333 & 0.6925 & 1.7469 & 0.7448 & 0.6557 \\ 
& & DMLFTS & 0.6107 & 0.8390 & 0.6602 & 0.6545 & 0.4408 & 0.6465 & 0.5101 & 0.5025 \\
& & CT11 & 0.6304 & 1.0513 & 0.6426 & 0.6716 \\
& & GG06 & 0.7738 & 1.5992 & 0.7844 & 0.7491 \\
& & RW & 0.4492 & 0.6483 & 0.5178 & 0.5105 \\
& & AR(1) & 0.4814 & 0.8123 & 0.5583 & 0.5134 \\
  \cmidrule{2-11}
  & 6M & FTS & 0.5524 & 0.7062 & 0.6281 & 0.5761 & 0.4932 & 0.5159 & 0.5802 & 0.5365 \\ 
& & MFTS & 0.4347 & 0.4427 & 0.5115 & 0.5022 & 0.4659 & 0.5291 & 0.5368 & 0.5241 \\ 
& & MLFTS & 0.5155 & 0.6291 & 0.5781 & 0.5655 & 0.3006 & 0.2205 & 0.3777 & 0.4046 \\
& & DFTS & 0.3269 & 0.2102 & 0.4180 & 0.4304 & 0.2411 & 0.1550 & 0.3391 & 0.3328 \\ 
& & DMFTS & 0.4459 & 0.3946 & 0.5215 & 0.5243 & 0.5072 & 0.5213 & 0.5467 & 0.5972 \\ 
& & DMLFTS & 0.4290 & 0.3126 & 0.5196 & 0.5148 & 0.2338 & 0.1542 & 0.3301 & 0.3252 \\
& & CT11 & 0.6920 & 1.2740 & 0.7615 & 0.6415 \\
& & GG06 & 0.8532 & 1.7669 & 0.8661 & 0.7945 \\
& & RW & 0.2419 & 0.1552 & 0.3373 & 0.3337 \\
& & AR(1) & 0.2501 & 0.1638 & 0.3451 & 0.3433 \\
  \cmidrule{2-11}
    & 2Y & FTS & 0.6179 & 0.5708 & 0.6973 & 0.6521 & 0.5953 & 0.5224 & 0.6835 & 0.6325 \\ 
& & MFTS & 0.5406 & 0.6739 & 0.6063 & 0.5791 & 0.7079 & 1.2069 & 0.7185 & 0.7215 \\ 
& & MLFTS & 0.7356 & 0.9265 & 0.7526 & 0.7673 & 0.6031 & 0.6936 & 0.6676 & 0.6393 \\
& & DFTS & 0.4794 & 0.2838 & 0.5493 & 0.6006 & 0.1544 & 0.0549 & 0.2556 & 0.2464 \\ 
& & DMFTS & 0.3110 & 0.2133 & 0.4054 & 0.3976 & 0.4279 & 0.7189 & 0.5089 & 0.4738 \\ 
& & DMLFTS & 0.3788 & 0.2617 & 0.4752 & 0.4581 & 0.1634 & 0.0599 & 0.2675 & 0.2561 \\
& & CT11 & 1.1444 & 1.9326 & 1.0844 & 1.0492 \\
& & GG06 & 1.0720 & 1.6410 & 1.0369 & 0.9965 \\
& & RW & 0.1582 & 0.0619 & 0.2582 & 0.2477 \\
& & AR(1) & 0.1611 & 0.0631 & 0.2673 & 0.2462 \\
  \cmidrule{2-11} 
  & Average & FTS &  0.5904 & 0.8663 & 0.6652 & 0.6065 & 0.5521 & 0.7197 & 0.6340 & 0.5805 \\
  & & MFTS & 0.5165 & 0.7364 & 0.5818 & 0.5597 & 0.5841 & 0.9473 & 0.6288 & 0.6155 \\
  & & MLFTS & 0.6032 & 0.8724 & 0.6496 & 0.6375 & 0.4629 & 0.5528 & 0.5305 & 0.5275 \\
  & & DFTS & 0.4242 & 0.3934 & 0.5001 & 0.5193  & 0.2800 & 0.2917 & 0.3697 & \textBF{0.3609} \\
  & & DMFTS & 0.4569 & 0.5475 & 0.5306 & 0.5184 & 0.5425 & 0.9957 & 0.6001 & 0.5756 \\
  & & DMLFTS & 0.4728 & 0.4711 & 0.5517 & 0.5425 & \textBF{0.2793} & \textBF{0.2869} & \textBF{0.3692} & 0.3613 \\
& & CT11 & 0.8223 & 1.4193 & 0.8295 & 0.7874 \\
&& GG06 & 0.8997 & 1.6690 & 0.8958 & 0.8467 \\
& & RW & 0.2831 & 0.2885 & 0.3711 & 0.3640 \\
& & AR(1) & 0.2975 & 0.3464 & 0.3902 & 0.3676 \\
\midrule
5 day & 1M & FTS & 1.0124 & 3.3659 & 1.0060 & 0.8683 & 1.0207 & 3.3148 & 1.0026 & 0.8865 \\ 
& & MFTS & 1.0519 & 3.4120 & 0.9989 & 0.9346 & 1.0581 & 3.4661 & 1.0037 & 0.9381 \\ 
& & MLFTS & 1.0388 & 3.3356 & 1.0157 & 0.8998 & 1.0437 & 3.2438 & 0.9957 & 0.9271 \\
& & DFTS & 1.0698 & 3.4298 & 0.9858 & 0.9744 & 1.0586 & 3.4123 & 0.9773 & 0.9634 \\ 
& & DMFTS & 1.0986 & 3.5501 & 1.0088 & 0.9917 & 1.1295 & 3.8937 & 1.0569 & 0.9877 \\ 
& & DMLFTS & 1.1539 & 3.6709 & 1.0515 & 1.0401 & 1.0622 & 3.4135 & 0.9813 & 0.9660 \\
& & CT11 & 1.1319 & 3.4981 & 1.0442 & 1.0070 \\
& & GG06 & 1.1574 & 3.6682 & 0.7978 & 0.7444 \\
& & RW & 1.0746 & 3.4042 & 0.9922 & 0.9804 \\
& & AR(1) & 1.0769 & 3.5476 & 1.0482 & 0.9594 \\
 \cmidrule{2-11}
   & 6M & FTS & 0.7126 & 1.0832 & 0.7695 & 0.6908 & 0.6729 & 0.9149 & 0.7321 & 0.6702 \\ 
& & MFTS & 0.6823 & 0.9267 & 0.7337 & 0.6854 & 0.7001 & 1.0005 & 0.7482 & 0.6954 \\ 
& & MLFTS & 0.7148 & 1.0846 & 0.7602 & 0.7024 & 0.5959 & 0.7409 & 0.6421 & 0.6385 \\
& & DFTS & 0.6078 & 0.7185 & 0.6484 & 0.6611 & 0.5647 & 0.6658 & 0.6148 & 0.6180 \\ 
& & DMFTS & 0.7310 & 1.0010 & 0.7563 & 0.7471 & 0.7426 & 1.0913 & 0.7461 & 0.7727 \\ 
& & DMLFTS & 0.6776 & 0.8342 & 0.7097 & 0.7156 & 0.5709 & 0.6918 & 0.6213 & 0.6178 \\
& & CT11 & 0.8767 & 1.7534 & 0.9210 & 0.7734 \\
& & GG06 & 1.0271 & 2.3684 & 0.8886 & 0.7948 \\
& & RW & 0.5702 & 0.6683 & 0.6191 & 0.6245 \\
& & AR(1) & 0.5807 & 0.6646 & 0.6348 & 0.6295 \\
 \cmidrule{2-11}
   & 2Y & FTS & 0.6817 & 0.7099 & 0.7345 & 0.7115 & 0.6588 & 0.6687 & 0.7217 & 0.6889 \\ 
& & MFTS & 0.6515 & 0.9375 & 0.6840 & 0.6790 & 0.8056 & 1.5061 & 0.7814 & 0.8119 \\ 
& & MLFTS & 0.8013 & 1.0887 & 0.7850 & 0.8336 & 0.6885 & 0.8884 & 0.7151 & 0.7222  \\
& & DFTS & 0.5696 & 0.4903 & 0.6225 & 0.6497 & 0.3720 & 0.2623 & 0.4620 & 0.4594 \\ 
& & DMFTS & 0.4726 & 0.4926 & 0.5411 & 0.5431 & 0.5701 & 1.1076 & 0.6227 & 0.5980 \\ 
& & DMLFTS & 0.5241 & 0.4709 & 0.5974 & 0.5848 & 0.3764 & 0.2720 & 0.4682 & 0.4597 \\
& & CT11 & 1.1904 & 2.1412 & 1.1195 & 1.0799 \\
& & GG06 & 1.1155 & 1.8423 & 1.0355 & 0.9927 \\
& & RW & 0.3764 & 0.2684 & 0.4661 & 0.4630 \\
& & AR(1) & 0.3760 & 0.2624 & 0.4739 & 0.4589 \\
  \cmidrule{2-11}
  & Average & FTS & 0.8022 & 1.7197 & 0.8367 & 0.7569 & 0.7841 & 1.6328 & 0.8188 & 0.7485 \\
  & & MFTS & 0.7953 & 1.7588 & 0.8055 & 0.7663 & 0.8546 & 1.9909 & 0.8444 & 0.8151 \\
  & & MLFTS & 0.8516 & 1.8363 & 0.8536 & 0.8119 & 0.7760 & 1.6244 & 0.7843 & 0.7626\\
 & & DFTS & 0.7491 & 1.5462 & 0.7522 & 0.7617 & \textBF{0.6651} & \textBF{1.4468} & \textBF{0.6847} & \textBF{0.6803} \\
    & & DMFTS & 0.7674 & 1.6812 & 0.7687 & 0.7606 & 0.8141 & 2.0309 & 0.8086 & 0.7861 \\
  & & DMLFTS & 0.7852 & 1.6586 & 0.7862 & 0.7802 & 0.6698 & 1.4591 & 0.6903 & 0.6812 \\
& & CT11 & 1.0663 & 2.4642 & 1.0282 & 0.9534 \\
& & GG06 & 1.1000 & 2.6263 & 0.9073 & 0.8440 \\
& & RW & 0.6737 & 1.4470 & 0.6925 & 0.6893 \\
& & AR(1) & 0.6779 & 1.4915 & 0.7190 & 0.6826 \\
  \midrule
  10 day & 1M & FTS  & 1.3987 & 6.6324 & 1.2971 & 1.1131 & 1.4210 & 6.8235 & 1.2999 & 1.1400 \\ 
& & MFTS & 1.5181 & 7.4034 & 1.3402 & 1.2334 & 1.5251 & 7.5188 & 1.3448 & 1.2380 \\
& & MLFTS & 1.4576 & 6.8783 & 1.3312 & 1.1622 & 1.4930 & 7.0896 & 1.3281 & 1.2110 \\
 & & DFTS & 1.5640 & 7.7719 & 1.3403 & 1.2954 & 1.5551 & 7.7560 & 1.3331 & 1.2872 \\ 
& & DMFTS & 1.5756 & 7.8232 & 1.3428 & 1.3079 & 1.5609 & 7.4343 & 1.3611 & 1.2762 \\
& & DMLFTS & 1.6294 & 8.2585 & 1.3894 & 1.3475 & 1.5610 & 7.7540 & 1.3388 & 1.2910 \\
& & CT11 & 1.5004 & 6.4889 & 1.3407 & 1.2248 \\
& & GG06 & 1.5381 & 6.9009 & 0.8071 & 0.7441 \\
& & RW & 1.5775 & 7.7307 & 1.3523 & 1.3100 \\
& & AR(1) & 1.5754 & 7.7677 & 1.3522 & 1.3763 \\
 \cmidrule{2-11}
   & 6M & FTS & 0.8745 & 1.5330 & 0.9098 & 0.7990 & 0.8548 & 1.3954 & 0.8876 & 0.7963 \\ 
& & MFTS & 0.9168 & 1.5954 & 0.9296 & 0.8476 & 0.9267 & 1.6577 & 0.9396 & 0.8503 \\
& & MLFTS & 0.9071 & 1.5980 & 0.9331 & 0.8264 & 0.8438 & 1.3370 & 0.8566 & 0.8141 \\
& & DFTS & 0.8527 & 1.3221 & 0.8501 & 0.8370 & 0.8246 & 1.2710 & 0.8293 & 0.8119 \\ 
& & DMFTS & 1.0029 & 1.8746 & 0.9696 & 0.9384 & 0.9963 & 1.9353 & 0.9451 & 0.9519 \\
& & DMLFTS & 0.9042 & 1.4410 & 0.8870 & 0.8795 & 0.8280 & 1.3002 & 0.8332 & 0.8122  \\
& & CT11 & 1.0275 & 2.3085 & 1.0590 & 0.8620 \\
& & GG06 & 1.2387 & 3.1616 & 0.9024 & 0.8016 \\
& & RW & 0.8299 & 1.2695 & 0.8359 & 0.8175 \\
& & AR(1) & 0.8350 & 1.2715 & 0.8472 & 0.8207 \\
 \cmidrule{2-11}
   & 2Y & FTS  & 0.7424 & 0.8474 & 0.7668 & 0.7716 & 0.7260 & 0.8131 & 0.7630 & 0.7520 \\ 
& & MFTS & 0.7627 & 1.2741 & 0.7713 & 0.7669 & 0.9067 & 1.9124 & 0.8589 & 0.8901 \\
& & MLFTS & 0.8721 & 1.2509 & 0.8291 & 0.8988 & 0.7826 & 1.0851 & 0.7770 & 0.8080 \\
& & DFTS & 0.6590 & 0.7010 & 0.6920 & 0.7127 & 0.5103 & 0.4741 & 0.5849 & 0.5700 \\ 
& & DMFTS & 0.6175 & 0.9030 & 0.6629 & 0.6531 & 0.7039 & 1.6297 & 0.7356 & 0.6989 \\
& & DMLFTS & 0.6509 & 0.7157 & 0.6971 & 0.6893 & 0.5142 & 0.4798 & 0.5901 & 0.5703 \\
& & CT11 & 1.2301 & 2.3295 & 1.1474 & 1.1061 \\
& & GG06 & 1.1648 & 2.0423 & 1.0396 & 0.9963 \\
& & RW & 0.5171 & 0.4794 & 0.5928 & 0.5770 \\
& & AR(1) & 0.5179 & 0.4838 & 0.5892 & 0.5757  \\
   \cmidrule{2-11}
   & Average & FTS & 1.0052 & \textBF{3.0043} & 0.9912 & 0.8946 & 1.0006 & 3.0107 & 0.9835 & 0.8961 \\
  & & MFTS & 1.0659 & 3.4243 & 1.0137 & 0.9493 & 1.1195  & 3.6963 & 1.0478 & 0.9928\\
  & & MLFTS & 1.0789 & 3.2424 & 1.0311 & 0.9625 & 1.0398 & 3.1706 & 0.9872 & 0.9444 \\
  & & DFTS &  1.0252 & 3.2650 & 0.9608 & 0.9484 & \textBF{0.9633} & 3.1670 & 0.9158 & 0.8897 \\
  & & DMFTS & 1.0653 & 3.5336 & 0.9918 & 0.9665 & 1.0870 & 3.6664 & 1.0139 & 0.9757 \\
  & & DMLFTS & 1.0615 & 3.4718 & 0.9912 & 0.9721 & 0.9677 & 3.1780 & 0.9207 & 0.8912 \\
& & CT11 & 1.2527 & 3.7090 & 1.1824 & 1.0643 \\
& & GG06 & 1.3139 & 4.0349 & \textBF{0.9157} & \textBF{0.8473} \\
& & RW & 0.9748 & 3.1599 & 0.9270 & 0.9015 \\
& & AR(1) & 0.9761 & 3.1743 & 0.9295 & 0.8909\\
  \bottomrule
\end{longtable}
\end{small}
\end{center}

In Table~\ref{tab:5_moneyness}, we present a comparison of out-of-sample accuracy for forecasting EUR-GBP IV for the five delta values considered. The errors are aggregated over the three maturities of one month, six months and two years.

\begin{center}
\tabcolsep 0.08in
\renewcommand{\arraystretch}{0.95}
\begin{small}
\begin{longtable}{@{}llrrrrrrrrrr@{}}
\caption{\small{This table presents one-step-ahead, five-step-ahead and ten-step-ahead out-of-sample forecast performance measures for forecasting EUR-GBP IV from January 1, 2015 to December 30, 2016, for the five delta values of 10, 25, 50, 75 and 90. The measures are aggregated over three maturities of one month, six months and two years.}}\label{tab:5_moneyness} \\
\toprule
& & \multicolumn{5}{c}{$\overline{\text{MAFE}}$} & \multicolumn{5}{c}{$\overline{\text{MSFE}}$} \\
\underline{Horizon} & & \multicolumn{5}{c}{Delta} & \multicolumn{5}{c}{Delta} \\
$K$ & Method & 10 & 25 & 50 & 75 & 90 & 10 & 25 & 50 & 75 & 90 \\ 
\endfirsthead
\midrule
\underline{1 day} \\
CPV &  FTS & 0.5406 & 0.5107 & 0.4022 & 0.6039 & 0.8948 & 0.5481 & 0.5135 & 0.3964 & 0.9083 & 1.9652 \\ 
&  DFTS & 0.4814 & 0.3773 & 0.4094 & 0.3937 & 0.4595 & 0.4364 & 0.2849 & 0.3284 & 0.3966 & 0.5210 \\ 
&  MFTS & 0.5563 & 0.3835 & 0.3136 & 0.4264 & 0.9026 & 0.6659 & 0.3087 & 0.2594 & 0.5442 & 1.9037 \\ 
&  DMFTS & 0.4941 & 0.4098 & 0.3212 & 0.4317 & 0.6281 & 0.5719 & 0.3625 & 0.2795 & 0.5015 & 1.0220 \\ 
&  MLFTS & 0.5520 & 0.5529 & 0.4447 & 0.5261 & 0.9402 & 0.5892 & 0.6488 & 0.4195 & 0.7051 & 1.9994 \\ 
&  DMLFTS & 0.5617 & 0.6176 & 0.3768 & 0.4391 & 0.3690 & 0.5399 & 0.5757 & 0.3194 & 0.4747 & 0.4458 \\ 
& CT11 & 0.8129 & 0.8611 & 0.4230 & 0.9382 & 1.0762 & 1.2835 & 1.3531 & 0.4743 & 1.7844 & 2.2012 \\ 
& GG06 & 0.8210 & 0.8512 & 0.5014 & 1.1189 & 1.2060 & 1.2043 & 1.2874 & 0.7364 & 2.1197 & 2.9972 \\ 
 & RW & 0.4988 & 0.4465 & 0.4473 & 0.4690 & 0.5039 & 0.4760 & 0.3999 & 0.4126 & 0.4788 & 0.6049 \\ 
 & AR(1) & 0.4971 & 0.4472 & 0.4573 & 0.4931 & 0.5431 & 0.4672 & 0.4022 & 0.4398 & 0.5727 & 0.8001 \\ 
\\
$K=4$ & FTS  & 0.4348 & 0.5595 & 0.4594 & 0.6780 & 0.6288 & 0.3967 & 0.5824 & 0.4965 & 1.1806 & 0.9421 \\ 
  & DFTS & 0.3008 & 0.2542 & 0.2580 & 0.2788 & 0.3083 & 0.2927 & 0.2079 & 0.2270 & 0.3046 & 0.4265 \\ 
  & MFTS & 0.6705 & 0.4099 & 0.3301 & 0.4387 & 1.0711 & 0.9249 & 0.3366 & 0.2750 & 0.5790 & 2.6211 \\ 
  & DMFTS & 0.6197 & 0.4436 & 0.3889 & 0.4854 & 0.7750 & 1.1699 & 0.4229 & 0.4231 & 0.8776 & 2.0849 \\ 
  & MLFTS & 0.4412 & 0.4254 & 0.3695 & 0.4082 & 0.6701 & 0.4476 & 0.3884 & 0.3338 & 0.4878 & 1.1066 \\ 
  & DMLFTS & 0.2996 & 0.2508 & 0.2582 & 0.2786 & 0.3093 & 0.2930 & 0.2032 & 0.2215 & 0.2998 & 0.4169 \\ 
\underline{5 day} \\
 CPV & FTS & 0.7367 & 0.7044 & 0.6264 & 0.8335 & 1.1101 & 1.0116 & 0.9430 & 1.0505 & 2.0263 & 3.5669 \\ 
 & DFTS & 0.7559 & 0.6291 & 0.6936 & 0.7852 & 0.8816 & 1.0738 & 0.8443 & 1.1829 & 1.9111 & 2.7188 \\ 
 & MFTS & 0.7932 & 0.6246 & 0.6122 & 0.7537 & 1.1926 & 1.3270 & 0.8558 & 1.0647 & 1.8413 & 3.7050 \\ 
 & DMFTS & 0.7644 & 0.6483 & 0.6409 & 0.7880 & 0.9952 & 1.2859 & 0.9733 & 1.1472 & 1.9375 & 3.0623 \\ 
 & MLFTS & 0.7566 & 0.7640 & 0.6932 & 0.8201 & 1.2243 & 1.0889 & 1.1209 & 1.1612 & 1.9802 & 3.8305 \\ 
 & DMLFTS & 0.8077 & 0.7988 & 0.6689 & 0.8144 & 0.8362 & 1.1742 & 1.1637 & 1.2108 & 2.0253 & 2.7192 \\ 
  & CT11 & 1.0381 & 1.0281 & 0.7039 & 1.1954 & 1.3662 & 1.9411 & 1.8221 & 1.2752 & 3.1227 & 4.1599 \\ 
  & GG06 & 1.0119 & 1.0125 & 0.7213 & 1.3346 & 1.4196 & 1.7904 & 1.7551 & 1.4430 & 3.3824 & 4.7605 \\ 
  & RW & 0.7566 & 0.6687 & 0.7039 & 0.8017 & 0.8879 & 1.0626 & 0.8730 & 1.1671 & 1.8347 & 2.7141 \\ 
  & AR(1) & 0.7460 & 0.6645 & 0.7097 & 0.8139 & 0.9053 & 0.9959 & 0.8511 & 1.1728 & 1.9193 & 2.9686 \\ 
\\
$K=4$ & FTS & 0.6854 & 0.7352 & 0.6550 & 0.8734 & 0.9718 & 0.9438 & 1.0031 & 1.1047 & 2.2196 & 2.8928 \\ 
  & DFTS & 0.6495 & 0.5655 & 0.6029 & 0.7100 & 0.7976 & 0.9393 & 0.7643 & 1.0679 & 1.7974 & 2.6652 \\ 
  & MFTS & 0.8888 & 0.6470 & 0.6221 & 0.7617 & 1.3536 & 1.6317 & 0.8918 & 1.0939 & 1.8929 & 4.4442 \\ 
  & DMFTS & 0.8431 & 0.6709 & 0.6701 & 0.8025 & 1.0838 & 1.9681 & 1.0327 & 1.2512 & 2.1203 & 3.7821 \\ 
  & MLFTS & 0.7335 & 0.6752 & 0.6478 & 0.7466 & 1.0770 & 1.0820 & 0.9127 & 1.0967 & 1.7923 & 3.2381 \\ 
  & DMLFTS & 0.6607 & 0.5700 & 0.6044 & 0.7140 & 0.8000 & 0.9803 & 0.7836 & 1.0825 & 1.7935 & 2.6555 \\
\underline{10 day} \\
 CPV & FTS & 0.9040 & 0.8609 & 0.8083 & 1.0650 & 1.3878 & 1.4232 & 1.3957 & 1.9656 & 3.8025 & 6.4343 \\ 
&  DFTS & 0.9625 & 0.8184 & 0.9237 & 1.1209 & 1.3008 & 1.6254 & 1.4160 & 2.3635 & 4.3021 & 6.6180 \\ 
&  MFTS & 1.0242 & 0.8368 & 0.8559 & 1.0798 & 1.5326 & 2.0804 & 1.5170 & 2.2579 & 4.0697 & 7.1963 \\ 
 & DMFTS & 1.0281 & 0.8705 & 0.8928 & 1.1367 & 1.3986 & 2.1600 & 1.7347 & 2.4419 & 4.3868 & 6.9444 \\ 
 & MLFTS & 0.9333 & 0.9343 & 0.8962 & 1.0996 & 1.5312 & 1.5263 & 1.6035 & 2.1620 & 3.9347 & 6.9855 \\ 
&  DMLFTS & 1.0126 & 0.9581 & 0.9075 & 1.1529 & 1.2765 & 1.7681 & 1.7749 & 2.4515 & 4.5203 & 6.8440 \\ 
& CT11 & 1.2307 & 1.1604 & 0.8778 & 1.3884 & 1.6059 & 2.5613 & 2.2352 & 2.1266 & 4.7604 & 6.8613 \\ 
  & GG06 & 1.1777 & 1.1450 & 0.9262 & 1.5922 & 1.7283 & 2.3543 & 2.2247 & 2.3602 & 5.3157 & 7.9198 \\ 
  & RW & 0.9993 & 0.8831 & 0.9631 & 1.1595 & 1.3193 & 1.6622 & 1.4666 & 2.3555 & 4.1809 & 6.5840 \\ 
  & AR(1) & 0.9824 & 0.8751 & 0.9354 & 1.1177 & 1.2818 & 1.5420 & 1.3792 & 2.1133 & 3.6932 & 5.9138 \\ 
\\
 $K=4$ & FTS & 0.8887 & 0.8870 & 0.8179 & 1.0798 & 1.3296 & 1.4272 & 1.4540 & 1.9576 & 3.8718 & 6.3428 \\ 
&  DFTS & 0.8885 & 0.7756 & 0.8536 & 1.0676 & 1.2314 & 1.5057 & 1.3327 & 2.2373 & 4.1716 & 6.5880 \\ 
&  MFTS & 1.1033 & 0.8573 & 0.8646 & 1.0850 & 1.6873 & 2.4749 & 1.5729 & 2.3023 & 4.1439 & 7.9874 \\ 
&  DMFTS & 1.0672 & 0.8728 & 0.9082 & 1.1261 & 1.4610 & 2.9047 & 1.7556 & 2.4268 & 4.2123 & 7.0328 \\ 
&  MLFTS & 0.9532 & 0.8615 & 0.8606 & 1.0567 & 1.4668 & 1.6570 & 1.4520 & 2.1241 & 3.7862 & 6.8335 \\ 
&  DMLFTS & 0.9042 & 0.7840 & 0.8538 & 1.0641 & 1.2325 & 1.5733 & 1.3794 & 2.2608 & 4.1317 & 6.5448 \\
  \bottomrule
\end{longtable}
\end{small}
\end{center}

Based on Tables~\ref{tab:5} and~\ref{tab:5_moneyness}, we observe overall, that the univariate functional time-series forecasting method is generally the most accurate method at the five-step-ahead and ten-step-ahead horizons. This result is achieved when we specify dynamic functional principal component analysis. At the one-step-ahead horizon, however, the multilevel functional time-series forecasting method further improves upon the forecast accuracy. As with EUR-USD, we again observe that the best-performing functional time-series method generally performs better than the benchmark approaches considered, especially at longer forecast horizons. Suppose we instead focus on static functional principal component analysis. In that case, the multivariate functional time-series method generally outperforms the others for horizons of one day and five days ahead. In contrast, the univariate functional time-series method gives the most accurate forecasts for longer horizons of ten days ahead. It can also be noted that as the forecast horizon increases, the forecasting accuracy difference between the three methods reduces. 

\subsubsection{EUR-JPY}

Similarly, a summary of the out-of-sample forecast measures calculated under a recursive estimation scheme and 522-day out-of-sample window length are presented for EUR-JPY in Table~\ref{tab:6}. In this case, similar to the static multivariate functional time-series method, the selected number of components based on the CPV criterion is also $K=4$. %As benchmark methods, we also consider the \citeauthor{CT11}'s \citeyearpar{CT11} method, the \citeauthor{GG06}'s \citeyearpar{GG06} method, RW and AR(1). 

\begin{center}
\tabcolsep 0.067in
\renewcommand{\arraystretch}{0.95}
\begin{small}
\begin{longtable}{@{}lllcccccccc@{}}
\caption{\small{This table presents one-step-ahead, five-step-ahead and ten-step-ahead out-of-sample performance measures for forecasting EUR-JPY IV from January 1, 2015 to December 30, 2016, for maturities of one month, six months, and two years. The measures are aggregated over the five delta values of 10, 25, 50, 75 and 90. FTS, MFTS and MLFTS represent the univariate, multivariate and multilevel functional time-series models, respectively, with a leading `D' indicating that dynamic principal component evaluation is used. When forecasting principal component scores, we use ARIMA$(p, d, q)$ with orders selected by the automatic algorithm of \cite{HK08}. CPV indicates that the number of principal components $K$ is determined as the minimum that reaches a cumulative percentage of variance level of 99\%, whereas $K=4$ fixes this number at four. The smallest averaged error measures are highlighted in bold for each maturity and each horizon.}}\label{tab:6}\\
\toprule
& &    & \multicolumn{4}{c}{CPV} & \multicolumn{4}{c}{$K = 4$} \\
Horizon & Maturity & Method& $\overline{\text{MAFE}}$ & $\overline{\text{MSFE}}$ & MME(U) & MME(O) & $\overline{\text{MAFE}}$ & $\overline{\text{MSFE}}$ & MME(U) & MME(O)  \\\hline
\endfirsthead
Horizon & Maturity & Method  & $\overline{\text{MAFE}}$ & $\overline{\text{MSFE}}$ & MME(U) & MME(O) & $\overline{\text{MAFE}}$ & $\overline{\text{MSFE}}$ & MME(U) & MME(O)   \\\hline
\endhead
\hline  \multicolumn{11}{r}{{Continued on next page}} \\ 
\endfoot
\endlastfoot
  1 day & 1M & FTS & 0.8663 & 1.1397 & 0.8665 & 0.8680 & 0.8130 & 1.0260 & 0.8268 & 0.8232 \\
& & MFTS & 0.6152 & 0.7601 & 0.6836 & 0.6375 & 0.7545 & 1.1776 & 0.7947 & 0.7374 \\
& & MLFTS & 0.7706 & 1.0435 & 0.8084 & 0.7664 & 0.6489 & 0.7386 & 0.7137 & 0.6727 \\
& & DFTS & 0.5432 & 0.6194 & 0.6107 & 0.5920 & 0.5071 & 0.5668 & 0.5798 & 0.5599 \\
& & DMFTS & 0.7211 & 1.0748 & 0.7268 & 0.7466 & 0.9972 & 1.9691 & 1.0284 & 0.8630 \\
& & DMLFTS & 0.5212 & 0.5844 & 0.5936 & 0.5718 & 0.4953 & 0.5516 & 0.5705 & 0.5492 \\
& & CT11 & 1.0553 & 2.1328 & 0.9665 & 1.0089 \\
& & GG06 & 1.0460 & 2.2203 & 1.0313 & 0.9286 \\
& & RW & 0.4948 & 0.5510 & 0.5781 & 0.5518  \\
& & AR(1) & 0.5253 & 0.5832 & 0.6013 & 0.5727 \\
  \cmidrule{2-11}
   & 6M & FTS & 0.7193 & 0.9478 & 0.7424 & 0.7423 & 0.7554 & 1.0153 & 0.7745 & 0.7683 \\
& & MFTS & 0.5023 & 0.5568 & 0.5593 & 0.5656 & 0.6524 & 1.0130 & 0.6639 & 0.6853 \\
& & MLFTS & 0.7733 & 1.0502 & 0.7908 & 0.7857 & 0.7803 & 1.0870 & 0.7754 & 0.8043  \\
& & DFTS & 0.5636 & 0.4616 & 0.6328 & 0.6369 & 0.2632 & 0.1699 & 0.3613 & 0.3556 \\
& & DMFTS & 0.3690 & 0.2954 & 0.4511 & 0.4558 & 0.6282 & 0.7706 & 0.6580 & 0.6823 \\
& & DMLFTS & 0.3790 & 0.2861 & 0.4774 & 0.4535 & 0.2423 & 0.1530 & 0.3415 & 0.3338 \\
& & CT11 & 1.5107 & 4.2911 & 1.4303 & 1.1884 \\
& & GG06 & 1.7647 & 6.0307 & 1.4782 & 1.4727 \\
& & RW & 0.2482 & 0.1597 & 0.3447 & 0.3401 \\
& & AR(1) & 0.2576 & 0.1663 & 0.3519 & 0.3529 \\
  \cmidrule{2-11}
     & 2Y & FTS & 0.7827 & 0.8548 & 0.8130 & 0.8010 & 0.7078 & 0.7865 & 0.7401 & 0.7348 \\
& & MFTS & 0.7179 & 0.9720 & 0.7196 & 0.7665 & 0.8712 & 1.3856 & 0.7934 & 0.9241 \\
& & MLFTS & 0.7479 & 0.9159 & 0.7607 & 0.7841 & 0.7604 & 0.8413 & 0.7899 & 0.7790 \\
& & DFTS & 0.6391 & 0.5292 & 0.7094 & 0.6858 & 0.1750 & 0.0641 & 0.2755 & 0.2730 \\ 
& & DMFTS & 0.3645 & 0.2558 & 0.4551 & 0.4458 & 0.7851 & 1.2771 & 0.7412 & 0.8278 \\
& & DMLFTS & 0.5991 & 0.7277 & 0.6682 & 0.6035 & 0.1720 & 0.0650 & 0.2752 & 0.2661 \\
& & CT11 & 0.5453 & 0.5210 & 0.6248 & 0.5886 \\
& & GG06 & 0.5253 & 0.5269 & 0.5906 & 0.5804 \\
& & RW & 0.1641 & 0.0648 & 0.2638 & 0.2557 \\
& & AR(1) & 0.1706 & 0.0683 & 0.2744 & 0.2598 \\
  \cmidrule{2-11}
  & Average & FTS & 0.7894 & 0.9808 & 0.8073 & 0.8038 & 0.7587 & 0.9426 & 0.7805 & 0.7754 \\
  & & MFTS & 0.6118 & 0.7630 & 0.6542 & 0.6565 & 0.7594 & 1.1921 & 0.7507 & 0.7823 \\
  & & MLFTS & 0.7639 & 1.0032 & 0.7866 & 0.7787 & 0.7299 & 0.8890 & 0.7597 & 0.7520 \\
  & & DFTS &  0.5820 & 0.5367 & 0.6510 & 0.6382 & 0.3151 & 0.2669 & 0.4055 & 0.3962 \\
  & & DMFTS & 0.4849 & 0.5420 & 0.5443 & 0.5494 & 0.8035 & 1.3389 & 0.8092 & 0.7910 \\
  & & DMLFTS & 0.4998 & 0.5327 & 0.5797 & 0.5429 & 0.3032 & \textBF{0.2565} & 0.3957 & 0.3830 \\
  & & CT11 & 1.0371 & 2.3150 & 1.0072 & 0.9286 \\
  & & GG06 & 1.1120 & 2.9260 & 1.0334 & 0.9939 \\ 
  & & RW & \textBF{0.3024} & 0.2585 & \textBF{0.3955} & \textBF{0.3825} \\
  & & AR(1) & 0.3178 & 0.2726 & 0.4092 & 0.3951 \\
\midrule
  5 day & 1M & FTS & 1.2709 & 2.8097 & 1.1074 & 1.1985 & 1.2500 & 2.7972 & 1.0982 & 1.1726 \\
& & MFTS & 1.1506 & 2.7260 & 1.0568 & 1.0486 & 1.2091 & 2.9791 & 1.1304 & 1.0636 \\
& & MLFTS & 1.2033 & 2.6999 & 1.0808 & 1.1159 & 1.1782 & 2.7409 & 1.0789 & 1.0787 \\
& & DFTS & 1.1508 & 2.8237 & 1.0647 & 1.0392 & 1.1325 & 2.7722 & 1.0490 & 1.0259 \\ 
& & DMFTS & 1.2433 & 3.0999 & 1.1048 & 1.1313 & 1.3760 & 3.7756 & 1.3274 & 1.1003  \\
& & DMLFTS & 1.1403 & 2.7536 & 1.0554 & 1.0333 & 1.1201 & 2.7139 & 1.0407 & 1.0169 \\
& & CT11 & 1.4569 & 4.1766 & 1.2299 & 1.3071\\
& & GG06 & 1.3244 & 3.6250 & 1.0393 & 0.9380 \\
& & RW & 1.1264 & 2.8010 & 1.0464 & 1.0329  \\
& & AR(1) & 1.2064 & 2.9359 & 1.0882 & 1.1028 \\
  \cmidrule{2-11}
  & 6M & FTS & 0.8497 & 1.2989 & 0.8510 & 0.8327 & 0.8861 & 1.3800 & 0.8792 & 0.8618 \\
& & MFTS & 0.7251 & 1.0054 & 0.7582 & 0.7310 & 0.8653 & 1.5037 & 0.8528 & 0.8400 \\
& & MLFTS & 0.9145 & 1.4826 & 0.9187 & 0.8647 & 0.9416 & 1.5520 & 0.9233 & 0.9005 \\
& & DFTS & 0.7572 & 0.9688 & 0.7740 & 0.7812 & 0.5794 & 0.6797 & 0.6293 & 0.6290 \\ 
& & DMFTS & 0.6417 & 0.7861 & 0.6909 & 0.6700 & 0.8115 & 1.2419 & 0.8336 & 0.7879 \\
& & DMLFTS & 0.6454 & 0.7850 & 0.6863 & 0.6817 & 0.5751 & 0.6717 & 0.6272 & 0.6227 \\
& & CT11 & 1.5685 & 4.6204 & 1.4750 & 1.2244 \\
& & GG06 & 1.8529 & 6.5308 & 1.4792 & 1.4774 \\
& & RW & 0.5790 & 0.6735 & 0.6293 & 0.6314 \\
& & AR(1) & 0.5838 & 0.6696 & 0.6369 & 0.6324 \\
  \cmidrule{2-11}
    & 2Y & FTS & 0.8229 & 0.9759 & 0.8306 & 0.8383 & 0.7682 & 0.9267 & 0.7816 & 0.7898 \\
& & MFTS & 0.7887 & 1.1675 & 0.7613 & 0.8259 & 0.9258 & 1.5515 & 0.8299 & 0.9648 \\
& & MLFTS & 0.8079 & 1.0808 & 0.7962 & 0.8346 & 0.8238 & 1.0481 & 0.8374 & 0.8240 \\
& & DFTS & 0.7125 & 0.7327 & 0.7557 & 0.7467 & 0.3741 & 0.2678 & 0.4657 & 0.4578 \\ 
& & DMFTS & 0.5037 & 0.4638 & 0.5701 & 0.5742 & 0.8495 & 1.4600 & 0.7836 & 0.8822 \\
& & DMLFTS & 0.7273 & 0.9212 & 0.7778 & 0.7177 & 0.3852 & 0.2868 & 0.4734 & 0.4696 \\
& & CT11 & 0.6607 & 0.7414 & 0.7074 & 0.6948 \\
& & GG06 & 0.6724 & 0.7751 & 0.6205 & 0.6028 \\
& & RW & 0.3748 & 0.2712 & 0.4655 & 0.4579 \\
& & AR(1) & 0.3758 & 0.2645 & 0.4726 & 0.4575 \\
  \cmidrule{2-11}
  & Average & FTS & 0.9812 & 1.6948 & 0.9297 & 0.9565 & 0.9681 & 1.7013 & 0.9197 & 0.9414 \\
 & & MFTS & 0.8881 & 1.6330 & 0.8588 & 0.8685 & 1.0001 & 2.0114 & 0.9377 & 0.9561 \\
 & & MLFTS & 0.9752 & 1.7544 & 0.9319 & 0.9384 & 0.9812 & 1.7803 & 0.9465 & 0.9344 \\
 & & DFTS & 0.8735 & 1.5084 & 0.8648 & 0.8557 & 0.6953 & 1.2399 & 0.7147 & 0.7042 \\
 & & DMFTS & 0.7962 & 1.4499 & 0.7886 & 0.7918 & 1.0123 & 2.1592 & 0.9815 & 0.9235 \\
 & & DMLFTS & 0.8376 & 1.4866 & 0.8398 & 0.8109 & 0.6935 & \textBF{1.2241} & 0.7138 & \textBF{0.7031} \\
 & & CT11 & 1.2287 & 3.1795 & 1.1374 & 1.0754 \\
 & & GG06 & 1.2832 & 3.6436 & 1.0463 & 1.0061 \\
 & & RW & \textBF{0.6934} & 1.2494 & \textBF{0.7137} & 0.7074 \\
 & & AR(1) & 0.7220 & 1.2900 & 0.7326 & 0.7309 \\
\midrule  
  10 day & 1M & FTS & 1.5709 & 4.4767 & 1.2832 & 1.4273 & 1.5730 & 4.5753 & 1.2956 & 1.4143 \\
& & MFTS & 1.5574 & 4.7588 & 1.3326 & 1.3436 & 1.5451 & 4.7949 & 1.3607 & 1.2952 \\
& & MLFTS & 1.5359 & 4.4135 & 1.2889 & 1.3699 & 1.5558 & 4.7530 & 1.3305 & 1.3480 \\
& & DFTS & 1.5794 & 5.0762 & 1.3587 & 1.3396 & 1.5685 & 5.0266 & 1.3505 & 1.3315 \\
& & DMFTS & 1.6501 & 5.2621 & 1.3914 & 1.4086 & 1.6821 & 5.6422 & 1.5518 & 1.2918 \\
& & DMLFTS & 1.5845 & 5.0507 & 1.3688 & 1.3386 & 1.5664 & 4.9803 & 1.3563 & 1.3250 \\
& & CT11 & 1.7673 & 6.2749 & 1.4031 & 1.5476 \\
& & GG06 & 1.5731 & 5.2184 & 1.0450 & 0.9449 \\
& & RW & 1.5715 & 4.9438 & 1.3607 & 1.3363 \\
& & AR(1) & 1.6394 & 5.2137 & 1.3562 & 1.4298 \\
  \cmidrule{2-11}
  & 6M & FTS & 0.9845 & 1.6961 & 0.9629 & 0.9245 & 1.0215 & 1.7913 & 0.9905 & 0.9530 \\
& & MFTS & 0.9093 & 1.4807 & 0.9129 & 0.8626 & 1.0263 & 2.0073 & 0.9887 & 0.9469 \\
& & MLFTS & 1.0573 & 1.9537 & 1.0510 & 0.9432 & 1.0946 & 2.0410 & 1.0618 & 0.9874 \\
& & DFTS & 0.9569 & 1.5494 & 0.9289 & 0.9248 & 0.8231 & 1.2615 & 0.8309 & 0.8084 \\
& & DMFTS & 0.8683 & 1.3400 & 0.8820 & 0.8306 & 0.9772 & 1.7689 & 0.9858 & 0.8818 \\
& & DMLFTS & 0.8643 & 1.3384 & 0.8604 & 0.8456 & 0.8155 & 1.2351 & 0.8271 & 0.8016 \\
& & CT11 & 1.6347 & 4.9465 & 1.5334 & 1.2572 \\
& & GG06 & 1.9369 & 6.9592 & 1.4785 & 1.4808 \\
& & RW & 0.8239 & 1.2501 & 0.8340 & 0.8111 \\
& & AR(1) & 0.8161 & 1.2391 & 0.8306 & 0.8044 \\
  \cmidrule{2-11}
    & 2Y & FTS & 0.8589 & 1.0847 & 0.8539 & 0.8686 & 0.8080 & 1.0538 & 0.8110 & 0.8170 \\
& & MFTS & 0.8466 & 1.3296 & 0.7942 & 0.8793 & 0.9755 & 1.6925 & 0.8604 & 1.0076 \\
& & MLFTS & 0.8627 & 1.2247 & 0.8331 & 0.8793 & 0.8758 & 1.2229 & 0.8776 & 0.8581 \\
& & DFTS & 0.7790 & 0.9276 & 0.7981 & 0.7976 & 0.4970 & 0.4627 & 0.5705 & 0.5594 \\ 
& & DMFTS & 0.6048 & 0.6487 & 0.6505 & 0.6583 & 0.9136 & 1.6328 & 0.8186 & 0.9444 \\
& & DMLFTS & 0.8125 & 1.0859 & 0.8437 & 0.7894 & 0.5103 & 0.4773 & 0.5781 & 0.5758 \\
& & CT11 & 0.7468 & 0.9641 & 0.7630 & 0.7692 \\
& & GG06 & 0.7608 & 0.9699 & 0.6372 & 0.6143 \\
& & RW & 0.5050 & 0.4781 & 0.5790 & 0.5691 \\
& & AR(1) & 0.4935 & 0.4612 & 0.5693 & 0.5578 \\
  \cmidrule{2-11}
  & Average & FTS & 1.1381 & 2.4192 & 1.0333 & 1.0735 & 1.1342 & 2.4735 & 1.0324 & 1.0614 \\
  & & MFTS & 1.1044 & 2.5230 & 1.0132 & 1.0285 & 1.1823 & 2.8316 & 1.0699 & 1.0832 \\
  & & MLFTS & 1.1519 & 2.5306 & 1.0577 & 1.0642 & 1.1754 & 2.6723 & 1.0900 & 1.0645 \\
  & & DFTS & 1.1051 & 2.5177 & 1.0286 & 1.0207 & \textBF{0.9629} & 2.2503 & \textBF{0.9173} & \textBF{0.8998} \\
  & & DMFTS & 1.0411 & 2.4169 & 0.9746 & 0.9658 & 1.1910 & 3.0146 & 1.1187 & 1.0393 \\
  & & DMLFTS & 1.0871 & 2.4916 & 1.0243 & 0.9912 & 0.9641 & 2.2309 & 0.9205 & 0.9008 \\
  & & CT11 & 1.3829 & 4.0618 & 1.2332 & 1.1913 \\
  & & GG06 & 1.4236 & 4.3825 & 1.0536 & 1.0133 \\
  & & RW & 0.9668 & \textBF{2.2240} & 0.9246 & 0.9055 \\  
  & & AR(1) & 0.9830 & 2.3047 & 0.9187 & 0.9307 \\
  \bottomrule
\end{longtable}
\end{small}
\end{center}

We present an out-of-sample accuracy comparison for forecasting EUR-JPY IV for the five delta values considered in Table~\ref{tab:6_moneyness}. The errors are aggregated over one-month, six-month and two-year maturities. 

\begin{center}
\tabcolsep 0.08in
\renewcommand{\arraystretch}{0.95}
\begin{small}
\begin{longtable}{@{}llrrrrrrrrrr@{}}
\caption{\small{This table presents one-step-ahead, five-step-ahead and ten-step-ahead out-of-sample performance measures for forecasting EUR-JPY IV from January 1, 2015 to December 30, 2016, for the five delta values of 10, 25, 50, 75 and 90. The measures are aggregated over three maturities one month, six month and two year maturities.}}\label{tab:6_moneyness} \\
\toprule
& & \multicolumn{5}{c}{$\overline{\text{MAFE}}$} & \multicolumn{5}{c}{$\overline{\text{MSFE}}$} \\
\underline{Horizon} & & \multicolumn{5}{c}{Delta} & \multicolumn{5}{c}{Delta} \\
$K$ & Method & 10 & 25 & 50 & 75 & 90 & 10 & 25 & 50 & 75 & 90 \\ 
\endfirsthead
\midrule
\underline{1 day} \\
CPV &  FTS & 0.9319 & 0.8862 & 0.4657 & 0.6709 & 0.9924 & 1.3369 & 1.0467 & 0.3901 & 0.6168 & 1.5132 \\ 
  & DFTS & 0.6762 & 0.6955 & 0.3326 & 0.6392 & 0.5662 & 0.7131 & 0.6829 & 0.2686 & 0.5454 & 0.4737 \\ 
  & MFTS & 0.7161 & 0.5676 & 0.3915 & 0.3677 & 1.0162 & 0.9449 & 0.5508 & 0.2826 & 0.2449 & 1.7916 \\ 
  & DMFTS & 0.7134 & 0.5505 & 0.2959 & 0.3103 & 0.5542 & 1.0905 & 0.6108 & 0.2356 & 0.2120 & 0.5610 \\ 
  & MLFTS & 0.9191 & 0.6660 & 0.4508 & 0.6894 & 1.0944 & 1.3720 & 0.6881 & 0.3424 & 0.6336 & 1.9798 \\ 
  & DMLFTS & 0.3938 & 0.6505 & 0.3033 & 0.8162 & 0.3351 & 0.4077 & 0.6867 & 0.2224 & 1.0846 & 0.2622 \\ 
  & CT11 & 1.2593 & 1.2015 & 0.5045 & 0.8457 & 1.3744 & 2.7692 & 2.3882 & 0.6080 & 1.2594 & 4.5500 \\ 
  & GG06 & 1.6162 & 1.1050 & 0.5652 & 0.9541 & 1.3195 & 5.3630 & 2.1169 & 0.6529 & 1.4974 & 4.9995 \\ 
  & RW & 0.5609 & 0.5077 & 0.4550 & 0.4475 & 0.4908 & 0.5773 & 0.4660 & 0.3881 & 0.3640 & 0.4303 \\ 
  & AR(1) & 0.5802 & 0.5241 & 0.4702 & 0.4642 & 0.5006 & 0.6061 & 0.4863 & 0.4043 & 0.3758 & 0.4406 \\ 
\\
$K=4$ & FTS & 0.8343 & 1.0061 & 0.4857 & 0.6435 & 0.8240 & 1.1247 & 1.3254 & 0.4703 & 0.5700 & 1.2228 \\ 
  & DFTS & 0.3713 & 0.3224 & 0.3284 & 0.2587 & 0.2947 & 0.4047 & 0.2885 & 0.2373 & 0.1717 & 0.2325 \\ 
  & MFTS & 0.9500 & 0.7271 & 0.4295 & 0.4048 & 1.2854 & 1.7320 & 0.9538 & 0.3454 & 0.2795 & 2.6495 \\ 
  & DMFTS & 1.1216 & 0.9055 & 0.5094 & 0.3495 & 1.1314 & 2.3895 & 1.3441 & 0.4575 & 0.2378 & 2.2656 \\ 
  & MLFTS & 0.9593 & 0.7249 & 0.3967 & 0.7085 & 0.8600 & 1.5214 & 0.7793 & 0.2958 & 0.6760 & 1.1724 \\ 
  & DMLFTS & 0.3727 & 0.3159 & 0.2702 & 0.2577 & 0.2995 & 0.4041 & 0.2773 & 0.2000 & 0.1700 & 0.2311 \\ 
\underline{5 day} \\
 CPV & FTS & 1.1859 & 1.0957 & 0.6879 & 0.8087 & 1.1277 & 2.6265 & 1.9214 & 0.9424 & 0.9900 & 1.9939 \\ 
  & DFTS & 1.0687 & 1.0220 & 0.6496 & 0.8254 & 0.8017 & 2.4337 & 1.8927 & 0.9858 & 1.0404 & 1.1894 \\ 
  & MFTS & 1.0569 & 0.8993 & 0.6787 & 0.5952 & 1.2104 & 2.4340 & 1.6339 & 0.9613 & 0.7091 & 2.4266 \\ 
  & DMFTS & 1.0743 & 0.8930 & 0.6239 & 0.5744 & 0.8156 & 2.6316 & 1.7215 & 0.9355 & 0.7126 & 1.2484 \\ 
  & MLFTS & 1.2045 & 0.9430 & 0.6925 & 0.8182 & 1.2180 & 2.7448 & 1.6578 & 0.9471 & 1.0099 & 2.4126 \\ 
  & DMLFTS & 0.8849 & 0.9776 & 0.6404 & 1.0123 & 0.6731 & 2.1084 & 1.7993 & 0.9458 & 1.5809 & 0.9986 \\ 
  & CT11 & 1.4586 & 1.4259 & 0.7432 & 0.9314 & 1.5843 & 4.0492 & 3.4186 & 1.2017 & 1.6183 & 5.6095 \\ 
  & GG06 & 1.7800 & 1.3031 & 0.7823 & 1.0717 & 1.4790 & 6.5110 & 3.0033 & 1.1918 & 1.9384 & 5.5735 \\ 
  & RW & 0.9548 & 0.8589 & 0.7104 & 0.6497 & 0.7432 & 2.0963 & 1.4829 & 0.9705 & 0.7734 & 1.0531 \\ 
  & AR(1) & 0.9889 & 0.8871 & 0.7450 & 0.6795 & 0.7595 & 2.2580 & 1.6028 & 1.0937 & 0.8720 & 1.0735 \\ 
\\
 $K=4$ & FTS & 1.1634 & 1.1875 & 0.7199 & 0.7727 & 0.9972 & 2.6467 & 2.1445 & 1.0374 & 0.9167 & 1.7614 \\ 
  & DFTS & 0.8785 & 0.7671 & 0.6369 & 0.5473 & 0.6469 & 2.1648 & 1.4678 & 0.9462 & 0.6703 & 0.9503 \\ 
  & MFTS & 1.2120 & 1.0168 & 0.7042 & 0.6177 & 1.4494 & 3.0898 & 1.9683 & 0.9939 & 0.7270 & 3.2781 \\ 
  & DMFTS & 1.3317 & 1.0977 & 0.7358 & 0.5836 & 1.3129 & 3.7456 & 2.3213 & 1.0964 & 0.7101 & 2.9223 \\ 
  & MLFTS & 1.3211 & 0.9984 & 0.6804 & 0.8520 & 1.0540 & 3.2753 & 1.7792 & 0.9481 & 1.0894 & 1.8097 \\ 
  & DMLFTS & 0.8778 & 0.7607 & 0.6159 & 0.5524 & 0.6605 & 2.1329 & 1.4335 & 0.9054 & 0.6785 & 0.9705 \\ 
\underline{10 day} \\
 CPV & FTS & 1.4051 & 1.2601 & 0.8521 & 0.9110 & 1.2623 & 3.9240 & 2.7695 & 1.4731 & 1.3503 & 2.5788 \\ 
  & DFTS & 1.3621 & 1.2723 & 0.8689 & 0.9639 & 1.0582 & 4.2030 & 3.1003 & 1.6952 & 1.5300 & 2.0601 \\ 
  & MFTS & 1.3365 & 1.1570 & 0.8878 & 0.7489 & 1.3919 & 3.9784 & 2.7038 & 1.6176 & 1.1507 & 3.1646 \\ 
  & DMFTS & 1.3707 & 1.1612 & 0.8517 & 0.7518 & 1.0700 & 4.3039 & 2.8629 & 1.6382 & 1.2039 & 2.0756 \\ 
  & MLFTS & 1.4511 & 1.1661 & 0.8782 & 0.9237 & 1.3407 & 4.1743 & 2.6378 & 1.5449 & 1.3635 & 2.9326 \\ 
  & DMLFTS & 1.2280 & 1.2117 & 0.8698 & 1.1562 & 0.9697 & 3.8965 & 2.9159 & 1.6625 & 2.0710 & 1.9123 \\ 
  & CT11 & 1.6299 & 1.6038 & 0.9069 & 0.9837 & 1.7905 & 5.3635 & 4.4427 & 1.6955 & 1.8797 & 6.9277 \\ 
  & GG06 & 1.9400 & 1.4691 & 0.9425 & 1.1725 & 1.5938 & 7.7808 & 3.9165 & 1.7208 & 2.3625 & 6.1318 \\ 
  & RW & 1.3082 & 1.1507 & 0.9417 & 0.8415 & 1.0419 & 3.9165 & 2.6593 & 1.6841 & 1.3134 & 1.9801 \\ 
  & AR(1) & 1.3381 & 1.1941 & 0.9906 & 0.8613 & 1.0284 & 4.0361 & 2.7898 & 1.8495 & 1.4279 & 1.8642 \\
\\
 $K=4$ & FTS & 1.4239 & 1.3283 & 0.8930 & 0.8674 & 1.1582 & 4.1834 & 2.9380 & 1.5818 & 1.2491 & 2.4149 \\ 
  & DFTS & 1.2342 & 1.0607 & 0.8601 & 0.7241 & 0.9352 & 3.9860 & 2.6367 & 1.6419 & 1.1565 & 1.8302 \\ 
  & MFTS & 1.4184 & 1.2392 & 0.8984 & 0.7606 & 1.5949 & 4.4566 & 2.9625 & 1.6138 & 1.1399 & 3.9849 \\ 
  & DMFTS & 1.5284 & 1.2821 & 0.9223 & 0.7400 & 1.4821 & 5.1521 & 3.3090 & 1.7377 & 1.1758 & 3.6984 \\ 
  & MLFTS & 1.5933 & 1.2020 & 0.8792 & 0.9606 & 1.2418 & 5.0171 & 2.7280 & 1.5701 & 1.4791 & 2.5674 \\ 
  & DMLFTS & 1.2266 & 1.0492 & 0.8462 & 0.7396 & 0.9586 & 3.9014 & 2.5735 & 1.6031 & 1.1839 & 1.8926 \\ 
   \bottomrule
\end{longtable}
\end{small}
\end{center}

In Table~\ref{tab:6} we observe that the dynamic functional principal component analysis estimated univariate functional time-series approach is generally the most accurate method for EUR-JPY. However, its outperformance over the multilevel functional time-series forecasting method is small. Again, the best-performing method typically outperforms the benchmark methods, especially at longer forecast horizons. When considering only static functional principal component analysis, the multivariate functional time-series method performs best, followed by the multilevel functional time-series method, as measured by our error criteria.

The results above were computed based on an out-of-sample forecast length of 522 days. However, the results are qualitatively similar for the alternative out-of-sample window lengths considered; namely, 131 and 261 days.

\subsection{Statistical significance tests}\label{sec:5.3}

In line with a plethora of forecasting literature seeking to establish statistical significance, we consider the model confidence set test \citep{HLN11}. The model confidence set procedure proposed by \cite{HLN11} consists of a sequence of tests leading to the construction of a set of ``superior" models, where the null hypothesis of equal predictive ability (EPA) is not rejected at a certain confidence level. The EPA test statistic can be evaluated for any arbitrary loss function, such as absolute or squared loss function. The model confidence set procedure has been employed in many economic and financial applications, including \cite{Audrino14}, \cite{CM14}, \cite{BGO16} and \cite{SSRF17}.

Let $M$ be some subset of full model $M^0$, and let $m$ be the number of models in $M$, and let $d_{\rho \xi, \ell}$ denote the loss differential between two models $\rho$ and $\xi$ that is
\begin{equation*}
d_{\rho \xi, \ell} = l_{\rho, \ell} - l_{\xi, \ell}, \qquad \rho, \xi=1,\dots,m, \quad \ell=1,\dots,N,
\end{equation*}
and calculate 
\begin{align*}
d_{\rho\cdot, \ell} = \frac{1}{m}\sum_{\xi\in M}d_{\rho \xi,\ell}, \qquad \rho = 1,\dots,m
\end{align*}
as the loss of model $\rho$ relative to any other model $\xi$ at time point $\ell$. Let $c_{\rho \xi} = \text{E}(d_{\rho \xi})$ and $c_{\rho.} = \text{E}(d_{\rho.})$ be finite and not time-dependent. The EPA hypothesis for a set of $M$ candidate models can be formulated in two ways:
\begin{align}
\text{H}_{0,\text{M}}: c_{\rho \xi}&=0, \qquad \text{for all}\quad \rho, \xi = 1,2,\dots,m\notag\\
\text{H}_{\text{A,M}}: c_{\rho \xi}&\neq 0, \qquad \text{for some}\quad \rho, \xi = 1,2,\dots,m.\label{eq:11}
\end{align}
or
\begin{align}
\text{H}_{0,\text{M}}: c_{\rho.}&=0, \qquad \text{for all}\quad \rho, \xi = 1,2,\dots,m\notag\\
\text{H}_{\text{A,M}}: c_{\rho.}&\neq 0, \qquad \text{for some}\quad \rho, \xi = 1,2,\dots,m.\label{eq:12}
\end{align}
Based on $c_{\rho \xi}$ or $c_{\rho.}$, we construct two hypothesis tests as follows:
\begin{equation}
t_{\rho \xi} = \frac{\overline{d}_{\rho \xi}}{\sqrt{\widehat{\text{Var}}(\overline{d}_{\rho \xi})}}, \qquad t_{\rho.} = \frac{\overline{d}_{\rho.}}{\sqrt{\widehat{\text{Var}}(\overline{d}_{\rho.})}}, \label{eq:13}
\end{equation}
where $\overline{d}_{\rho.} = \frac{1}{m}\sum_{\xi\in M}\overline{d}_{\rho \xi}$ is the sample loss of the $\rho^{\text{th}}$ model compared to the averaged loss across models, and $\overline{d}_{\rho \xi} =\frac{1}{m}\sum^m_{\ell=1}d_{\rho\xi,\ell}$ measures the relative sample loss between the $\rho^{\text{th}}$ and $\xi^{\text{th}}$ models. Note that $\widehat{\text{Var}}\left(\overline{d}_{\rho.}\right)$ and $\widehat{\text{Var}}\left(\overline{d}_{\rho\xi}\right)$ are the bootstrapped estimates of $\text{Var}\left(\overline{d}_{\rho.}\right)$ and $\text{Var}\left(\overline{d}_{\rho\xi}\right)$, respectively. \cite{BC14} perform a block bootstrap procedure with 5,000 bootstrap samples by default, where the block length $p$ is given by the maximum number of significant parameters obtained by fitting an AR($p$) process on all the $d_{\rho\xi}$ terms. For both hypotheses in~\eqref{eq:11} and~\eqref{eq:12}, there exist two test statistics:
\begin{equation*}
T_{\text{R}, \text{M}} = \max_{\rho,\xi\in M}\left|t_{\rho\xi}\right|,\qquad T_{\max, \text{M}} = \max_{\rho\in M} t_{\rho.},
\end{equation*} 
where $t_{\rho\xi}$ and $t_{\rho.}$ are defined in~\eqref{eq:13}.
 
The MCS procedure is a sequential testing procedure, which eliminates the worst model at each step until the hypothesis of equal predictive ability is accepted for all the models belonging to a set of superior models. The selection of the worst model is determined by an elimination rule that is consistent with the test statistic, 
\begin{equation*}
e_{\text{R},\text{M}} = \argmax_{\rho\in M}\left\{\sup_{\xi \in M}\frac{\overline{d}_{\rho\xi}}{\sqrt{\widehat{\text{Var}}\left(\overline{d}_{\rho\xi}\right)}}\right\}, \qquad
e_{\max, \text{M}} = \argmax_{\rho\in M}\frac{\overline{d}_{\rho.}}{\widehat{\text{Var}}\left(\overline{d}_{\rho.}\right)}.
\end{equation*} 

We use the $\overline{\text{MAFE}}$ error measure to examine statistical significance among the three functional time-series methods in Table~\ref{tab:7}. The MCS test results using the $\overline{\text{MSFE}}$ error measure are given in the supplement, with qualitatively similar results obtained. In terms of point forecast accuracy, we identify the superior method(s) below.

\begin{center}
\tabcolsep 0.16in
\renewcommand{\arraystretch}{1}
\begin{longtable}{@{}lllll@{}}
\caption{\small{The table presents the results of the MCS procedure with a 95\% level of confidence applied to the out-of-sample forecast error measure $\overline{\text{MAFE}}$ for the nine methods, when forecasting the IV of the three currency pairs from January 1, 2015 to December 30, 2016, for 1-month, 6-month and 2-year maturities. The measures are aggregated over the five delta values of 10, 25, 50, 75 and 90. FTS, MFTS and MLFTS represent the univariate, multivariate and multilevel functional time-series models, respectively, with a leading `D' indicating that dynamic principal component evaluation is used. `CPV' indicates that the number of principal components $K$ is determined as the minimum that reaches a cumulative percentage of variance level of 99\%, whereas $K=4$ fixes this number at four.}}\label{tab:7}\\
\toprule
Currency & $h$ & Maturity & \multicolumn{2}{c}{Superior models}   \\
& & & $T_{\max}$ & $T_{\text{R}}$ \\ 
\hline
\endfirsthead
Currency & $h$ & Maturity & \multicolumn{2}{c}{Superior models}   \\\hline
\endhead
\hline  \multicolumn{5}{r}{{Continued on next page}} \\ 
\endfoot
\endlastfoot
EUR-USD & 1 day 	& 1M &  DFTS & DFTS		\\
		 & 		& 6M &  DMLFTS & DMLFTS		\\
		& 		& 2Y &   DMLFTS & DMLFTS		\\
		\cmidrule{3-5} 
		& 5 day 	& 1M & DFTS & DFTS		\\
		&		& 6M & DMLFTS & DMLFTS		\\
		&		& 2Y & DFTS & DFTS		\\
		\cmidrule{3-5}
		& 10 day   & 1M & FTS & FTS		\\
		&		& 6M & DFTS & DFTS		\\
		&		& 2Y & DFTS, DMLFTS, AR(1) & AR(1)		\\
		\cmidrule{3-5}
EUR-GBP & 1 day 	& 1M & DMLFTS & DMLFTS	\\
		& 		& 6M & DMLFTS & DMLFTS		\\
		&		& 2Y & DFTS &	 DFTS	\\
		\cmidrule{3-5}
		& 5 day 	& 1M & FTS & FTS		\\
		&		& 6M & DFTS & DFTS		\\
		&		& 2Y & DFTS &	 DFTS	\\
		\cmidrule{3-5}
		& 10 day 	& 1M & FTS & FTS		\\
		&		& 6M & DFTS & DFTS		\\
		&		& 2Y & DFTS &	 DFTS	\\
				\cmidrule{3-5}
EUR-JPY & 1 day	& 1M & RW & RW	\\
		&		& 6M & DMLFTS & DMLFTS		\\
		& 		& 2Y & RW &	 RW	\\
				\cmidrule{3-5}
		& 5 day	& 1M & DMLFTS & DMLFTS		\\
		&		& 6M & DMLFTS & DMLFTS		\\
		&		& 2Y & DFTS & DFTS		\\
				\cmidrule{3-5}
		& 10 day	& 1M & FTS, DFTS, MFTS, MLFTS,  & FTS, DFTS, MFTS, MLFTS, 		\\
		&		& 	& DMLFTS, GG, RW & DMLFTS, GG, RW \\
		&		& 6M & DMLFTS & DMLFTS		\\
		&		& 2Y & DFTS &	 DFTS	\\
		\bottomrule
\end{longtable}
\end{center}

We observe from Table~\ref{tab:7} that using the $T_{\max}$ test statistic, the dynamic univariate functional time-series method ranks as a superior model in 13 out of 35 cases. The dynamic multilevel functional time-series method ranks as a superior model in 11 out of 35 cases, and the static univariate functional time-series method ranks as a superior model in 4 out of 35 cases. Similarly, using the $T_{\text{R}}$ test statistic, the dynamic univariate functional time-series method ranks as a superior model in 12 out of 33 cases. The dynamic multilevel functional time-series method ranks as a superior model in 10 out of 33 cases, and the static univariate functional time-series method ranks as a superior model in 4 out of 33 cases.

\subsection{Trading strategy}\label{sec:trading}

While the forecasting performance of the dynamic univariate functional time-series models is notable, such an ability to forecast may not necessarily translate into an ability to exploit the identified inefficiencies from a trading perspective. Thus, in this section, we seek to shed light on the economic significance of the forecasting accuracy of the functional time-series model though implementing a stylised trading strategy similar to \cite{BG14} and \cite{KCM18}.

Following the trading rule approach of \cite{BG14}, we use the forecasts obtained from the models as trading signals to go long or short straddle positions. Taking straddle positions gives the market participant exposure to the volatility of the underlying futures contract, but avoids the need to take a directional view on the market. Specifically, if the model's forecast is for IV to increase, then we go long the ATM (50 delta) straddle and if the forecast is for IV to decrease we go short the ATM straddle position. The ATM delta values are chosen as they are the most actively traded. We focus on a simple trading strategy to illustrate the benefit of the functional time-series approaches proposed here, namely the dynamic univariate functional time-series model. We calculate this investment strategy over the 1-day ahead trading horizon to aid tractability. However, it could be extended to other trading horizons. We construct a portfolio for each day consisting of the 1-month, 6-month, and 2-year option maturities where we calculate the straddle transaction costs using the \cite{GK83} option pricing model with foreign and domestic interest rates obtained from Bloomberg. Trades are rebalanced in line with the one-day-ahead horizon, and the return on our strategy is aggregated and evaluated across the January 2015 to December 2016, 522 day out-of-sample period.

Results are presented in Table~\ref{tab:Trading_Srategy} for each of the three currencies considered. It outlines the daily percentage return for each specification over the out-of-sample period. It also presents \textit{t}-test results for the null hypothesis that the trading strategy profits are from a normal distribution with zero mean and unknown variance, against the alternative that the daily mean is greater than zero. In line with \cite{KCM18} we also highlight the `Trimmed Samples' where we remove 5\% of the returns observed for each currency, split equally between the left and right tails of the returns. Sharpe and Sortino ratios are also presented to convey the risk-adjusted returns arising from the strategies.

\begin{table}[!htbp]
\centering
\tabcolsep 0.18in
\caption{\small{The table presents the results of the straddle trading strategies implemented in line with \cite{BG14} and \cite{KCM18} and as described in Section 5.4. The strategy utilises ATM options and takes a long straddle position when the 1-day ahead dynamic univariate FTS forecast is for IV to increase and a short position when the forecast is for IV to decrease. 1-month, 6-month and 2-year maturities are all simultaneously considered in the strategy with the \citeauthor{GK83}'s \citeyearpar{GK83} method used to price the options. Results for the USD, GBP and JPY currency pairs versus the Euro are shown. The period used is 522 days from January 2015 to December 2016. The table reports the mean daily return of the strategy with the t-statistic given in brackets underneath. The t-statistic is based on the null hypothesis that the daily trading strategy returns are derived from a normal distribution with mean zero and unknown variance, against an alternative that the daily mean is greater than zero. ***, ** and * indicate significance at 1\%, 5\% and 10\% levels, respectively. The lower panel presents risk-adjusted return results from calculated Sharpe and Sortino ratios.}}\label{tab:Trading_Srategy}
\begin{tabular}{@{}llllllll@{}}
\toprule
 & \multicolumn{7}{c}{Daily mean return} \\
 & \multicolumn{3}{c}{Full sample} &  & \multicolumn{3}{c}{5\%-Trimmed sample} \\
\midrule
 & USD & GBP & JPY &  & USD & GBP & JPY \\
Portfolio & 0.02\% & 0.55\%*** & -0.27\% &  & 0.02\% & 0.56\%*** & -0.27\% \\
 & (0.26) & (5.43) & (-6.88) &  & (0.19) & (6.17) & (-7.59) \\
1 month & 0.20\%** & 0.73\%*** & -0.90\% &  & 0.20\%** & 0.73\%*** & -0.91\% \\
 & (1.80) & (6.06) & (-9.44) &  & (1.87) & (6.61) & (-10.42) \\
6 month & 0.02\%* & 0.00\% & -0.02\% &  & 0.02\%* & 0.00\% & -0.01\% \\
 & (1.38) & (-0.02) & (-1.03) &  & (1.50) & (0.31) & (-0.87) \\
12 month & 0.01\% & -0.02\% & -0.02\% &  & 0.01\% & -0.03\% & -0.01\% \\
 & (0.52) & (-1.33) & (-1.02) &  & (0.66) & (-1.94) & (-0.87)\\
\midrule
& & \multicolumn{6}{c}{Risk adjusted return ratios}                                            \\
          &  & \multicolumn{2}{c}{USD} & \multicolumn{2}{c}{GBP} & \multicolumn{2}{c}{JPY} \\
          &  & Sharpe     & Sortino    & Sharpe     & Sortino    & Sharpe     & Sortino    \\ \midrule
Portfolio &  & 0.01       & 0.02       & 0.24       & 0.39       & -0.30      & -0.37      \\
1 month   &  & 0.08       & 0.12       & 0.27       & 0.43       & -0.41      & -0.45      \\
6 month   &  & 0.06       & 0.09       & 0.00       & 0.00       & -0.04      & -0.06      \\
2 year    &  & 0.03       & 0.04       & -0.06      & -0.08      & -0.04      & -0.05      \\
\bottomrule
\end{tabular}
\end{table}

The trading strategy exercise indicates that after accounting for option premium incurred in constructing the straddle positions, the dynamic univariate functional time-series model identifies profitable trading positions for both EUR-GBP and EUR-USD currency pairs. These returns are statistically significantly greater than zero. These returns are accompanied by positive Sharpe and Sortino ratios too. Unfortunately, despite the model showing good performance when forecasting EUR-JPY options using this functional time-series model, the associated trading strategy does not prove profitable when the option premiums are accounted for. When we drill into the results further, we see that the shorter-term maturity option trades prove more profitable, with a greater difficulty obtaining statistically significant profitable returns using 6-month and 2-year forecast horizons. We must highlight, however, that this is a stylised trading strategy and that many additional idiosyncratic considerations are required for a practitioner to implement such trades in practice. Examples include maximum drawdowns, margin calls, liquidity and hedging concerns, possible funding costs, idiosyncratic risk, slippages, latency and other market frictions. However, that being said the results of this simple illustrative trading strategy are promising; they indicate that there could be an economic significance in the use of our proposed dynamic functional time-series models as trading signals for two of the three currencies considered.

\subsection{SPX data}

To showcase the use of the proposed functional time-series approaches for another asset class, we employ the use of an Optionmetrics dataset of SPX S\&P500 equity options. This data set consists of best bid and offers, volume, strike, forward price, IV, and days to expiry for over 15 million traded call and put SPX options between January 2013 and December 2019. Using the above information, we calculate midpoint price, moneyness and maturity for the options. We impose several exclusionary criteria to the data in line with prior literature, including \cite{GG06} and \cite{BG14}. These filters include only using observations with
\begin{inparaenum}
\item[1)] moneyness values between 90\% and 110\%,
\item[2)] midprices greater than 0.375,
\item[3)] IVs greater than 0\% and less than 70\%,
\item[4)] more than six days to expiry, in line with \cite{BG14}, and finally,
\item[5)] traded volume greater than 25.
\end{inparaenum}
We initially imposed a filter requiring trading volume to be greater than 100, in line with \cite{GG06}; however, this resulted in too few SPX observations to produce functional time-series estimates.

\begin{figure}[!htbp]
\centering
\includegraphics[width=8.9cm]{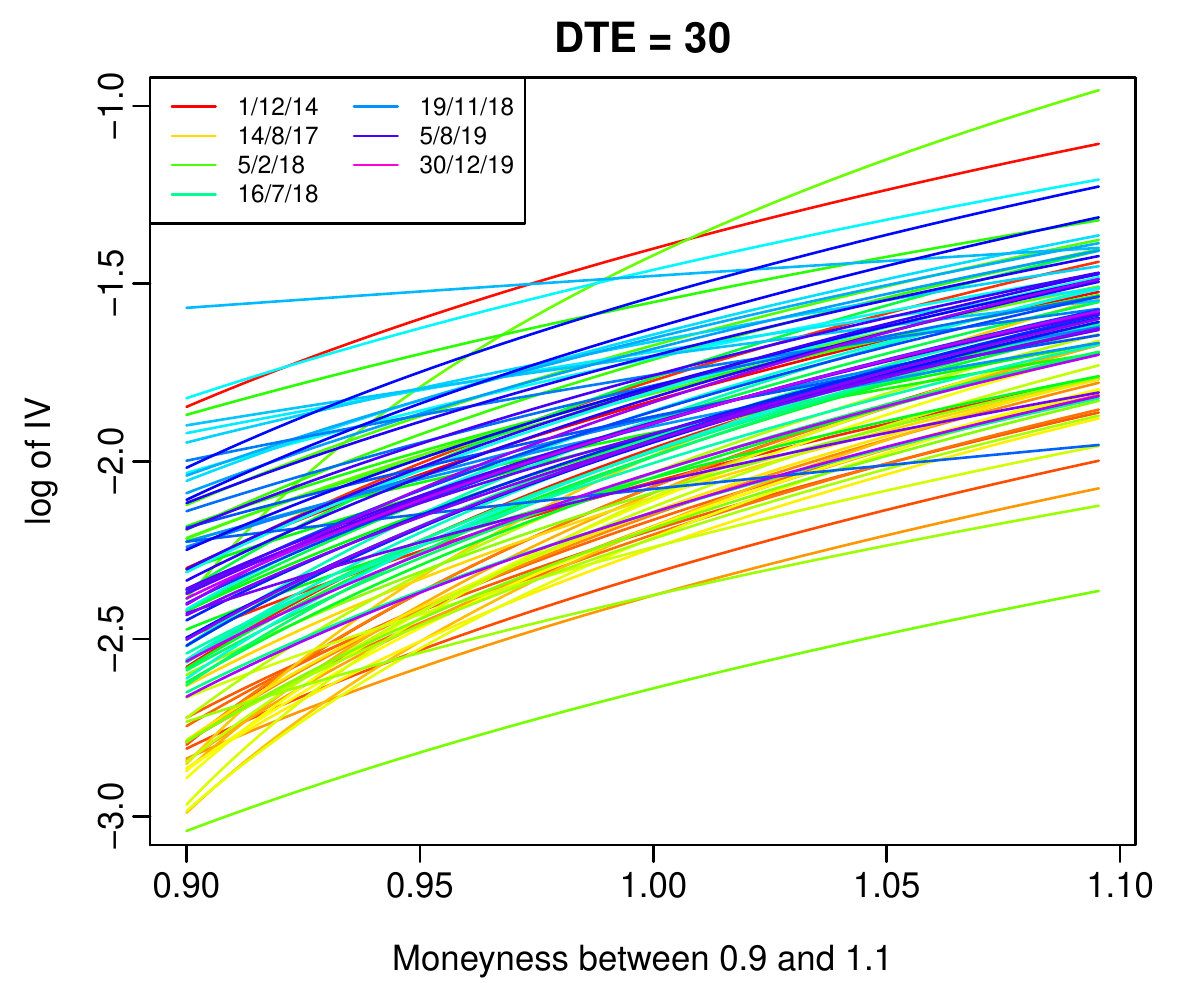}
\quad
\includegraphics[width=8.9cm]{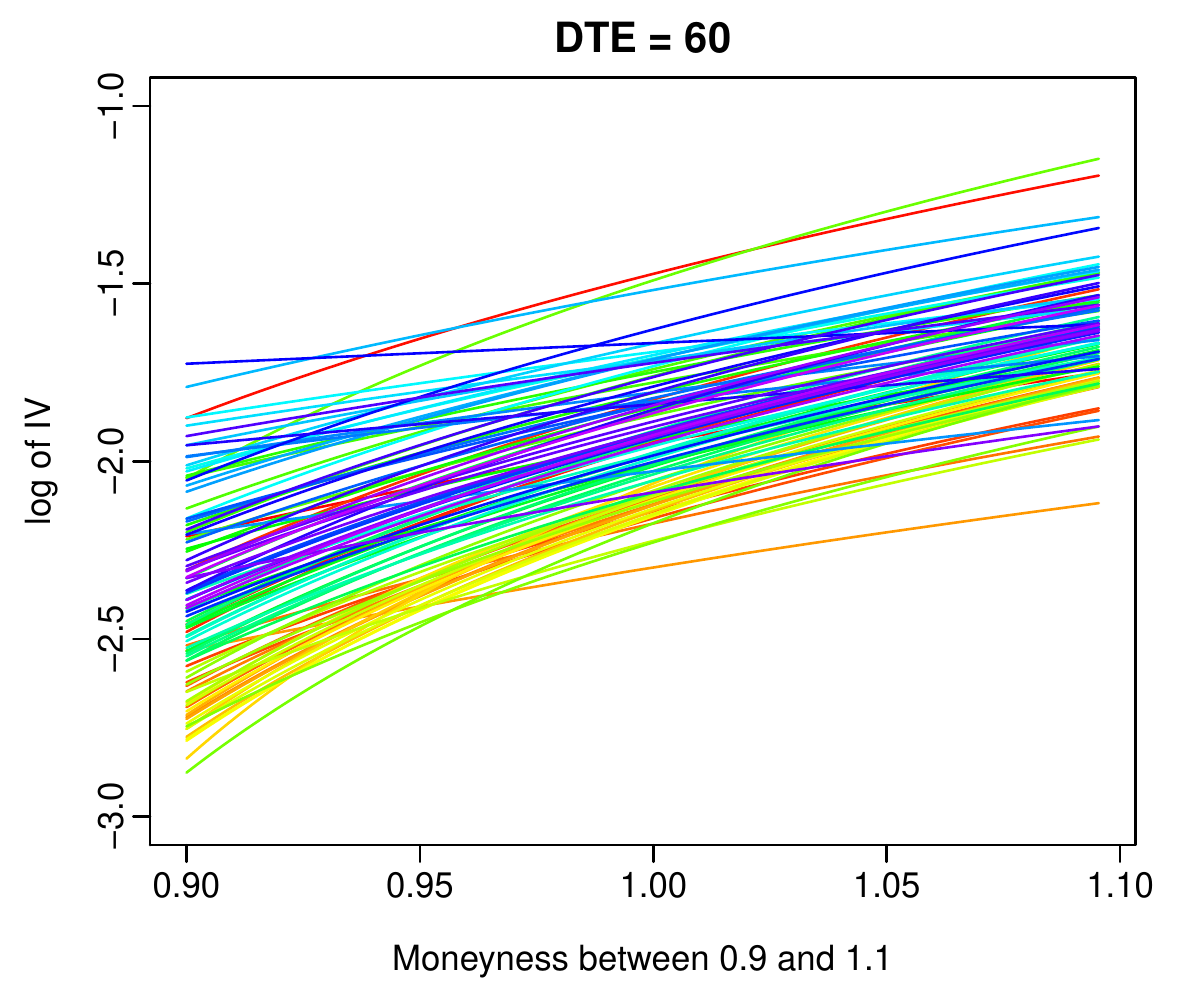}
\caption{\small{The figure presents functional time-series plots of SPX S\&P500 index data from Optionmetrics for IV smiles associated with 30 and 60 days to expiry (DTE), respectively. Moneyness is filtered to be between 90\% and 110\% only, which is mapped by a grid of 132 discrete points in the functional time-series representation.}}\label{fig:data_SPX}
\end{figure}

To construct the multivariate and multilevel functional time-series plots of the equity options data, we are required to specify constant maturity points at which to extract the observations. However, specifying one-month, six-month, and 2-year maturities, as we did for the currency options was unfortunately not an option, as despite the 15 million observations we begin with, there are too few days where options of all of these maturities are traded simultaneously. For this reason, we chose to focus on the more commonly occurring option maturities of 30 and 60 days, shown in Figure~\ref{fig:data_SPX}. Filtering the data in this manner to produce the bivariate functional time-series estimates leads to 97 daily curves. In comparison utilising 30, 60 and 91 days to expiry led to only five observed curves. The initial training sample we use contains the first 77 curves, while the remaining 20 curves are designated as forming our testing sample. With the initial training sample, we produce a one-step-ahead forecast for the 78\textsuperscript{th} curve. We then increase the training sample by one observation and produce another one-step-ahead forecast for the 79\textsuperscript{th} curve. We iterate this procedure until the training sample contains all 97 curves. This gives us 20 one-step-ahead forecasts to evaluate and compare point forecast accuracy, as measured by $\overline{\text{MAFE}}$ and $\overline{\text{MSFE}}$. It should be noted that due to the relatively low number of days with both a sufficient volume of 30 and 60-day expiries traded, the iterative one-step-ahead forecasts are not continuously `observed' in a chronological equally-spaced distance. This lack of trading on 30 and 60-day expiries is particularly prevalent earlier in the training sample period. However, during the testing period, which starts on 15 July 2019, the observed testing curves are mostly spaced one week apart with observations occurring on a Monday. From Table~\ref{tab:SPX}, it can be seen that the dynamic univariate functional time-series method performs the best with the smallest errors. In this bivariate functional time-series equity options approach, the inferior performance of the multivariate and multilevel functional time-series methods suggests that the forecasting improvements from incorporating the correlation structure of multiple maturity curves is more likely to be observed when utilising more than two simultaneous curves, as we have shown for currency options.

\begin{table}[!htbp]
\tabcolsep 0.3in
\centering
\caption{\small{This table presents one-step-ahead forecast accuracy results of functional time-series methods and two benchmark methods for SPX S\&P500 index data from Optionmetrics. Bivariate functional time-series are produced, consisting of IV smiles associated with 30 and 60 days to expiry (DTE). The moneyness is filtered to be between 90\% and 110\% only, with the testing sample covering the 15 July 2019 to 30 December 2019 period.}}\label{tab:SPX}
\begin{tabular}{@{}lllll@{}}
\toprule
Method & \multicolumn{2}{c}{DTE = 30} & \multicolumn{2}{c}{DTE = 60} \\
& $\overline{\text{MAFE}}$ & $\overline{\text{MSFE}}$ & $\overline{\text{MAFE}}$ & $\overline{\text{MSFE}}$ \\ \midrule
FTS & 0.0984 & 0.0186 & 0.0952 & 0.0165 \\
DFTS & 0.0979 & 0.0182 & 0.0946 & 0.0164 \\
MFTS & 0.1108 & 0.0209 & 0.0965 & 0.0175 \\
DMFTS & 0.1007 & 0.0207 & 0.0937 & 0.0173 \\
MLFTS & 0.0979 & 0.0183 & 0.1068 & 0.0210 \\
DMLFTS & 0.0985 & 0.0186 & 0.1040 & 0.0204 \\
RW & 0.1075 & 0.0217 & 0.1061 & 0.0207 \\
AR(1) & 0.1054 & 0.0191 & 0.0964 & 0.0174 \\
\bottomrule
\end{tabular}
\end{table}

\section{Conclusions and future research}\label{sec:6}

We propose dynamic, multivariate and multilevel functional time-series extensions to model and forecast currency option IV. When focusing on in-sample fit, we find that incorporating correlation among maturities leads to improved accuracy, with the multivariate approach showing high $R^2$ values. However, when we move to an out-of-sample forecasting environment, the use of dynamic estimation when producing functional principal components leads to the greatest benefit, as measured by our adopted $\overline{\text{MAFE}}$, $\overline{\text{MSFE}}$, MME(U) and MME(O) criteria. The dynamic univariate functional time-series approach better captures the time-series momentum present in the daily IV surface observations, as opposed to the static approach used by \cite{KCM18}. Further, the best functional time-series approaches also generally outperform the leading benchmark models employed. The model confidence set procedure of \cite{HLN11} confirms the statistical significance of improvements in point forecast errors. Further, to enhance the robustness of our empirical results, various currencies and out-of-sample windows are considered. Finally, to highlight a possible economic benefit of the functional time-series models, a simple straddle based stylised trading strategy is employed, which shows significant returns for two of the three currencies considered.

The current paper can be extended in a number of ways. We briefly mention three. First, the multivariate and multilevel functional time-series methods could be applied to simultaneously model and forecast other financial data, for example, related futures price curves. Further, if the IV smile of a single maturity is of interest, the frameworks could be adapted to focus on dependencies between currencies in a multi-currency portfolio, as opposed to across various maturities as we have done. Second, we treat moneyness as a continuum of our functional time-series for each of the three maturities. A possible extension is to also allow maturity to be a continuous variable. To some extent, this problem leads to high-dimensional functional time-series forecasting \citep[see, e.g.,][]{GSY19, TNH20, CCQ20}. Finally, future studies could consider a sieve bootstrap procedure of \cite{PS20} to construct prediction intervals of the three functional time-series methods, and to compare their interval forecast accuracy.

%\section*{Acknowledgments}

%The authors are grateful for the insightful comments and suggestions of two reviewers, which led to a much-improved version of the manuscript. Thanks also go to the comments and suggestions of the participants at the $37^{\text{th}}$ International Symposium on Forecasting, and those of the seminar participants at the Department of Econometrics and Business Statistics at Monash University. The first author acknowledges the financial support of a research grant from the College of Business and Economics at the Australian National University.

%\newpage
\bibliographystyle{agsm}
\bibliography{currency_exchange}

\end{document}